\RequirePackage{fix-cm}
\documentclass[smallextended]{svjour3}    
\smartqed  
\usepackage{graphicx}
\usepackage{url}
\usepackage{latexsym}
\usepackage{color}
\usepackage[round]{natbib}
\usepackage{txfonts}
\usepackage{amssymb}

\definecolor{aurometalsaurus}{rgb}{0.43, 0.5, 0.5}
\definecolor{ao(english)}{rgb}{0.0, 0.5, 0.0}
\definecolor{seagreen}{rgb}{0.180392,0.545098,0.341176}
\definecolor{forestgreen}{rgb}{0.133333,0.545098,0.133333}

\newcommand{\etal}{et al.}
\newcommand{\degree}{$^\circ$~}

\newcommand{\wse}{{\it WISE}}
\newcommand{\chn}{{\it Chandra}}
\newcommand{\swf}{{\it Swift}}
\newcommand{\suz}{{\it Suzaku}}
\newcommand{\fer}{{\it Fermi}}
\newcommand{\aap}{Astron. Astrophys.}
\newcommand{\aapr}{Astron. Astrophys. Rev.}
\newcommand{\aaps}{Astron. Astrophys. Suppl. Ser.}
\newcommand{\aj}{Astronom.~J.}

\newcommand{\apj}{Astrophys.~J.}
\newcommand{\apjl}{Astrophys.~J. Lett.}
\newcommand{\apjs}{Astrophys.~J. Suppl. Ser.}
\newcommand{\apss}{Astrophys. Space Sci.}

\newcommand{\mnras}{Mon. Not. Roy. Astr. Soc.}

\newcommand{\nat}{Nature}

\newcommand{\pasp}{Pub. Astron. Soc. Pac.}

\newcommand{\pr}{Phys. Rev.}
\newcommand{\prl}{Phys. Rev. Lett.}

\newcommand{\rpp}{Rep. Prog. Phys.}

\newcommand{\sci}{Science}
\newcommand{\ssr}{Space Sci. Rev.}

\journalname{Astron. Astrophys. Rev.}
\begin{document}

\title{The Extragalactic Gamma-ray Sky in the \fer\ era}
\titlerunning{The Extragalactic Gamma-ray Sky in the \fer\ era}  

\author{Francesco Massaro  \and David J. Thompson \and Elizabeth C. Ferrara}
\institute{Francesco Massaro \at
Dipartimento di Fisica, Universit\`a degli Studi di Torino, via Pietro Giuria 1, 10125 Torino, Italy.  \\
 Tel.: +39-011-670-7462\\
 Fax: +39-011-670-7020\\
\email{f.massaro@unito.it} 
\and David J. Thompson \at
NASA Goddard Space Flight Center, Greenbelt, MD 20771 USA. \\
Tel.: +1-301-286-8168\\
Fax: +1-301-286-1682\\
\email{david.j.thompson@nasa.gov} 
\and Elizabeth C. Ferrara \at
NASA Goddard Space Flight Center, Greenbelt, MD 20771 USA. \\
Tel.: +1-301-286-7057\\
\email{elizabeth.c.ferrara@nasa.gov}
 }

\date{Accepted version.}

\maketitle

\begin{abstract}
The Universe is largely transparent to $\gamma$ rays in the GeV energy range, making these high-energy photons valuable for exploring energetic processes in the cosmos.  After seven years of operation, the \fer\ {\it Gamma-ray Space Telescope} has produced a wealth of information about the high-energy sky.  This review focuses on extragalactic $\gamma$-ray sources:  what has been learned about the sources themselves and about how they can be used as cosmological probes. Active galactic nuclei (blazars, radio galaxies, Seyfert galaxies) and star-forming galaxies populate the extragalactic high-energy sky. \fer\ observations have demonstrated that these powerful non-thermal sources display substantial diversity in energy spectra and temporal behavior.  Coupled with contemporaneous multifrequency observations, the \fer\ results are enabling detailed, time-dependent modeling of the energetic particle acceleration and interaction processes that produce the $\gamma$ rays, as well as providing indirect measurements of the extragalactic background light and intergalactic magnetic fields.  Population studies of the $\gamma$-ray source classes compared to the
extragalactic $\gamma$-ray background place constraints on some models of dark matter.  Ongoing searches for the nature of the large number of $\gamma$-ray sources without obvious counterparts at other wavelengths remains an important challenge.
\end{abstract}
\keywords{gamma rays; extragalactic astronomy; active galactic nuclei; quasars, BL Lac objects; background light}

\tableofcontents


\section{Introduction}
\label{sec:intro}

As the most energetic form of electromagnetic radiation, $\gamma$ rays can probe many of the most powerful, non-thermal, often explosive phenomena in the Universe, potentially including exotic processes like quantum gravity and dark matter self-annihilation. Unaffected by magnetic fields, these high-energy photons bring information directly from their sources, and, at energies even as great as 100 GeV, reach Earth from cosmological distances.  Direct detection of $\gamma$ rays must, however, be carried out from space platforms, because the atmosphere is opaque to photons at these energies.  For more than seven years, the international \fer\ {\it Gamma-ray Space Telescope}, originally known as Gamma-Ray Large Area Space Telescope (GLAST), has surveyed the entire $\gamma$-ray sky on a daily basis, vastly increasing our knowledge of the high-energy Universe.  Sources in our Galaxy seen by \fer\ include the Sun and Moon, pulsars, supernova remnants, high-mass binaries, and novae. While such objects are bright, they do not have sufficient intrinsic luminosity to be seen individually in other galaxies by the \fer\ instruments.  The extragalactic $\gamma$-ray regime that is the subject of this review is therefore dominated by the galaxies themselves, including normal, starburst, and active galactic nuclei (AGN). All these sources are seen against diffuse $\gamma$-ray backgrounds:  cosmic-ray particle interactions with the interstellar medium in our own Galaxy (including the ``Fermi Bubbles,'' huge structures that appear to be related to extreme outflows from the region of the Galactic Center), and an isotropic extragalactic $\gamma$-ray background whose origin is not fully understood. 

Before the launch of  the \fer\ {\it Gamma-ray Space Telescope} the previous window on the high-energy $\gamma$-ray sky was provided by  the 1991$-$2000 {\it Compton Gamma-Ray Observatory (CGRO)}, in particular with its Energetic Gamma Ray Experiment Telescope (EGRET) observing between 20 MeV and $\sim$30 GeV. EGRET showed the high-energy $\gamma$-ray sky to be surprisingly dynamic and diverse, with source classes ranging from the Sun and Moon to pulsars and supermassive massive black holes hosted in AGN at high redshifts. At that epoch our knowledge of the source classes known as $\gamma$-ray emitters was mainly limited to pulsars, among the Galactic objects, the nearby galaxy the Large Magellanic Cloud, and {\it blazars} in the extragalactic sky. In the zoo of active galaxies, blazars (flat-spectrum radio quasars (FSRQs) and BL Lac objects) are radio-bright  objects whose emission, at all frequencies, is due to interactions of high-energy particles accelerated in a relativistic jet pointing toward the Earth, making them the rarest AGN class. EGRET detected $\gamma$-rays from about 60 blazars. As expected, \fer\ discovered and is still discovering high-energy emission from more than 1000 blazars. Many of the $\gamma$-ray sources detected by EGRET remain unidentified even years after the release of its last catalog. 

In 2008, with the advent of \fer\, a new $\gamma$-ray era began \citep{abdo10a}. The \fer\ primary instrument, the Large Area Telescope (LAT), has a much larger effective area and solid angle than EGRET, a broader energy range, no consumables, vastly smaller dead time per trigger, and significantly better angular resolution, particularly at  higher energies.  While EGRET took a year and a half to map the whole $\gamma$-ray sky once, \fer\ has produced an all-sky map about every three hours since the beginning of the mission. The discoveries collected in the last seven years by \fer\ have been revolutionary and in several cases also completely unexpected. Thanks to the \fer\ observing strategy that enables access to the whole sky every $\sim$ 3 hours, we have had the unique opportunity to investigate the nature of $\gamma$-ray source populations on a statistical basis.  \fer--LAT has detected more than an order of magnitude more sources than all previous $\gamma$-ray missions.  In addition, since the $\gamma$-ray sky is  known to be the most  variable across the whole electromagnetic spectrum, \fer\ has permitted us to study the temporal behavior of the $\gamma$-ray sources on a wide variety of time scales.

The study of the non-thermal $\gamma$-ray Universe is intrinsically a multiwavelength endeavor. A vast suite of multifrequency observations has enriched the data sets available to compare the spectral and temporal evolution of the $\gamma$-ray sources with those at lower energies. Moreover this assembly of multifrequency data collected to investigate the nature of the \fer\ sources improved the associations of $\gamma$-ray sources with their lower-energy counterparts.These campaigns were motivated by the large fraction of unidentified sources seen by EGRET, and thus many of them started even before the \fer\ launch.  \fer\ is also not alone in studying the $\gamma$-ray sky:  the small Italian satellite {\it Astro-rivelatore Gamma a Immagini LEggero (AGILE)}, launched a year before \fer\, uses similar technology to the LAT \citep{tavani08} {\it AGILE} results have contributed to a number of the studies of extragalactic $\gamma$-ray sources. 

This review is focused on the results achieved by \fer\  for the extragalactic sky during its first seven years of operation. We discuss the source classes that populate the high energy sky in the MeV-GeV range, except for the transient objects that belong to the $\gamma$-ray burst (GRB) class.

\subsection{The extragalactic $\gamma$-ray sky before \fer: an historical overview}
\label{sec:history}

The first indication of extragalactic high-energy photons came from the {\it Orbiting Solar Observatory 3 (OSO-3)}, which found E$>$50 MeV  $\gamma$-rays arriving from all directions on the sky \citep{kraushaar72}.  Data from the {\it Small Astronomy Satellite 2 (SAS-2)} confirmed this result, showing that an isotropic, apparently extragalactic component could be distinguished from the diffuse $\gamma$-ray emission originating in our Galaxy \cite[e.g.][]{thompson82}.  In the same time frame, the {\it Celestial Observation Satellite (COS-B)} telescope discovered the first extragalactic $\gamma$-ray source, the nearby quasar 3C 273 \citep{swanenburg78}, suggesting the possibility that the diffuse extragalactic radiation might actually consist of unresolved active galaxies, particularly Seyfert galaxies, BL Lac objects, and quasars \citep{bignami79}.  

The first all-sky high-energy $\gamma$-ray survey was carried out by the EGRET instrument on the {\it CGRO}. Results for $\gamma$-ray sources seen with EGRET are summarized in the Third EGRET Catalog \cite[3EG][]{hartman99} and an alternate catalog \cite[EGR][]{casandjian08}, and an overview of all EGRET results can be found in \citet{thompson08}.  Some principal results about the extragalactic $\gamma$-ray sky can be summarized as follows:
\begin{enumerate}
\item Nearly all the EGRET extragalactic sources are blazar-class AGN.  These $\gamma$-ray blazars are highly variable on time scales down to less than one day and are seen out to distances beyond redshift $z$=2. 
\item No Seyfert galaxies, starburst galaxies, or clusters of galaxies appear in the 3EG catalog.
\item The Large Magellanic Cloud and Centaurus A (NGC 5128), a normal satellite galaxy of the Milky Way and a nearby radio galaxy, were also detected in  high-energy 
 $\gamma$ rays.
\item A diffuse extragalactic $\gamma$-ray background is seen, with a spectrum that depends on the details of how the diffuse Galactic radiation is subtracted. 
\item GRBs, although not covered by this review, are primarily a hard X-ray phenomenon but are also seen in the high-energy $\gamma$-ray sky.
\end{enumerate}

\subsection{The \fer\ mission}
\label{sec:technical}

\begin{figure}[]
\includegraphics[height=7.cm,width=13.cm,angle=0]{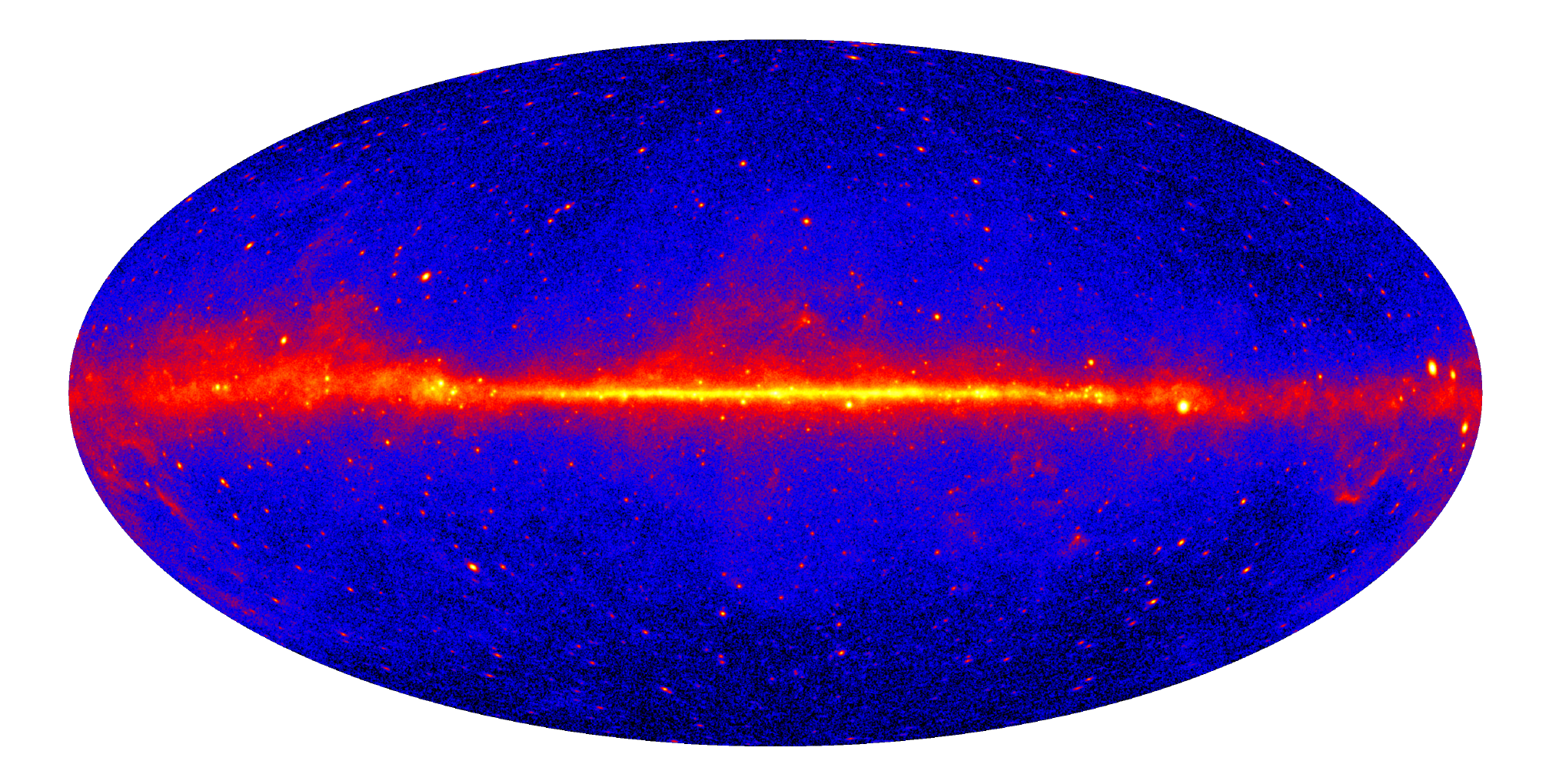}
\caption{The $\gamma$-ray sky seen by \fer\ after 60 months of operation at energies grater than 1 GeV, shown in Galactic coordinates in a Hammer-Aitoff projection. Brighter colors indicate higher $\gamma$-ray intensities. (Figure courtesy of the \fer\ Large Area Telescope Collaboration).}
\label{fig:LATsky}
 \end{figure}

The {\it Fermi Gamma-ray Space Telescope}, launched into a low-Earth orbit 11 June 2008, is an international and multi-agency space mission built to investigate the cosmos in the energy range 10 keV to $>$300 GeV \citep[see e.g.,][]{michelson10}. The {\it Fermi}-LAT, the principal instrument,  operates from $20$ MeV to $>300$ GeV. Similar to detectors used at particle accelerators, the LAT is an array of $4 \times 4$ identical towers, each  consisting of a silicon-strip tracker (where the $\gamma$-ray interacts by pair production,  $\gamma$ $\rightarrow$  $e^+$ +  $e^-$, and the resulting particles are tracked) and a cesium iodide calorimeter (where the energy of the particles and hence the pair-converted photon is measured). The  instrument is covered by a plastic scintillator  anticoincidence detector for rejecting the huge charged-particle background found in the low-Earth orbit of the satellite. The sensitivity of the LAT is greatest around 1 GeV, where the effective area is about $0.8$\,m$^2$, the field of view (FoV) is about $ 2.4$\,sr, the energy resolution is better than $10\%$, and the single-photon angular resolution ($68\%$ containment angle) is better than 1$^{\circ}$. 
Further details about the  LAT instrument are given by \citet{atwood09}.  The other instrument on board, the Gamma-ray Burst Monitor \cite[GBM;][]{meegan09}, has a FoV several times larger than the LAT and provides spectral and temporal coverage of transients, mostly GRBs, from $\sim$10 keV $-$ 40 MeV, overlapping the LAT energy range. \fer\ normally operates in sky-survey mode,  in which the whole sky is observed every 2 orbits ($\sim$3 hours).  Figure \ref{fig:LATsky} shows the $\gamma$-ray sky as observed by the LAT.  The bright horizontal strip is the plane of the Milky Way, which is a bright source of diffuse $\gamma$-ray emission from interactions of cosmic rays with interstellar gas and photons. 

Since the beginning of \fer\ operations the LAT instrument team has released one bright $\gamma$-ray source list \cite[0FGL;][]{abdo09b} and three full source catalogs: 1FGL, 2FGL and 3FGL, standing for first, second and third \fer\ Gamma-ray LAT source catalog \cite[see][respectively]{abdo10a,nolan12,acero15}.  Each represents a cumulative analysis of the \fer--LAT data since the beginning of the mission:  0FGL, 3 months; 1FGL, 11 months; 2FGL, 24 months; and 3FGL, 48 months.  In addition to more data, each new release has incorporated improved analysis methods and a refined model of the diffuse  $\gamma$-ray backgrounds \citep[e.g., ][]{ackermann12a}. 

Each of these publications has been accompanied by a corresponding publication focused on  AGN detected in $\gamma$ rays: \cite[LAT Bright AGN Sample; LBAS][]{abdo09a} and the first, second and third \fer--LAT AGN Catalogs: 1LAC, 2LAC and 3LAC \cite[e.g.][respectively]{abdo10b,ackermann11,ackermann15a}.   Analysis of \fer--LAT sources in the LAT AGN catalogs concentrates on those at high Galactic latitudes ($|b|>$10\degree) to reduce the contamination from Galactic diffuse emission and sources.

\begin{figure}[]
\includegraphics[height=9.cm,width=12.cm,angle=0]{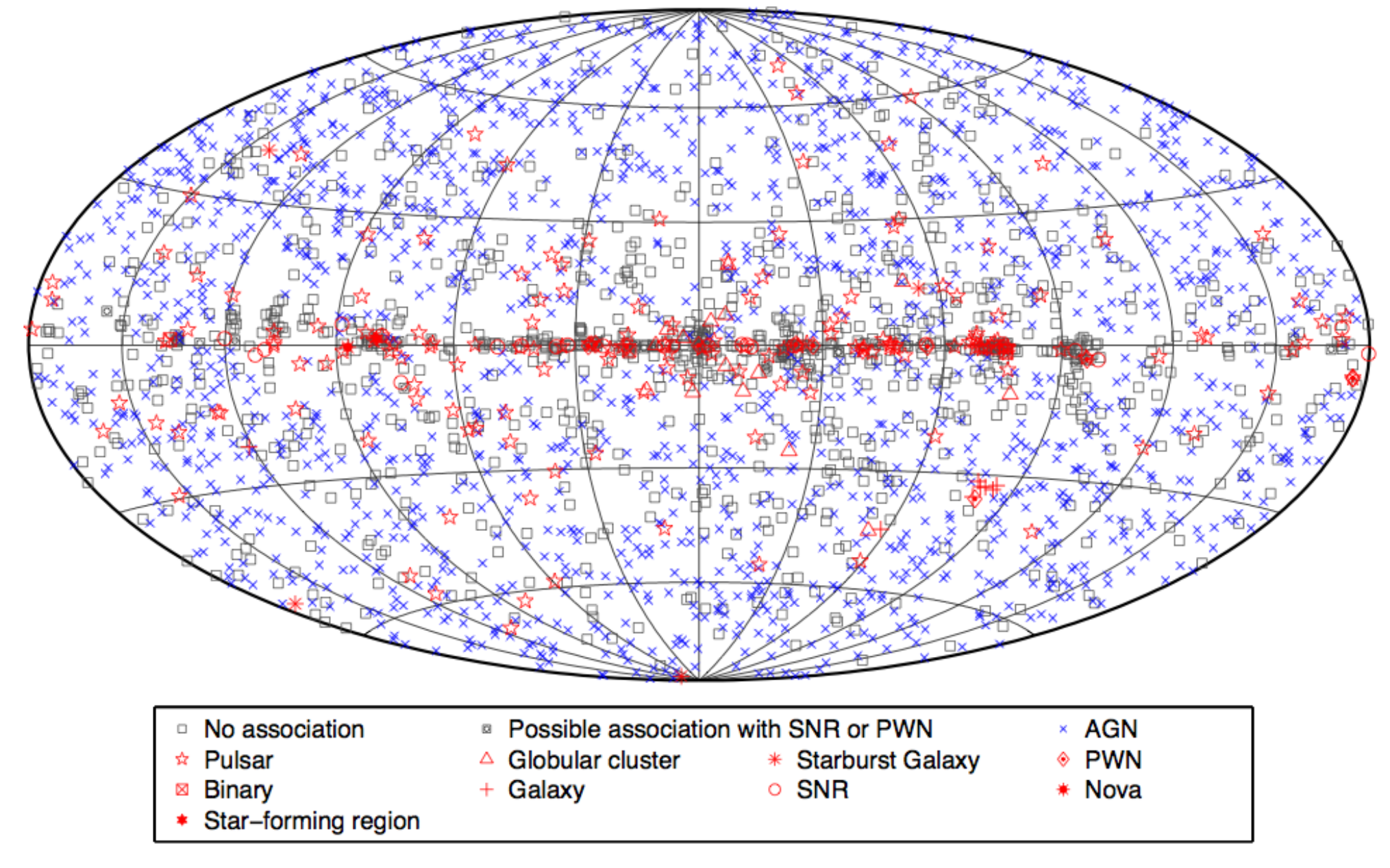}
\caption{The sky distribution of the sources detected by \fer--LAT and listed in the 3FGL catalog (Hammer-Aitoff projection), shown in Galactic coordinates \citep{acero15}. Labels for the different source classes are also reported. (Figure courtesy of the \fer\ Large Area Telescope Collaboration).}
\label{fig:3FGL}
 \end{figure}

\begin{figure}[]
\includegraphics[height=7.cm,width=13.cm,angle=0]{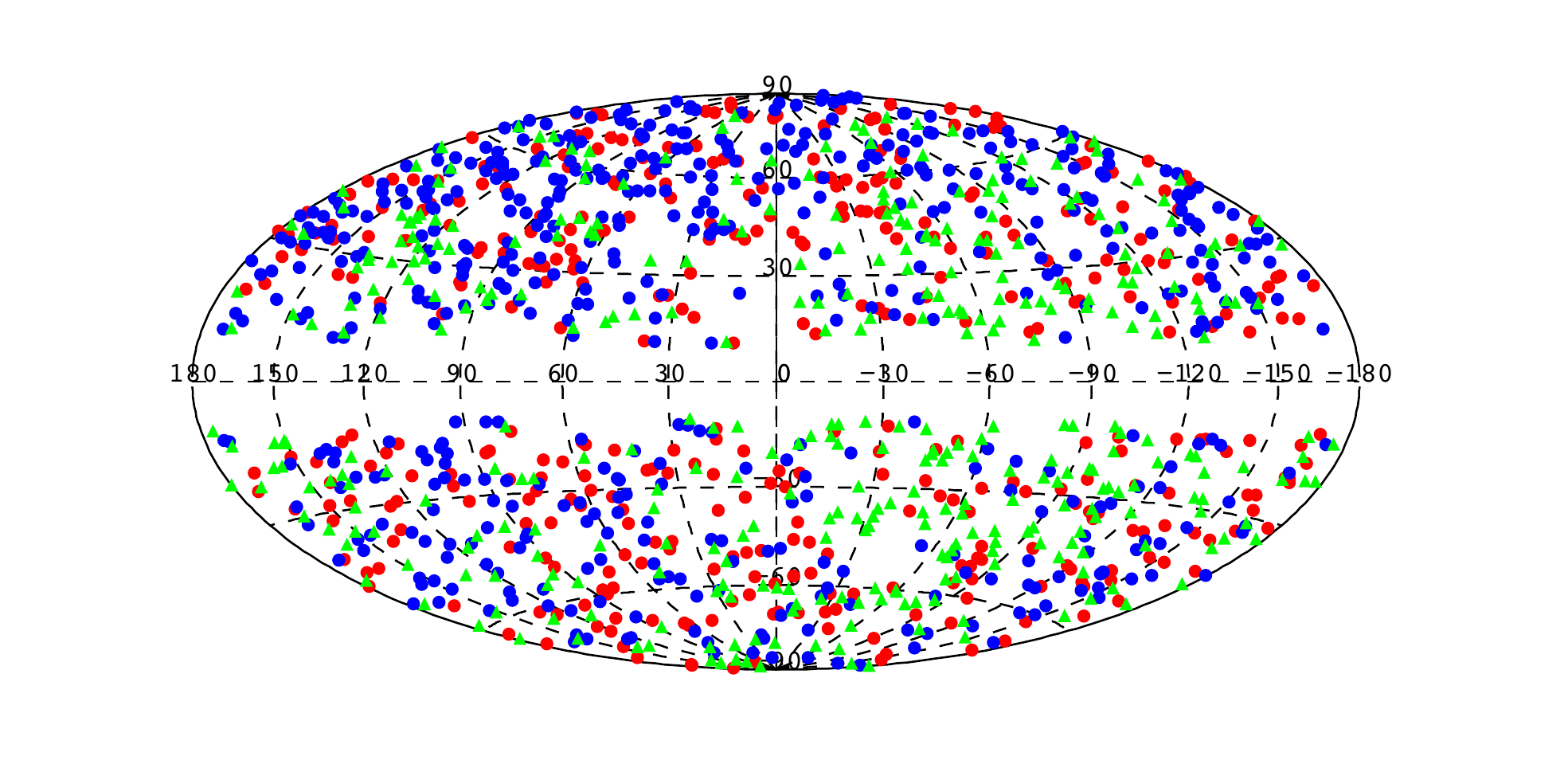}
\caption{The sky distribution of the blazars listed in the 3LAC (Hammer-Aitoff projection), shown in Galactic coordinates  \citep{ackermann15a}.  Red circles: FSRQs; blue circles: BL Lacs; green triangles: blazar candidates of uncertain type (BCUs). (Figure courtesy of \fer\ Large Area Telescope Collaboration).}
\label{fig:3LAC}
 \end{figure}

The most recent results are the Third \fer--LAT Source Catalog \cite[3FGL][]{acero15}, containing 3033 sources, and the Third Catalog of Active Galactic Nuclei Detected by the \fer--LAT \cite[3LAC;][]{ackermann15a}.  The 3LAC lists 1591 AGN  at high Galactic latitudes (i.e., $|b|>$10\degree), corresponding to a $\sim$70\% increase in the number of AGN compared to the 2LAC. The distribution of the sources listed in both the 3FGL and the 3LAC, shown in Figures~\ref{fig:3FGL} and ~\ref{fig:3LAC}, provides the latest information on the extragalactic content of the \fer\ survey.  For comparison, the 3EG catalog \cite{hartman99} had a total of 271 sources, at least 60 of which were blazars, and an alternate EGRET catalog, released well beyond the end of the {\it CGRO} operations \citep[3EGR;][]{casandjian08} listed 188 $\gamma$-ray sources. 

\subsection{Hunting low-energy counterparts of the \fer\ sources}
\label{sec:hunt}

A systematic search for lower-energy counterparts of the $\gamma$-ray sources has long been a challenging task \cite[e.g.][]{mattox97}. The  underlying reason is that $\gamma$ rays cannot be reflected or refracted, thus detection via the pair-production interaction in a converter-tracker results in unavoidably limited precision in measurement of the incident direction. The result is that the positional uncertainties of the celestial objects detected in $\gamma$ rays can be orders of magnitudes larger than those in the X-ray, optical-infrared(IR) or radio bands. Despite the improvements achieved by the \fer--LAT compared to previous instruments, these uncertainties are too large to identify sources strictly by positional coincidence on the sky. The typical positional uncertainty in the 3EG catalog was $\sim$0.7 degrees while that in the 3FGL is about 0.1 degrees, thus the \fer\ improvement is about a factor of 7 in diameter, or almost a factor of 50 in area. 

\begin{figure}[]
\includegraphics[height=6.1cm,width=6.3cm,angle=0]{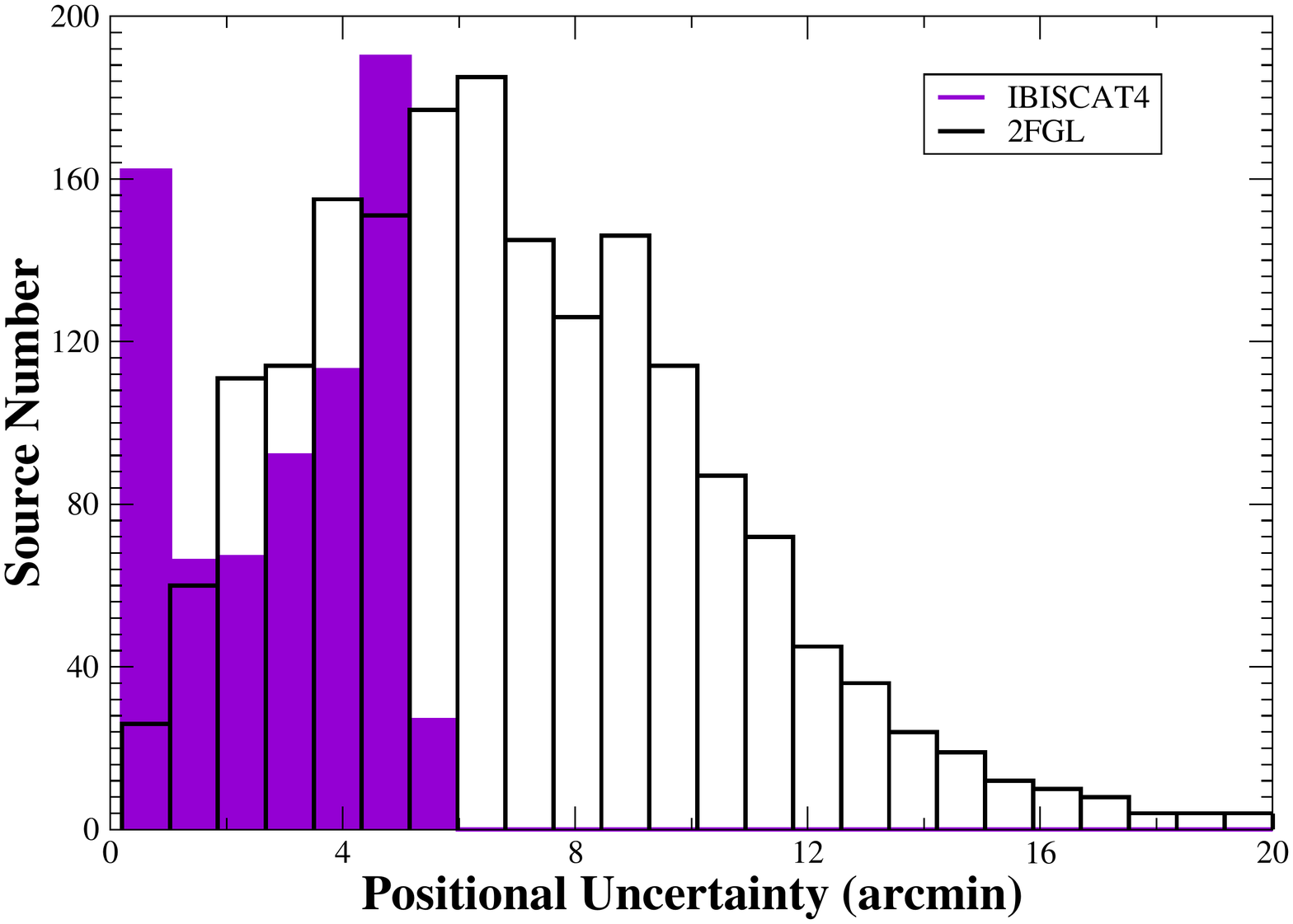}
\includegraphics[height=6.1cm,width=6.3cm,angle=0]{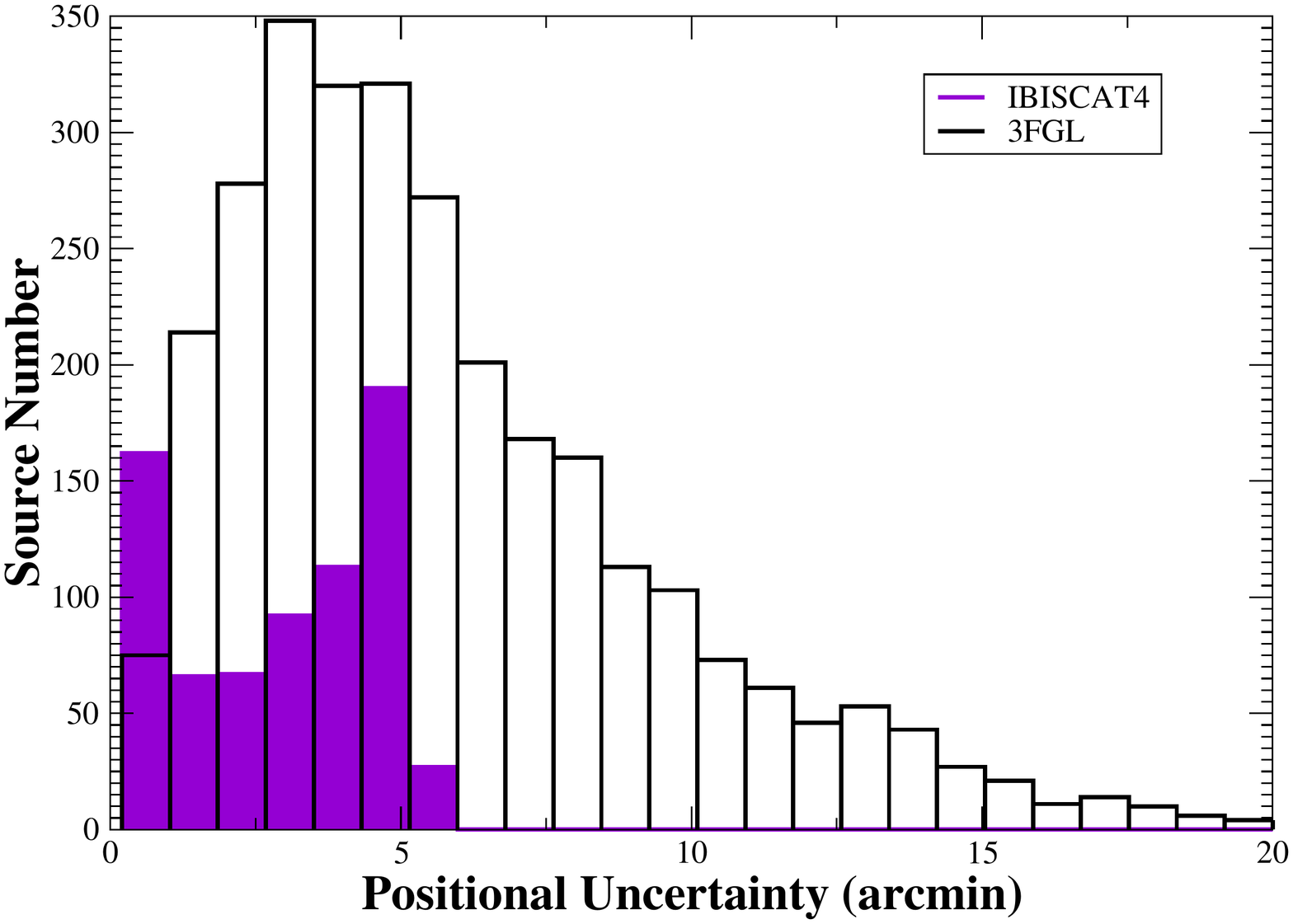}
\caption{Distributions of the positional uncertainty (semi-major axis) at the 95\% level of confidence for the \fer--LAT sources (black) that belong to the 2FGL (left panel) and those of the 3FGL (right panel) catalog \cite[see][respectively]{nolan12,acero15} and the positional error circles of the hard X-ray sources listed in the 4$^{th}$ {\it INTEGRAL}-IBIS catalog (violet).}
\label{fig:posunc}
 \end{figure}

\begin{figure}[]
\includegraphics[height=8.cm,width=12.cm,angle=0]{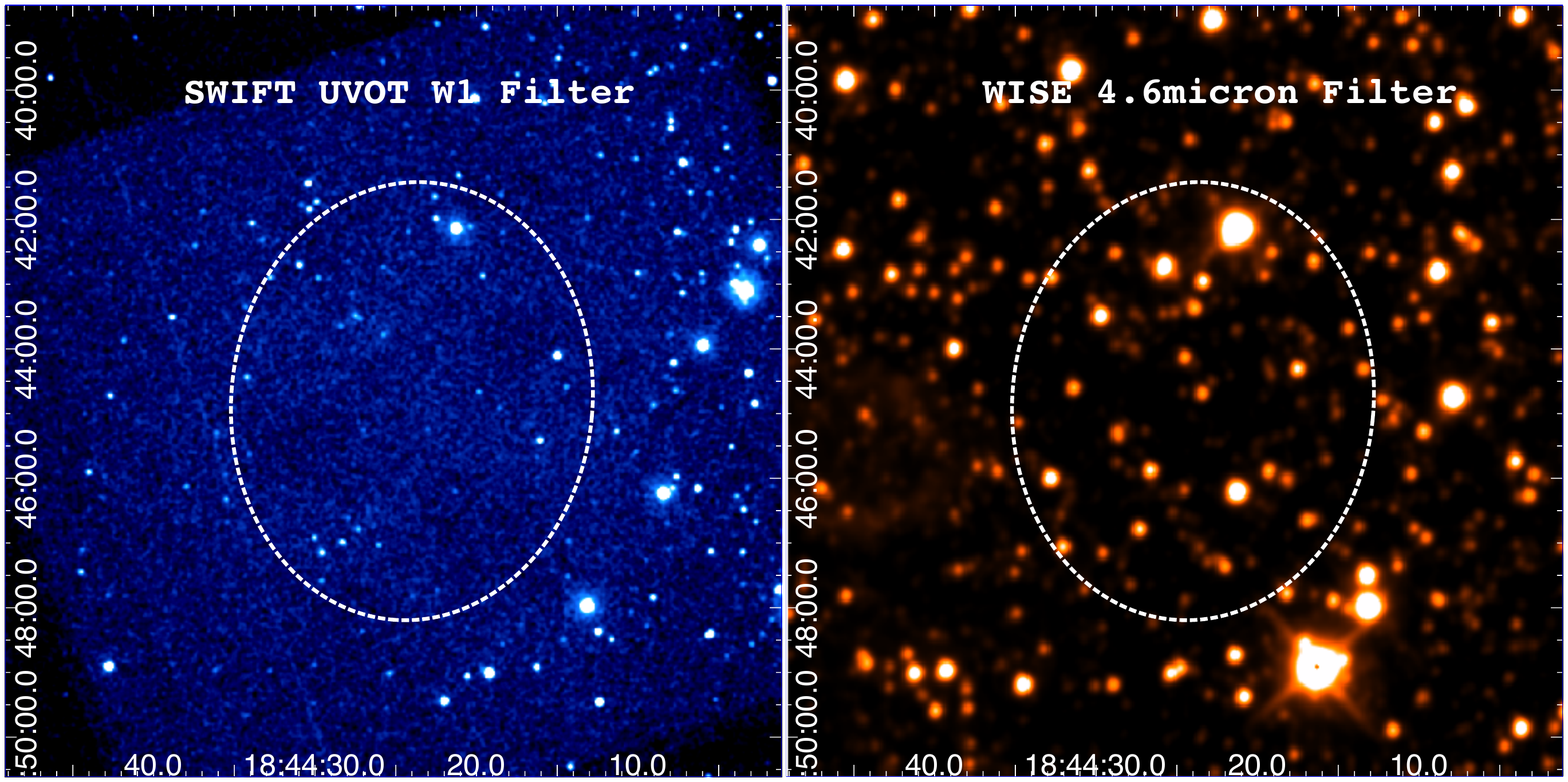}
\caption{Ultraviolet and mid-IR images of the unidentified $\gamma$-ray source  3FGL J1844.3$-$0344. The \fer\ positional uncertainty region at the 95\% level of confidence (white dashed ellipse) is overlaid on the {\it Swift}-UVOT image taken with the W1 filter (left panel) and to the {\it WISE} image at 4.6$\mu$m (right panel) to highlight the large number of potential low-energy counterparts of the \fer\ source that appear in the UV and IR bands.  It could be possible that the $\gamma$-ray emission is due to a source not even detected at those frequencies, as for example could occur for pulsars.}
\label{fig:3FGLJ1844}
 \end{figure}

The position of a $\gamma$-ray source listed in the \fer--LAT catalogs is reported with an associated uncertainty that corresponds to an elliptical region \cite[e.g., 1FGL,][]{abdo10a}. The semi-major and semi-minor axes of the ellipse together with the positional angle are listed in these $\gamma$-ray catalogs at the 95\% level of confidence. The distributions of the semi-major axes for the 2FGL \citep{nolan12}  and the 3FGL \citep{acero15} sources in comparison with those listed in the 4$^{th}$ catalog of the {\it INTErnational Gamma-Ray Astrophysics Laboratory (INTEGRAL)} performed with the Imager on-Board the {\it INTEGRAL} Satellite (IBIS) instrument \citep{bird10} are shown in Figure~\ref{fig:posunc}.  Even with the visible improvement between 2FGL and 3FGL, the uncertainties are larger than those of hard X-ray surveys. 
The number of potential counterparts  within a typical $\gamma$-ray positional uncertainty region can be huge, from tens up to hundreds in optical/IR frequencies (e.g. Figure~\ref{fig:3FGLJ1844}). This number can increase significantly for regions close to the Galactic plane, since both the $\gamma$-ray positional uncertainties and the optical/IR source densities increase due to uncertainties on the knowledge of the Galactic $\gamma$-ray background. Evaluiating associations of $\gamma$-ray sources with their low-energy counterparts becomes an arduous task \citep[e.g.,][]{reimer05}.

The challenge of finding counterparts is somewhat mitigated by several considerations. Most importantly, any high-energy $\gamma$-ray object must have a significant non-thermal energy source.  Our Sun, for example, is only visible to \fer--LAT because it is so close by.  No other types of normal star are plausible candidates as a counterparts of \fer\ sources.  Many IR and X-ray sources are similarly dominated by thermal emission, making them highly unlikely to produce significant $\gamma$-ray radiation. The implication is that relatively few known astrophysical source classes are expected to be $\gamma$-ray emitters \citep[e.g.,][]{reimer05,reimer07a}, although we cannot dismiss exotic phenomena like annihilating dark matter (DM) clumps \citep[e.g., ][]{baltz08}. The spatial density (i.e., number of sources per square degree) of this restricted number of potential $\gamma$-ray source classes is also relatively low \cite[e.g.][]{abdo10a}. This set of considerations suggests that if one of these objects lies within the $\gamma$-ray positional uncertainty there is a reasonable chance that this is the low-energy counterpart of the $\gamma$-ray source. 

Since the time of the {\it CGRO} mission, $\gamma$-ray source association analyses have applied statistical methods. These techniques, together with new procedures, were also improved and used to build all the \fer--LAT catalogs available to date. However in all these catalogs there is an important distinction between {\it identification} of low-energy counterparts for the sources and {\it association} \cite[e.g.][]{abdo10a,nolan12}. Identification is based on i) correlated variability at at least two other wavelengths, or on ii) spin or orbital periodicity, or on iii) the consistency between the measured angular sizes in $\gamma$ rays and at lower energies.

The {\it association} designation  depends on the specific procedure adopted.  Examples used in the \fer--LAT catalogs are:
\begin{enumerate}
\item {\it The Bayesian Association Method}: Initially applied to associate EGRET sources with flat-spectrum radio sources \citep{mattox97}, this method assesses the probability of association between a $\gamma$-ray source and a candidate counterpart taking into account their local densities. The local density is estimated by counting candidates in a nearby region of the sky.  The most detailed description of how this method is applied to \fer--LAT catalogs, including the choices of candidate source classes, is found in the 1FGL paper \citep{abdo10a}.
\item {\it The Likelihood Ratio (LR) method}: Used to search for possible counterparts in uniform surveys in the radio and in X-ray bands, the LR approach adds source flux information to the analysis.  It was first applied to $\gamma$-ray data as a secondary association method specifically for AGN in the 2FGL and 2LAC catalogs \citep{nolan12,ackermann11}.
\item {\it The logN-logS association method}: This is a modified version of the Bayesian method for blazars, used in the 2LAC catalog, taking into account the distribution of radio continuum fluxes of blazars \cite[e.g.][and references therein]{ackermann11}.
\end{enumerate}

These procedures assign a value for the probability that the source is the ``real'' counterpart for each \fer--LAT object. Then only high-confidence associations, selected above a certain threshold, are listed in the catalogs. The 80\% confidence level adopted by the LAT team is fairly conservative.  Some real associations will be missed, and not every counterpart is guaranteed to be correct.  Taken as a whole, however, these associations provide a well-defined sample of $\gamma$-ray AGN that can be used for population studies. 

The above methods to assign low-energy counterparts depend on the densities of sources in the catalogs of potential counterparts.  Many of these catalogs are growing as new observations are made. A better estimate of the counterpart density leads to a better estimate of the false positive associations and of the association probability, and this implies that a previously unassociated source could have an assigned counterpart in a new release of the \fer--LAT catalog.  To increase the number of associated \fer--LAT sources, different multifrequency follow-up campaigns have also been carried out and are continuing \citep[e.g.,][]{massaro15b}.  Most of these studies search for potential counterparts directly in the \fer--LAT source error regions (see Section~\ref{sec:ugs}).  The combination of improved counterpart catalogs and dedicated searches has led to larger numbers of associated sources in each iteration of the \fer--LAT catalogs.  The 3FGL catalog, for example, contains more associated sources (2024) than the total number of sources in the 2FGL catalog (1873).  Nevertheless, a rather steady $\sim$30\% of the sources in each \fer--LAT catalog have no plausible associations: 1010 in 3FGL. 

Sources with no assigned low-energy counterpart are historically called unidentified $\gamma$-ray sources (UGS), however given the  distinction between identification and association introduced during the \fer\ era, a more appropriate definition is unassociated $\gamma$-ray sources (UGS, same acronym). These UGS represent an ongoing challenge and opportunity for $\gamma$-ray astrophysics \citep[e.g.,][]{reimer05,thompson08}.  The sky distribution of these UGS, although showing some excess in the Galactic plane, is mostly uniform, making these  potentially extragalactic sources.
Section~\ref{sec:ugs} describes some of the searches for counterparts other than those used to find blazars, the most likely candidate for any \fer\ source. 

Motivated by the large fraction of unidentified EGRET sources, astronomers started a number of campaigns to create catalogs that could be simplify the association task, particularly for blazars. In 2005 the first release of the Multifrequency Catalog of blazars, Roma-BZCAT was presented. This catalog meant to serve as a list of carefully checked blazars to be used for selecting potential counterparts of $\gamma$-ray sources. The largest number of associations in the 3FGL comes from the latest versions of this catalog \citep[see e.g.,][for more details]{massaro09,massaro11,massaro15c}. 

Since a flat radio spectrum is generally believed to be a signature of jet emission from radio galaxies and/or blazar-like sources, during the EGRET mission \citet{mattox97} used this feature to associate sources, a procedure that was extended by Sowards-Emmerd et al. (2003,2005) to expand the number of EGRET blazars. \citet{healey07} combined radio data from the major surveys with new observations obtained at the Very Large Array (VLA) and at the Australia Telescope Compact Array (ATCA) to produce the Combined Radio All-Sky Targeted Eight GHz Survey (CRATES). This catalog listed flat-spectrum radio sources with nearly uniform distribution at Galactic latitudes greater than 10 degrees for objects brighter than 65 mJy at $\sim$5 GHz.  It has been extensively used in all the \fer\ catalogs for blazar associations. A restricted sample of the CRATES sources, generated on the basis of multifrequency observations available in the literature such as X-ray and/or optical spectroscopic data, has also been used for the same purpose \citep{healey08}. The only limitation on the use of flat-spectrum sources is that this spectral signature is not sufficient to identify blazar-like objects, and thus additional information to identify the source nature is necessary, motivating large, ongoing optical campaigns  to characterize such sources more completely \cite[e.g.,][]{shaw13a,shaw13b}. 

An additional problem occurring when low-energy counterparts are assigned to \fer\ sources is the presence of duplicate associations. These are rare cases where two  or more celestial objects (Galactic or extragalactic) belonging to a known class of $\gamma$-ray emitters lie within the positional uncertainty of the a \fer\ source. In these cases the $\gamma$-ray emission/spectrum could be due to one of them or both, thus showing a unusual $\gamma$-ray behavior. 

In conclusion, the large suite of multifrequency databases built before and during the \fer\ mission, combined with the improvements achieved in the association procedures, has been successful, particularly in finding AGN associations.  The fraction of unassociated sources has decreased substantially compared to the EGRET era.

\subsection{The extragalactic component of the $\gamma$-ray sky seen by \fer}
\label{sec:sky}

Our current view of the extragalactic \fer\ sky includes only a handful of source classes. Among AGN, $\gamma$-ray emission is detected for blazars (actually $\sim$98\% of the associated \fer\ objects), some special types of Seyfert galaxies, radio galaxies, and a few steep-spectrum radio quasars. On the other hand, for more normal galaxies, whose  emissions come primarily from star formation, \fer--LAT has detected only those in our neighborhood: Andromeda (i.e., M\,31) and the Small and the Large Magellanic Clouds, as well as a few nearby starburst galaxies such as M\,82 and NGC\,253. Table \ref{tab:summary} summarizes the contents of the \fer--LAT catalogs. 

Among the $\gamma$-ray blazar sample, the number of \fer--LAT BL Lac objects is larger than that of FSRQs. BL Lac objects generally show harder (flatter) energy spectra in the MeV-GeV energy range \citep[e.g.,][]{acero15} and can be detected more easily than FSRQs at a given significance limit when increasing the exposure. About half  the BL Lac objects  in  3FGL/3LAC still lack a redshift estimate, a usual problem for this class of AGN. The most distant 3LAC blazar detected in $\gamma$ rays is the same one found in the 1LAC catalog: PKS\,0537-286 at z=3.104 \citep{osmer94}. The growing fraction of $\gamma$-ray BL Lac objects compared to FSRQs going from 2FGL to 3FGL  was predicted on the basis of the shape of the logN-logS distributions.

Since most of the 3FGL associations have been found in radio, IR and X-ray catalogs where a large fraction of sources is not yet spectroscopically classified, a significant number of the \fer\ classifications remain somewhat uncertain.  Nearly 30\% of the entire 3LAC AGN sample is classified as blazar candidate of uncertain type (BCU), a change from the terminology of 2LAC that referred to AGN of Uncertain type \citep[AGU;][]{ackermann11}. These are sources associated using the standard procedures and considered as blazar-like but for which insufficient multifrequency information is available to make a more definite classification. Optical spectroscopic measurements are required to verify their true nature and distances.

\begin{figure}[]
\includegraphics[height=8.cm,width=12.cm,angle=0]{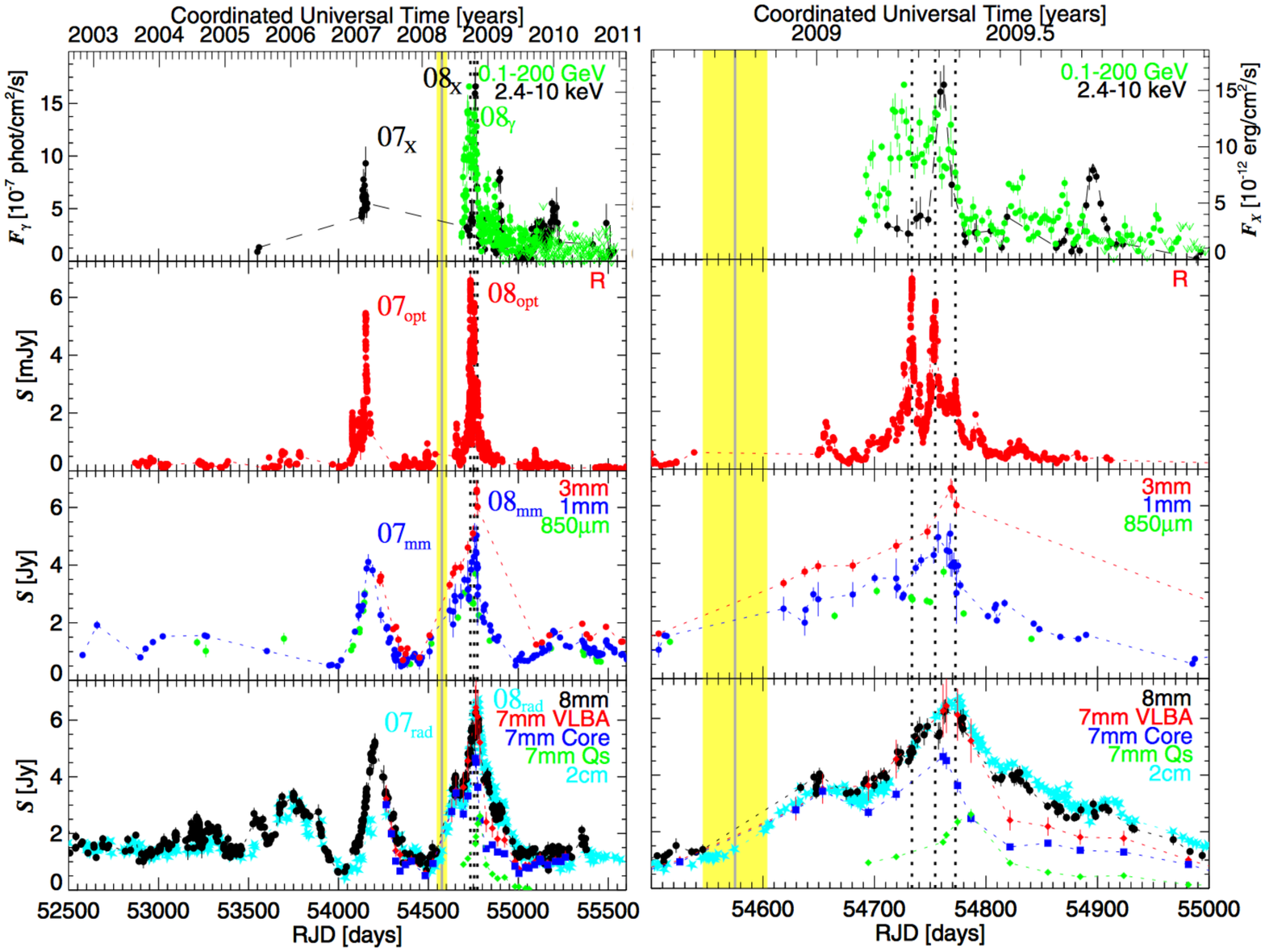}
\caption{Left: light curves of 0235+164 from $\gamma$-rays (top) to millimeter (bottom) energies. Vertical dotted lines mark the three most prominent optical peaks. The yellow area represents the epochs of ejection of radio knot features visible on parsec scale. RJD = Julian Date - 2400000.0. Right: same as left panel for RJD in the range [54500, 55000] \citep[see][for more details]{agudo11}. (Figure courtesy of Dr. I. Agudo, under the copyright permissions of the Astrophysical Journal).}
\label{fig:ao0235}
 \end{figure}

Gamma-ray variability is detected for many $\gamma$-ray blazars.  These variable blazars are often targets of multifrequency follow-up campaigns at lower energies  \citep[e.g.,][and Section~\ref{sec:optgamma} for more details]{marscher10,jorstad10}. Some radio galaxies also show flaring activity in the GeV energy range correlated with emission at lower energies. This peculiar behavior has been detected for example in NGC 1275 \citep{kataoka10}, PKS 0521$-$36 \citep{dammando15c} and 3C120 \citep{tanaka15}. Figure~\ref{fig:ao0235}  shows the extreme multifrequency behavior of AO0235+164 \citep{agudo11} between 2007 and 2008, close to the beginning of the \fer\ mission.

Variability may account for the fact that  several non-blazar, \fer--LAT AGN that were clearly detected in the 2LAC are now not included in the 3LAC. In addition the recent 3LAC analysis also showed that there are no strong differences in the radio, optical and X-ray flux distributions between $\gamma$-ray detected and non-detected blazars. This result suggests that all known blazars could potentially be detected by \fer--LAT.
\begin{table*}
\caption{Summary of associated extragalactic sources for each class as listed in the \fer--LAT catalogs. Numbers in parenthesis indicate the sources {\it identified}.}
\label{tab:summary}
\begin{tabular}{|lrrr|}
\hline
Source class  & 1FGL & 2FGL & 3FGL \\
\hline 
\noalign{\smallskip}
Normal galaxy$^1$                      &   2 (2)  &   2  (2) &   3  (2) \\  
Star forming galaxy                    &   0      &   4  (0) &   4  (0) \\
BL Lac object                          & 295 (0)  & 436  (7) & 660 (18) \\
Flat spectrum radio quasar             & 274 (4)  & 370 (17) & 484 (38) \\
Blazar candidate of Uncertain type     &  92 (0)  & 257  (0) & 573  (5) \\
Non-blazar active galaxy               &  28 (0)  &  11  (1) &   3  (0) \\
Radio galaxy                           &   0      &  12  (2) &  15  (3) \\
Seyfert galaxy$^2$                     &   0      &   6  (1) &   6  (2) \\
Compact steep spectrum radio quasar    &   0      &   0      &   1  (0) \\
Steep spectrum radio quasar             &   0      &   0      &   3  (0) \\
\noalign{\smallskip}
\hline 
\noalign{\smallskip}
Unassociated                           & 630      & 575      & 1010     \\
\noalign{\smallskip}
\hline
\end{tabular}\\
$^1$ In the 1FGL and in the 2FGL catalogs the $\gamma$-ray emission from  the Magellanic Clouds was associated to more than one \fer\ source ,while here we only report one.\\
$^2$ Includes a number of Narrow Line Seyfert 1 (see Section~\ref{sec:sey} for more details).
\end{table*} 

One emission process widely thought to be responsible for the production of high-energy $\gamma$-rays in   extragalactic sources is inverse Compton scattering. Compton scattering is an inelastic scattering of a photon by a charged particle, typically an electron, where the frequency of the scattered light decreases while part of its energy is transferred to the recoiling particle. If the particle has kinetic energy grater than the photon energy, the reverse process can occur, with net energy transferred from the particle to the photon. This process is known as inverse Compton scattering.

As will be discussed in following sections, in the extragalactic $\gamma$-ray sky, the inverse Compton process can take place within the relativistic jets of blazars and radio galaxies, up-scattering low frequency photons to the MeV-GeV and even TeV energies.  This same process can produce high-energy photons from cosmic rays interacting with  interstellar photon fields as, for example, occurs in the Milky Way and in nearby star-forming galaxies. Although the physical process is conventional,  \fer\ blazar data coupled with  multifrequency campaigns are challenging our theoretical view of where and how these interactions occur, leading to different interpretations/models to describe their broad-band emission (see Section~\ref{sec:theor} and references therein).

\section{A multifrequency view of the $\gamma$-ray blazars}
\label{sec:blazar}

\subsection{An overview of the $\gamma$-ray blazar phenomenon}
\label{sec:overview}

Because blazars are such a dominant feature of the $\gamma$-ray sky, we focus on their properties and the contributions \fer\ has made to understanding these powerful jet sources. Blazars are inherently multifrequency objects, and the availability of observations across the electromagnetic spectrum has been critical to interpreting the $\gamma$-ray results. 

Since \fer\ operates in survey mode, for the first time astronomers have the opportunity to perform statistical studies on both the spectral and temporal behavior of the $\gamma$-ray blazar population rather than  observations biased towards ``favorite'' bright sources or objects with well-known interesting features and/or peculiarities. \fer\ offers the unique opportunity to study fundamental aspects of the jet physics by investigating the multifrequency emission of blazars, including variability studies in the MeV-GeV band and cross-band correlations of unprecedented quality for a large sample of blazars. 

Blazars are the rarest type of AGN, with only a few thousand known \citep[e.g. the Roma-BZCAT][]{massaro09,massaro11,massaro15c}. Their emission is dominated by non-thermal radiation, extending from radio frequencies up to TeV energies. According to the unification scenario of AGN \citep{urry95}, blazars are radio-loud sources in which a relativistic jet is pointed close to the line of sight, as originally proposed by \citet{blandford78}.  They come in two flavors: BL Lac objects and FSRQs. They are distinguished by the equivalent width of their rest frame optical emission/absorption lines, narrower than 5${\AA}$ in the former, broader in the latter \citep[see][]{stickel91,stocke91,falomo14}. Both blazar classes also feature a compact radio morphology coupled with flat radio spectra  \citep{ugs3,massaro13d}, apparent superluminal motions, variable and high polarization from the radio to the optical band, and, as recently discovered, peculiar IR colors \citep{paper1}. Blazars exhibit a double humped spectral energy distribution (SED) characterized by a low-energy  peak between the IR and the optical band and a high-energy peak in  $\gamma$ rays.

It is widely agreed that the low-energy component is produced by synchrotron radiation of ultrarelativistic electrons in the jet. In the synchrotron self-Compton scenario \citep[SSC; e.g.,][]{marscher81,dermer95} the second component comes from inverse-Compton scattering of the synchrotron photons by the same electron population. Alternatively, these relativistic electrons could interact with an ambient photon field, upscattering external photons to higher energies, again via inverse Compton scattering. This is the so-called external Compton (EC) scenario \citep[e.g.,][]{reynolds82,dermer02}. External photons can come directly from the accretion disk surrounding the supermassive black hole in the center of the blazar, from the broad line region, and/or the dusty torus \citep[e.g.,][]{sikora94}. While the SSC models are generally adopted to described the BL Lac SEDs, the EC scenario seems to be more appropriate for the FSRQs, where the inverse Compton component has more power than the one peaking at lower energies (i.e., Compton dominance). 

\begin{figure}[]
\includegraphics[height=8.cm,width=12.cm,angle=0]{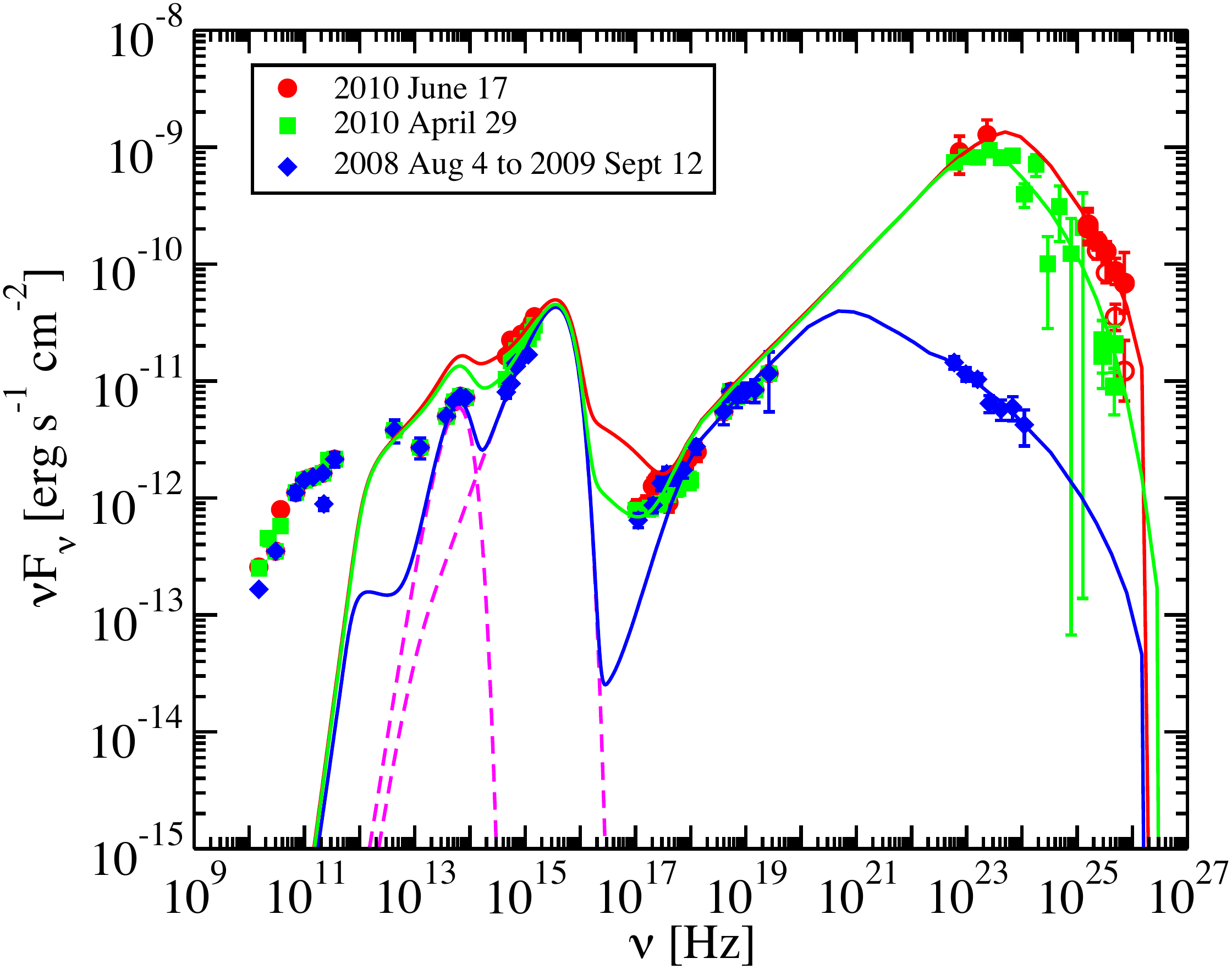}
\caption{Spectral energy distribution of FSRQ 4C +21.35 in three epochs: 2010 June 17 (red circles), 2010 April 29 (green squares), and 2008 August 4 to 2009 September 12 (blue diamonds). Dashed magenta lines indicate the dust torus and accretion disk emission components. The MAGIC data have been corrected for extragalactic background light (EBL) absorption using the model of \cite{finke10b}. Empty symbols refer to non-EBL-corrected data, filled symbols to EBL-corrected ones \citep{ackermann14a}. (Figure courtesy of Dr. J. Finke, under the copyright permissions of the Astrophysical Journal).}
\label{fig:4C21.35}
 \end{figure}

Many blazar features are illustrated in Figure~\ref{fig:4C21.35}, including the strong $\gamma$-ray variability, the dominance of the Compton component, and one possible model for the broadband emission. 

One of the first major results found by \fer\ was the discovery that the brightest blazars show flux variations with the $\gamma$-ray luminosity during high states exceeding that in quiescence by as much as 3 orders of magnitude \citep{abdo09c,jorstad10}. The multifrequency campaigns carried out as follow-up programs of $\gamma$-ray flares showed that many $\gamma$-ray flares are closely related to events at lower frequencies. However this scenario is not ubiquitous, since there are $\gamma$-ray flares having no low-frequency counterparts, e.g., the recent observational campaigns on PKS 0208$-$512 \citep{chatterjee13a,chatterjee13b} or on PKS 0537$-$441 \citep{dammando13a}. In $\gamma$ rays, blazars also exhibit extraordinary rapid variability characterized short timescales lasting even few minutes and with fluxes that increase by a few orders of magnitude above their quiescent states \citep{abdo09c,jorstad10}. These $\gamma$ ray flares occur randomly and can last for hours to days with or without a corresponding counterpart at lower energies.

A VLBI long-term monitoring program, carried out since 1994,  Monitoring Of Jets in Active galactic nuclei with VLBA Experiments (MOJAVE) recently started a direct investigation of \fer\ blazars \citep{lister09,lister11,lister15} \footnote{http://www.physics.purdue.edu/MOJAVE/}. The main goals of the program are searching for radio brightness and polarization variations in jets associated with active galaxies visible in the Northern Hemisphere to improve our knowledge of the the evolution of jet structures on parsec scales and how this activity is linked with the $\gamma$-ray emission detected by \fer. These VLBI investigations have been also combined with radio polarization studies of \fer\ active galaxies \citep[see e.g.,][]{hovatta10,hovatta12} and optical follow-up campaigns \citep[see e.g.][]{arshakian12}. A similar southern hemisphere program is called Tracking Active Galactic Nuclei with Austral Milliarcsecond Interferometry (TANAMI) \citep[see e.g.,][]{ojha10}.  Among other results, MOJAVE/TANAMI have shown that  $\gamma$-ray blazars are typically those AGN with the fastest jets. 

Flaring activity  from \fer\ blazars has been also used in different ways. Following a hint of a $\gamma$-ray gravitational lensing signal in the distant blazar PKS 1830$-$211 \citep{barnacka11},  in September 2012, \fer\ caught a series of bright $\gamma$-ray flares from the blazar B0218+357, a known double-image lensed system \citep{cheung14}. The temporal delay measured between flares from the gravitationally lensed images of B0218+357 during a period of enhanced $\gamma$-ray activity allowed the first conclusive $\gamma$-ray measurement of a gravitational lens. 

The broad energy range explored by \fer, from a few tens of MeV up to several hundreds of GeV, has allowed also the finding that $\gamma$-ray spectra of blazars can be curved. As occurs at lower energies, for example in  X-rays, some $\gamma$-ray spectra are well described as having a log-parabolic shape \citep{massaro04,massaro06}. On the other hand, several blazars, all belonging to the FSRQ class, have also shown spectral breaks in their $\gamma$-ray spectra around a few GeV \citep[e.g., 3C 454.3][]{abdo09g}, but since this peculiar feature was not found in all the bright sources for which a detailed analysis was performed, it does not appear to be ubiquitous.  

A significant fraction of the BL Lacs detected by \fer\  show hard $\gamma$-ray spectra and are detected up to 300 GeV. Thus \fer\ has identified these objects as potential TeV sources, and several of them have been  discovered by ground-based follow-up observations (e.g., the Very Energetic Radiation Imaging Telescope Array VERITAS, the High Energy Stereoscopic System H.E.S.S., the Major Atmospheric Gamma-ray Imaging Cherenkov Telescope MAGIC).  A summary of the hard-spectrum sources seen by \fer\--LAT  is the First {\it Fermi}-LAT Catalog of Sources above 10 GeV \citep{ackermann13a}. 

\begin{figure}[]
\includegraphics[height=12.cm,width=8.cm,angle=-90]{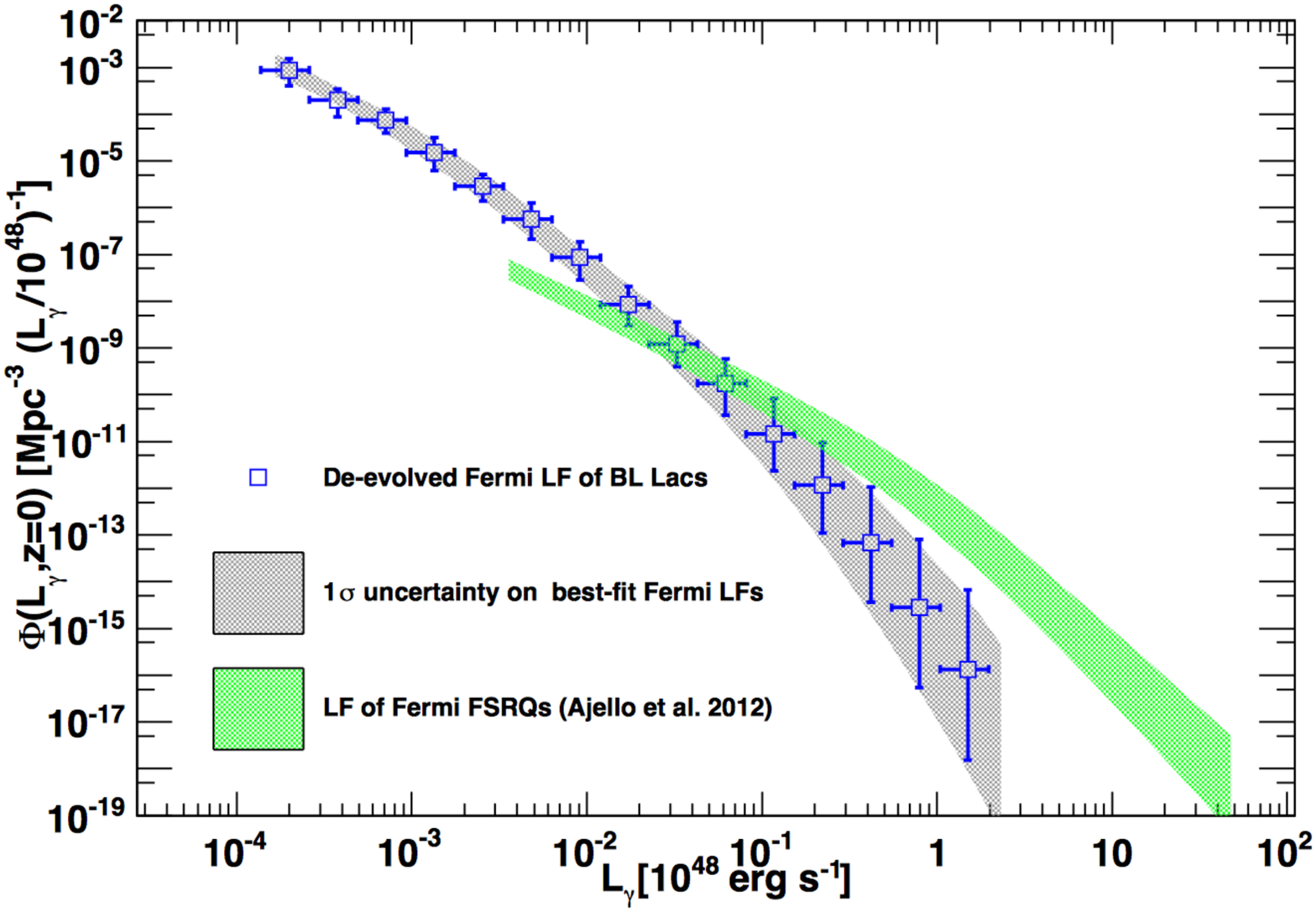}
\caption{Local (z=0) luminosity function derived for the \fer\ BL Lacs \citep[see][for details on the sample selection and the luminosity function calculations]{ajello14}. (Figure courtesy of Prof. M. Ajello, under the copyright permissions of the Astrophysical Journal).}
\label{fig:BLLF}
 \end{figure}

\begin{figure}[]
\includegraphics[height=12.cm,width=8.cm,angle=-90]{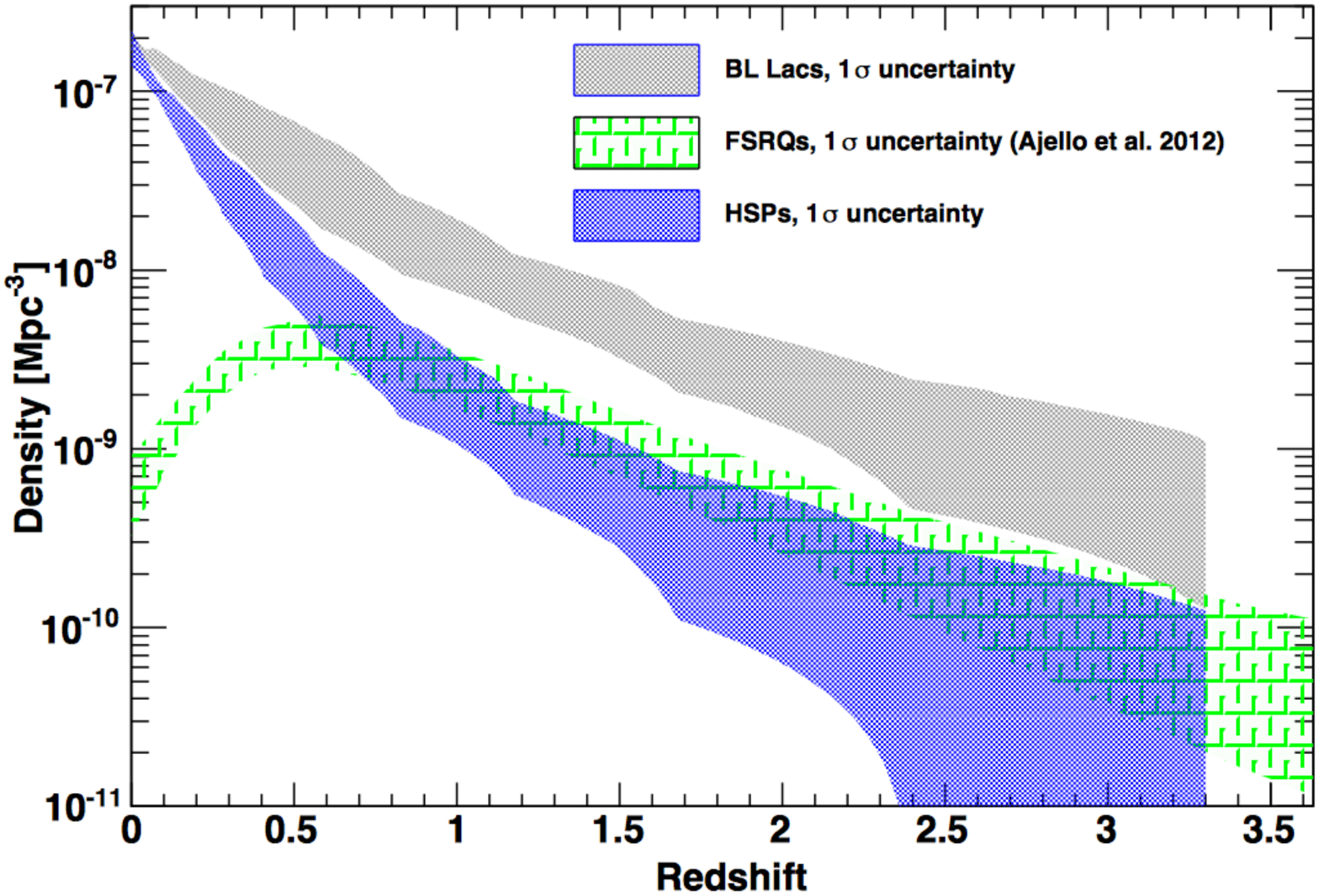}
\caption{Source number per unit co-moving volume for the BL Lacs and the FSRQs. The density of HSPs is also highlighted separately \citep[see][for additional details]{ajello14}. (Figure courtesy of Prof. M. Ajello, under the copyright permissions of the Astrophysical Journal).}
\label{fig:BLdensity}
 \end{figure}

The redshift range of $\gamma$-ray blazars extends  to $z=3.1$. The non-detection of higher-redshift FSRQs could be associated with a change of SED properties with redshift, suggesting a cosmological evolution in their high-energy emission. The latest results on the cosmological evolution of the BL Lac population have been recently presented by \citet{ajello14}. Using the largest and most complete sample of $\gamma$-ray BL Lacs available in the literature, they found that  their cosmological evolution is positive with a space density peaking at modest redshift $\sim$1 (see Figure \ref{fig:BLLF}). In particular selecting only those BL Lacs that show the low-frequency component peaking between the UV and X-rays (so-called ``high frequency peaked BL Lacs'' or ``high synchrotron peaked BL Lacs'') \citet{ajello14} found that they evolve negatively, with their number density increasing for redshifts below $\sim$0.5 (see Figure \ref{fig:BLdensity}). This rise corresponds to a drop-off in the density of FSRQs, suggesting a possible interpretation in terms of an accretion-starved end-state of an earlier merger-driven gas-rich phase for sources belonging to this subclass. Blazar population studies are also crucial to  estimate  their contribution to the extragalactic $\gamma$-ray background, as discussed in the following. Such population studies are also important to determine the distribution of  beaming factors in blazar jets independently of radio observations, which were previously the only way to estimate them.

In addition to FSRQ there are also several steep spectrum radio quasars (SSRQ) that have been associated to \fer\ sources. Two SSRQ,  3C 207 and 3C 380, were  reported in the 1LAC, the latter also indicated as a compact steep spectrum (CSS) radio source. These objects show a blazar-like behavior,  classified in the Roma-BZCAT as blazar of uncertain type (BZU). Three new SSRQ have been  recently added with the 3LAC: 3C 275.1 , TXS 0348+013  and 4C +39.26  in addition to 4C +04.40, which is part of a double association with the FSRQ MG1 J120448+0408 for the \fer\ source 3FGL J1205.4+0412.

\subsection{The radio $-$ $\gamma$-ray connection}
\label{sec:radiogamma}

\begin{figure}[]
\includegraphics[height=12.cm,width=8.cm,angle=-90]{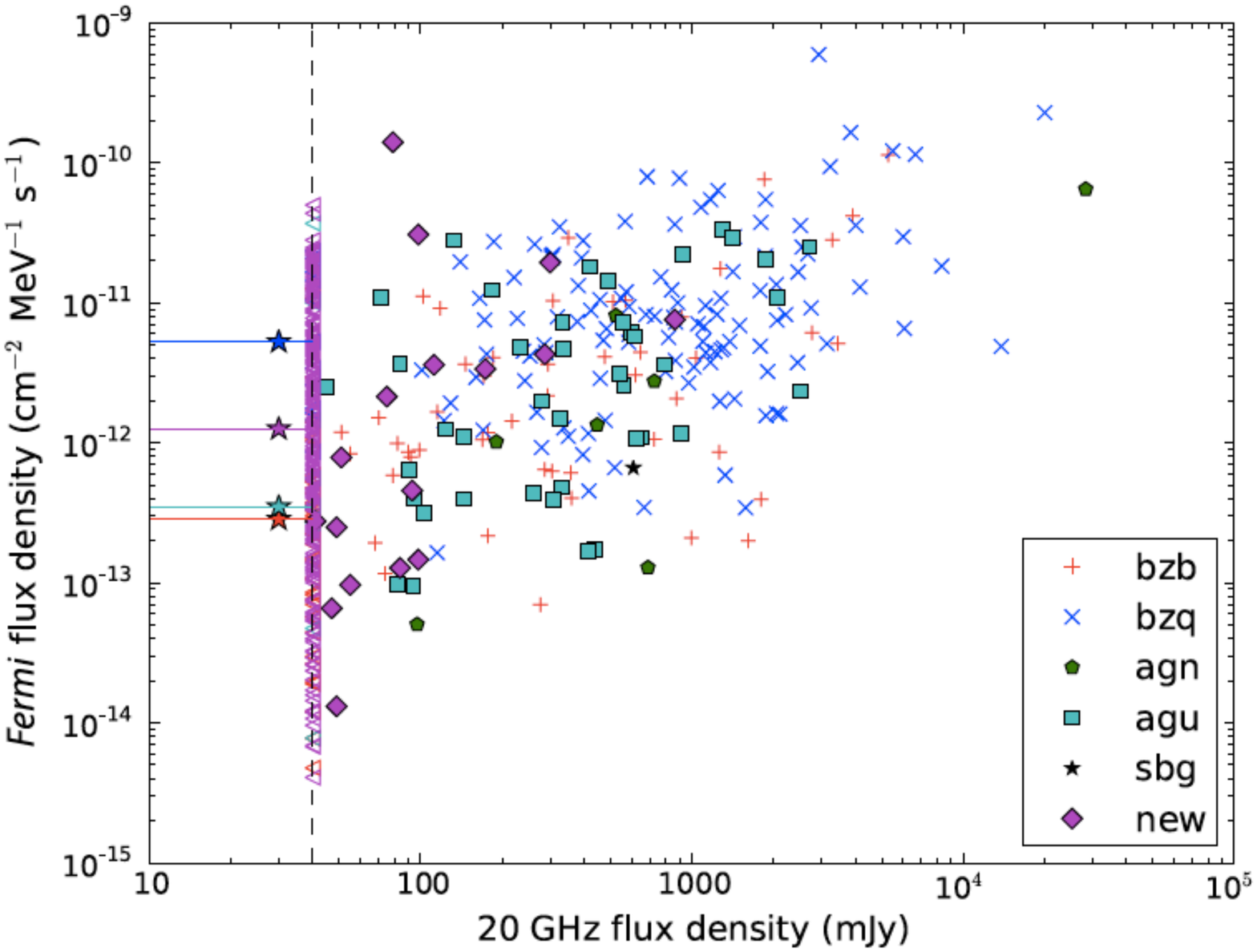}
\caption{20 GHz flux against, the \fer\ flux density for different classes of sources given in the 1FGL catalog and visible in the Southern Hemisphere \citep{mahony10}. (Figure courtesy of Dr. E. Mahony, under the copyright permissions of the Astrophysical Journal).}
\label{fig:radiog1}
 \end{figure}

\begin{figure}[]
\includegraphics[height=12.cm,width=8.cm,angle=-90]{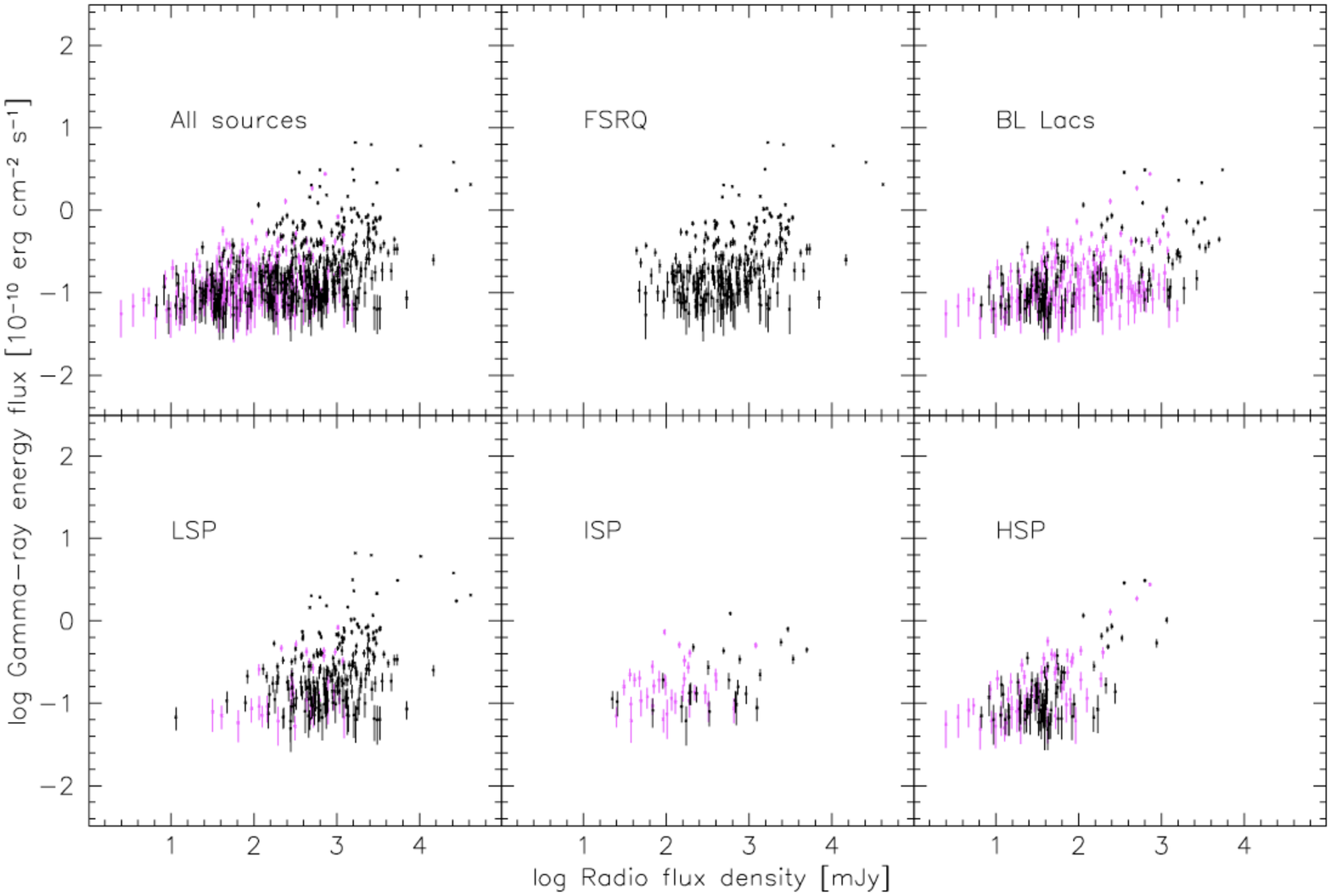}
\caption{Broadband $\gamma$-ray energy flux vs. 8 GHz archival radio flux density for the 1LAC sample, divided by  source type. (Figure courtesy of the \fer\ Large Area Telescope Collaboration, under the copyright permissions of the Astrophysical Journal).}
\label{fig:radiog2}
 \end{figure}

Early attempts to investigate correlations between radio and $\gamma$-ray luminosities of AGN, specifically for blazars, \cite[e.g.,][]{padovani93,stecker93,salamon98,taylor07}  appeared inconclusive, since biases and selection effects played an important role. Thanks to the unprecedented sample of \fer\ blazars it is now possible to establish with high accuracy the statistical significance of the radio-$\gamma$-ray correlation. Taking into account the `common-distance' bias and the effect of a limited dynamical range in the observed quantities, such investigations \cite[e.g.,][]{ghirlanda10,ghirlanda11,mahony10,ackermann11} found a relationship between the centimeter radio and the broadband $\gamma$-ray energy flux above 100 MeV, with a chance probability of $\sim$10$^{-7}$ for both BL Lacs and FSRQs. The statistical significance of the radio-$\gamma$-ray connection does not have a simple dependence on the apparent correlation strength. despite the tightness of the observed correlation. As highlighted by \cite{ackermann11}, there could be various factors affecting its significance but not its existence. For example the $\gamma$-ray energy range used to compute the $\gamma$-ray flux appears to affect the strength of the correlation. Different types of blazars show a different apparent strength of the radio-$\gamma$-ray correlation, with the highest apparent correlation strength shown by the HBL/HSP. Figure \ref{fig:radiog1} shows the preliminary results obtained for the 1FGL sources in the Southern Hemisphere \citep{mahony10} while Figure \ref{fig:radiog2} shows the more complete analysis performed subsequently on the sources in the 1LAC sample \citep{ackermann11}.

This link between the low and the high-energy emission of blazars has been extended during the \fer\ era to both higher and lower radio frequencies, strengthening and improving the associations \citep{ugs3}. Low-frequency radio observations (i.e., below $\sim$1~GHz) revealed a new spectral behavior that allowed us to search for blazar-like sources lying within the positional uncertainty regions of the UGSs. No correlation between the radio and the $\gamma$-ray spectral shapes has been found using data at $\sim$300MHz or even below at $\sim$70 MHz \citep{massaro13d}. The $\gamma$-ray blazars appear to have flatter radio spectra than the whole population. The presence of flat radio spectra below $\sim$1~GHz was completely unexpected and challenges the unification scenario of active galaxies \citep{massaro13d}. 

At higher radio frequencies, the opportunity to obtain data simultaneously from both Planck and Fermi satellites allowed us to search for connection between the (sub-)millimeter and $\gamma$-ray emission in blazars \citep{leontavares12}. This trend appears to exist since these luminosities are correlated over five orders of magnitude. However, this correlation is not significant when simultaneous observations, restricted to some energy ranges are considered. On the other hand, blazars with an approximate spectral turnover in the millimeter energy range tend to be stronger $\gamma$-ray emitters. Additional investigatios of these blazar SED have been addressed by different groups \citep[see e.g.,][]{fuhrmann14} also considering optical, UV and X-ray observations performed with SWIFT \citep{giommi12} .

\subsection{\fer\ blazars in  infrared light}
\label{sec:irgamma}

During the second year of \fer\ operation, at the end of 2009 and beginning of 2010, a survey of the mid-IR sky was performed by the 
{\it Wide-field Infrared Survey Explorer (WISE)} \citep{wright10}.  {\it WISE} mapped the whole sky in 4  bands, namely at 2.4, 4.6 12 and 22 $\mu$m, with an angular resolution of 6.1, 6.4, 6.5, and 12.0 arcseconds. In these four bands it achieved  5$\sigma$ point source sensitivities better than 0.08, 0.11, 1 and 6 mJy in unconfused regions on the ecliptic, respectively.

\begin{figure}[]
\includegraphics[height=9.6cm,width=11.cm,angle=0]{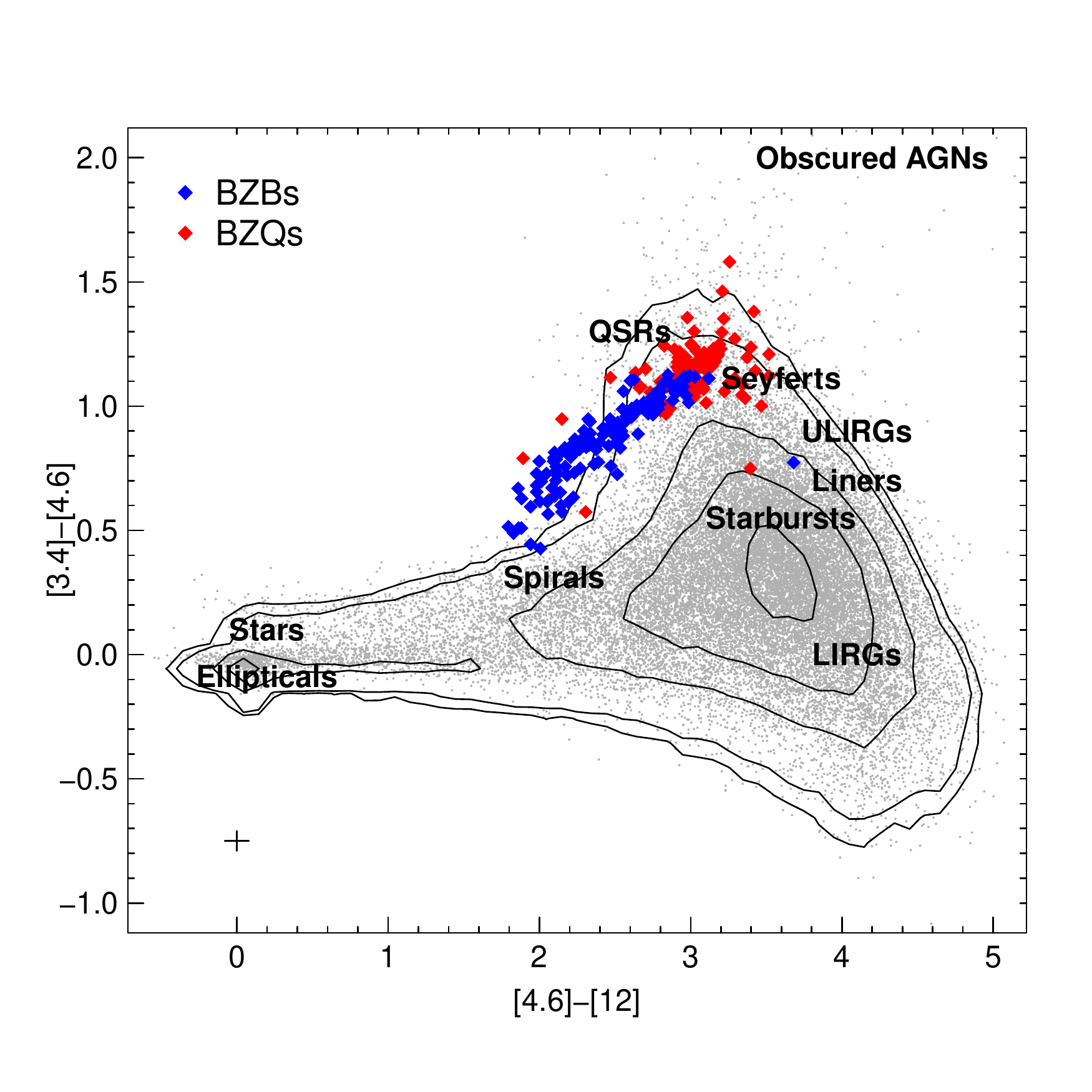}
\caption{  {\it WISE} IR colors of \fer\ blazars compared with those of sources belonging to different classes
in the [3.4]-[4.6]-[12] $\mu$m color-color diagram.  These preliminary results  were confirmed with subsequent analyses performed with the latest release of the {\it WISE} all-sky survey.
The two blazar classes of BZBs (blue) and BZQs (red) are shown overlaid on the background of grey dots corresponding to 
453420 \wse\ sources detected in a region of 56 deg$^2$ at high Galactic latitude. The isodensity 
curves for the \wse\ sources, corresponding to 50, 100, 500, 2000 sources per {\bf unit} area in the color-color 
plane respectively, are shown. The location of different classes of objects is also shown. (Figure courtesy of Dr. R. D'Abrusco, under the copyright permissions of the Astrophysical Journal).}
\label{fig:color2color1}
 \end{figure}

In 2011 a  comparative analysis of the IR colors of the $\gamma$-ray blazars  in the 2FGL catalog was performed using the {\it WISE} Preliminary release \citep{paper1}. A tight connection was immediately evident between the IR and the $\gamma$-ray spectral shapes for the whole \fer\ blazar population \citep{paper2}. In particular, among the blazars those emitting in $\gamma$-rays were clearly distinguished from other classes of galaxies and/or AGN and/or Galactic sources using their IR color distributions (see Figure \ref{fig:color2color1}). They delineate a narrow, distinct region of the IR color-color plots, originally designated the {\it WISE} Gamma-ray Strip \citep{paper3}. Based on this finding, a preliminary method to search for blazar-like counterparts of the UGS was  developed \citep{paper4}. A refined analysis performed by \citet{ugs1} indicated a 3-dimensional region occupied by $\gamma$-ray emitting blazars as the $locus$, and its 2-dimensional projection in the [3.4]-[4.6]-[12] $\mu$m color-color diagram as the \wse\ Gamma-ray Strip. Peculiar IR colors were then used to develop a new procedure to recognize $\gamma$-ray blazar candidates among the UGS \citep{ugs1,ugs2,ugs5}. Moreover the \wse\ fluxes were often included in the broad band SEDs of \fer\ blazars. The analysis of the IR sources as potential UGS counterparts has been recently concluded once a catalog of IR radio-loud sources selected for having not only IR colors similar to the \fer\ blazars but also a radio counterpart in one of the major radio surveys was released \citep[i.e., WIBRaLs;][]{wibrals}. This IR-based catalog allowed us to adopt the same procedures for associating $\gamma$-ray sources with their low energy counterparts as done for other databases in the released \fer\ catalogs.

IR colors have ben recently used to assemble a sample of $\sim$1000 HSP candidates toward a better understanding their jet properties and to investigate their cosmological evolution \citep{arsioli15}. This sample have been also used to search for counterparts in the latest release of the \fer\ catalog.

To compare the radio-$\gamma$-ray connection and the one recently found for the IR, note that the former one not only indicates a relation between emitted powers at  energy ranges separated by $\sim$ 14 orders of magnitude, but it is also related to their spectral shapes \citep{paper2,ugs1,ugs2}. To highlight this link a significant correlation was found between the IR and the $\gamma$-ray spectral indices. In addition the IR blazar fluxes allowed us to compute the ``Compton dominance''  that again appears to be significantly greater than 1 for the FSRQs.

\subsection{Optical properties of the $\gamma$-ray blazars}
\label{sec:optgamma}

Despite the established radio-to-$\gamma$-ray connection and the newly discovered IR-to-$\gamma$-ray one, there is not yet an established relation between their optical emission and that at higher energies for any class of $\gamma$-ray sources, Galactic or extragalactic.  For AGN, the optical often includes emission from the underlying galaxy or the accretion disk, unrelated to the jet where the $\gamma$ rays are produced. Nevertheless, optical spectroscopic observations are the real breakthrough to improve  $\gamma$-ray source associations \citep{shaw13a,shaw13b,massaro15b}. The underlying reason is that optical spectroscopic observations are crucial to disentangle and/or confirm the natures of the low-energy counterparts selected with different methods. Since blazars are the largest population of $\gamma$-ray sources, the largest fraction of these associations come from the comparison with the Roma-BZCAT \citep{massaro15c}, but sources can be inserted therein and classified as BL Lacs only if there is an optical spectrum that can confirm the classification. 

Optical follow-up observations, mainly photometric, are often carried out when $\gamma$-ray flares occur. 
The theoretical motivation underlying this quick ``race'' to perform follow-up observations is the possibility to obtain a constraint on the size of the emitting region and/or the distance of the $\gamma$-ray emitting site from the supermassive black hole at the center of the blazar \citep[e.g.,][]{marscher08}. Such estimates can be determined from the time delay between different bands and/or using for example microlensing \citep{neronov15}, although these scenarios require several assumptions. 
The detection of flaring activity by these optical campaigns  has highlighted the same behavior seen at other frequencies: there are many cases of quasi-simultaneous $\gamma$-ray flares with an optical counterpart, but ``orphan'' flares are also often detected.  They reinforce the idea that $\gamma$-ray AGN are diverse in their behavior, not easily explained by simple models. 

Extensive optical campaigns have been carried out and are still ongoing thanks to the GLAST-AGILE Support Program (GASP) organized within the Whole Earth Blazar Telescope (WEBT)\footnote{http://www.to.astro.it/blazars/webt/} \citep[e.g.,][]{villata09}. This collaboration provides optical-to-radio long-term continuous monitoring of a selected sample of $\gamma$-ray-loud blazars during the operation of the \fer\ satellite, such as the multifrequency campaigns done for 3C 454.3 \citep{raiteri11}, 4C 38.41 \citep{raiteri12} and BL Lacertae \citep{raiteri13}. Optical monitoring campaigns are also carried out at the University of Arizona\footnote{http://james.as.arizona.edu/$\sim$psmith/Fermi/} \citep{smith07}. This program  observes optical linear polarization of a blazar sample at the Steward Observatory \citep[see e.g.,][]{jorstad13}. Spectropolarimetry of the blazars yields measurements of the brightness and spectral index of the optical synchrotron light and are crucial since they provide the only direct information about the magnetic field direction within the region producing the optical synchrotron emission. An additional follow-up optical program, in the Southern Hemisphere, is carried out by the blazar group at Yale University with the Small and Medium Aperture Research Telescope System (SMARTS) facility\footnote{http://www.astro.yale.edu/smarts/glast/home.php} \citep[see e.g.,][]{bonning12,isler15}.

Optical spectroscopic observations have been also used to determine jet parameters. The emission lines  can be interpreted as a proxy for the disc luminosity to be compared with the luminosities of the high-energy emission for the jet power \citep{sbarrato12,ghisellini15}. These observations were used to search for an explanation of the transition between BL Lacs and FSRQs driven by line luminosity in Eddington units.
A larger sample is needed to confirm this result.

\subsection{X-raying the blazars in the \fer\ sky}
\label{sec:xraygamma}

A substantial effort has been invested in X-ray follow-up campaigns of $\gamma$-ray sources, either in flaring states or to search for their low-energy counterparts \cite[e.g.,][]{mirabal09,paggi13,takeuchi13,acero13}. As high-energy sources emitting over the whole electromagnetic spectrum blazars are also expected to be detected in X-rays at some level. The X-ray sky, however, like the optical sky, is filled with thermal sources that are unlikely to emit non-thermal $\gamma$-rays, so a search of a $\gamma$-ray error region for an X-ray counterpart is likely to produce spurious candidates. At the  moment there is no evidence of a simple X-ray - $\gamma$-ray connection. A large fraction of the \fer\ blazars have X-ray counterparts, but the X-ray band often lies in the valley between the two humps of their SED, so the X-ray emission is often faint and cannot  always be detected.  

In the hard X-ray sky (10-150 keV), a large fraction (i.e., $\sim$75\%) of the blazars associated in the {\it Swift}--BAT 70-month survey \citep{baumgartner08} have also a counterpart in the 3FGL, showing a trend between their spectral indices.

\begin{figure}[]
\begin{center}
\includegraphics[height=12.4cm,width=10.cm,angle=0]{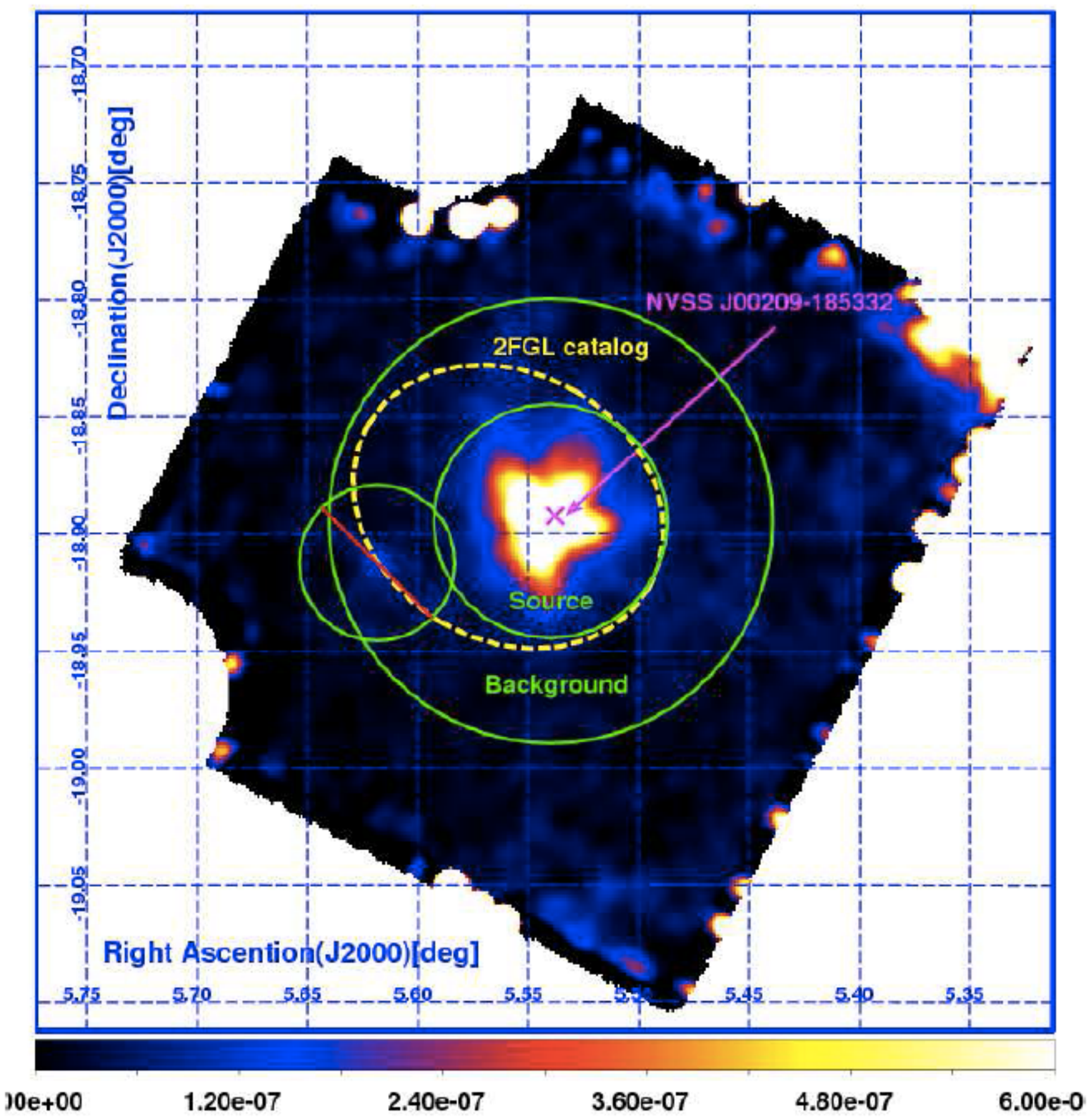}
\caption{Suzaku X-ray image of 1FGL J0022.2-1850 (alias 2FGL J0022.2-1853) in the 0.4-10 keV energy range. The magenta cross points to the position of radio counterpart NVSS J00209-185332, and the yellow dotted ellipse indicates the positional uncertainty region of the \fer\ source at 95\% level of confidence. Green ellipses are those chosen for the X-ray source extraction and the background \citep[][for additional details on the X-ray analysis]{takeuchi13}. (Figure courtesy of Prof. J. Kataoka, under the copyright permissions of the Astrophysical Journal).}
\label{fig:xraysuz}
\end{center}
\end{figure}

As with the optical, some $\gamma$-ray flares have corresponding X-ray flares, but some do not. These difficulties should not and have not discouraged the search for a possible X-ray view of the $\gamma$-ray sky.  Finding an X-ray flare matching one seen in $\gamma$-rays is a strong indication that both are coming from the same object, helping us toward firm identifications of some $\gamma$-ray blazars when combined with detected simultaneous variability in at least another band.  Equally important, potential X-ray counterparts have far better localizations than the original $\gamma$-ray sources, enabling a more efficient search for radio, IR or optical matches.  Thus a \swf\ X-ray survey for all the UGSs listed in the \fer\ source catalogs is still ongoing$^1$ \citep[see e.g.][]{stroh13}, and additional X-ray observations  performed with {\it XMM-Newton}, {\it Chandra} \citep{cheung12} and {\it Suzaku} have improved our knowledge about the UGSs \citep{kataoka12,takahashi12,takeuchi13}. In Figure \ref{fig:xraysuz} we show an example on the use of {\it Suzaku} observation to unveil the nature of the 1FGL J0022.2-1850 \fer\ source as a BL Lac object.

\subsection{Theoretical insights from Fermi observations of blazars}
\label{sec:theor}

Astrophysical jets are important examples of particle acceleration and energy transport.  The powerful, large-scale jets of blazars offer a remarkable opportunity to learn about these processes, because these jets are pointed nearly in our direction.  The \fer\ observations have clearly demonstrated the diversity of jet phenomena in the $\gamma$-ray universe.  Although an exhaustive study of blazar theory is beyond the scope of the review, in this section we survey some of the insights that have emerged.


One of the major open questions about the origin of the $\gamma$-ray emission from blazars, posed after the EGRET era, is the location of the  $\gamma$-ray emitting region. Two possible sites have been identified: the first is close to the supermassive black hole, where the jets originate (i.e, $<\sim$0.1-1 pc), while the second is much farther from the central engine (i.e., $>>$ 1pc), where the jet starts to decelerate and can be visible at millimeter frequencies with Very-Long-Baseline Interferometry (VLBI). This is a long-standing and highly controversial issue, still poorly constrained \citep[e.g.,][]{dermer14,nalewajko14}.

\begin{figure}[]
\includegraphics[height=10.cm,width=12.cm,angle=0]{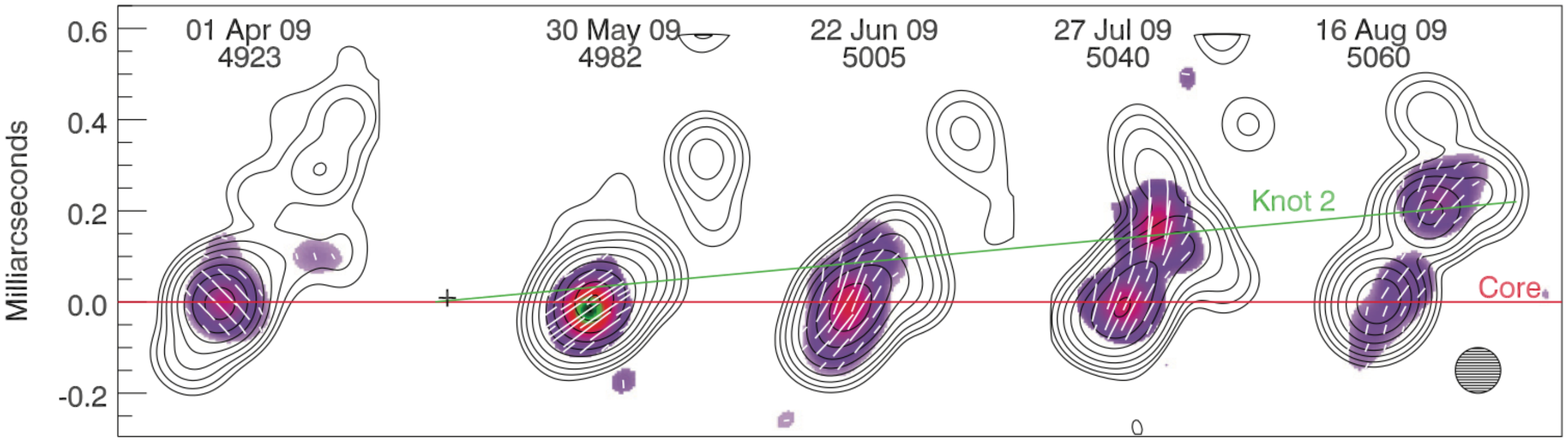}
\caption{Sequence of 43 GHz VLBA images showing ejections of a superluminal jet knot moving away from the core of PKS 1510$-$089 with a proper motion 0.97 ± 0.06 mas yr$^{-1}$, during an intense $\gamma$-ray outburst seen by the \fer\--LAT  in 2009. Images are convolved with a circular Gaussian beam of FWHM = 0.1 mas (the shaded circle on bottom right) \citep[see][for additional details on the multifrequency campaign]{marscher10}. (Figure courtesy of Prof. A. Marscher, under the copyright permissions of the Astrophysical Journal).}
\label{fig:knot}
 \end{figure}

Extensive radio VLBI campaigns have been carried out to investigate if the appearance of new radio knots on parsec and sub-parsec scale is related to $\gamma$-ray flaring states\footnote{see e.g., https://www.bu.edu/blazars/VLBAproject.html}. Several such observations appear to support the second scenario for some $\gamma$-ray blazars, as for example those obtained on PKS 1510$-$089, 3C 454.3, AO0235+164 and OJ287  \citep[see][respectively]{marscher10,jorstad10,agudo11,agudo12}, using light-curve correlations between optical, millimeter and $\gamma$-ray bands coupled with ultrahigh-resolution VLBI imaging. 
Figure~\ref{fig:knot} shows the Very Long Baseline Array (VLBA) images obtained in 2009 while monitoring the multifrequency activity of PKS 1510$-$089 \citep{marscher10} during an extreme $\gamma$-ray flaring period. The radio campaign revealed a new jet knot  separating from the radio core at apparent superluminal speed during the $\gamma$-ray flaring state.

The origin of the spectral breaks discovered in FSRQs such as 3C 454.3 has been interpreted either as due to intrinsic absorption or to the spectral shape of the emitting particles. The lack of $\gamma$-ray spectral breaks in all FSRQs has been used to constrain the origin of the seed photons for the EC models \citep{finke10a,stern11}. In particular, its absence could imply that the $\gamma$-ray emission in at least some sources arises from regions external to the inner jet.

On the other hand, recent evidence of  $\gamma$-ray microlensing in two blazars has been also used to estimate the projected size of the $\gamma$-ray jet \citep{neronov15,vovk15}. These measurements are consistent with constraints derived by the intrinsic variability and showed that $\gamma$-ray emission in these sources arises from regions extremely close the central supermassive black hole.

Particle acceleration via shocks in relativistic jets of blazars have been considered for many years the main acceleration mechanism to obtain high-energy electrons (and protons) that can radiate photons from radio up to TeV frequencies \citep[see e.g.,][and references therein]{paggi09,ghisellini10,linford12}. The emitted spectra are then amplified by relativistic effects so resulting in the non-thermal double-humped SED characteristic of the broad-band emission in blazars. The high values of the emitted power observed at very high energies by \fer\ over relatively short timescale (i.e., of the order of hours) has proven to be a challenge for the shock acceleration scenario and has encouraged theoreticians to work on models based on magnetic reconnection \citep{massaro11d,giannios13,sironi15}, structured jets \citep{ghisellini05,tavecchio08}, and Poynting-dominated jets \citep[see e.g.,][]{nalewajko12a}. Models including plasma instability-induced short variability behavior have also been considered \citep{nalewajko12b}, as well as turbulent cell models \citep{marscher14}. Second-order acceleration mechanisms have been extensively considered to describe both the $\gamma$-ray spectra and the broad-band blazar SEDs generally described with a log-parabolic shape \citep{massaro06}. Such spectral shapes can arise from both statistical and stochastic acceleration mechanisms for the emitting particles \citep[see ][for a recent review]{tramacere11}.

Multifrequency observations of bright BL Lacs have revealed several problems of the standard single-zone SSC models, with power-law injection and radiative cooling unable to explain the whole SED behavior.  In particular, \fer\ found that FSRQs can emit variable $\gamma$ radiation on hour timescales, as recently observed for BL Lacs in the TeV energy range, and exceeding the Eddington luminosity for brief periods of time. Such observed behavior has clearly challenged previous theoretical models, highlighting the need for new and more sophisticated physical scenarios \citep{giommi12,boettcher13}. 

Short timescales observed during blazar flares in $\gamma$-rays and $\gamma$-ray ``orphan'' flares (i.e., with no simultaneous counterpart at low energies) have also challenged the single zone SSC emission models, and hadronic scenarios have been considered, in particular to explain the hardening of  $\gamma$-ray spectra observed during these extreme episodes \citep[see][e.g.]{boettcher13}. Distinguishing hadronic and leptonic models for blazars was one of the major pre-launch goals of \fer\, and the possibility to combine MeV-to-GeV observations with those already available at TeV energies motivated  new perspectives on this theoretical scenario \citep[see ][and references therein]{reimer12}. Assuming that both protons and electrons are efficiently accelerated in the blazar jets, $\gamma$-rays can be produced by synchrotron emission of protons and ions up to the GeV-TeV energies \citep[see e.g.,][]{aharonian00}; then photon-lepton cascades induced by photo-pion production would be again visible in $\gamma$-rays, with no direct effect on the synchrotron emission.  This could potentially explain ``orphan'' flares \citep{mannheim92}. These hadronic scenarios, however, seem to require highly super-Eddington jet power, in conflict with the estimates provided by other methods \citep{zdziarski15}, although this is still an open issue \citep{cerruti15}. Additional hadronic scenarios have been  considered, as for example the proton/ion synchrotron radiation that could be directly seen in the MeV-GeV energy range.  That model seems disfavored because it requires a highly magnetized jet. The $\gamma$-ray emission could also arise from interactions occurring outside the blazar jet \citep[][e.g.,]{essay10a,essay10b}. If ultra-relativistic protons escape and reach the intergalactic medium, they interacti with either the cosmic microwave background on a Mpc length scale via photo-pion process or with the EBL photons to produce photo-pairs on Gpc scale. A breakthrough measurement in favor of hadronic processes occurring in \fer\ blazars would be the detection of correlated neutrino events, a result not yet confirmed \citep{aartsen14}.

More details on how the $\gamma$-ray data provided by the \fer\ all-sky survey and multifrequency observations have been crucial to improve our knowledge of the theoretical models for their emission can be found in \cite{finke13b} and \cite{dermer15}. The wide variety of theoretical models still being considered for $\gamma$-ray blazar jets illustrates the complexity that has emerged from having extensive, simultaneous multifrequency data.

\begin{figure}[]
\includegraphics[height=12.cm,width=8.cm,angle=-90]{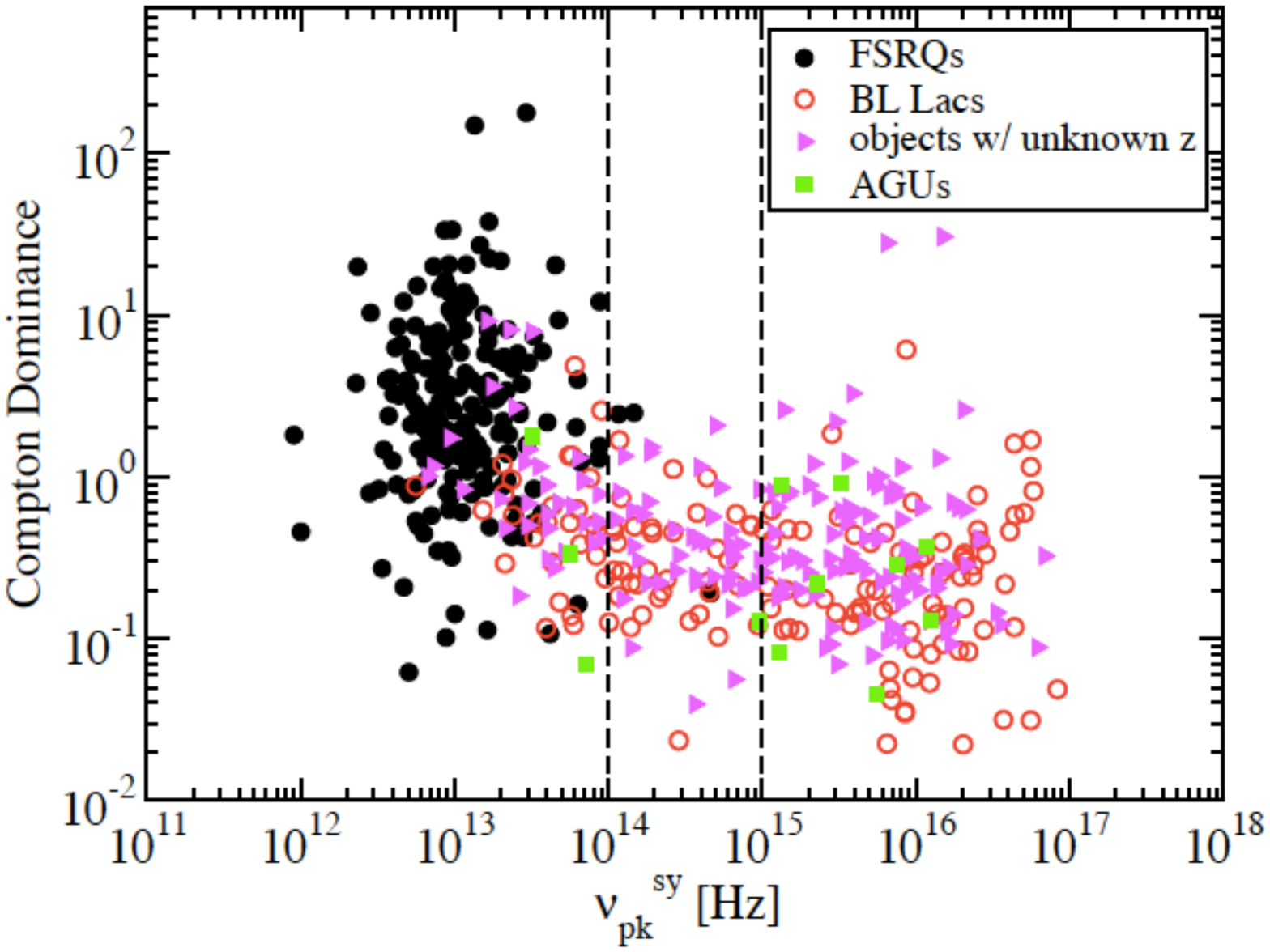}
\caption{The Compton dominance versus peak frequency for the synchrotron component in the \fer\ blazars listed in the 2FGL \citep{finke13a}. Symbols corresponding to the different source classes are indicated in the legend. (Figure courtesy of Dr. J. Finke, under the copyright permissions of the Astrophysical Journal).}
\label{fig:CD}
 \end{figure}

One idea that emerged from the {\it CGRO} era was the ``blazar sequence,'' an apparent pattern with higher-luminosity blazars having peaks in their SEDs shifted to higher energies \citep[e.g.][]{fossati98}. Population studies on \fer\ blazars allowed us to determine robustly the jet powers and their energetics and thus permitted improved tests on the existence of the ``blazar sequence,'' in particular trying to determine if it is due to biases and selection effects or if it is a real trend between observational properties \citep[e.g.,][]{giommi13}. Two theoretical investigations were carried out recently supporting its existence. The first performed by \citet{meyer11} attempts to refine the ``blazar sequence" by including a relation between efficiency of accretion and the kinetic power of the jet, while the second, carried out by \citet{finke13a}, showed that selection effects, such as the lack of redshift estimate for a large fraction of the BL Lacs, could be mitigated by using the Compton dominance parameter, defined as the ratio between the luminosity peak of the two components in the blazar SEDs, instead of the pure bolometric luminosity (Figure \ref{fig:CD}). On the other hand, a recent analysis carried out by \cite{giommi15} shows that calculations based on the blazar sequence, combined with the observed luminosity function of blazars, tend to over-predict the blazar contribution to the extragalactic $\gamma$-ray background compared to the observed value. At the moment this debate is still open, and \fer\ has not yet provided a definitive answer. The hope is that in the near future this will be achieved once more $\gamma$-ray blazars are included in the current samples. There is still a lack of BL Lacs having the low-energy component peaking in the X-rays, and thus with hard $\gamma$-ray spectra at high redshifts or at high luminosities.

\begin{figure}[]
\includegraphics[height=8.cm,width=12.cm,angle=0]{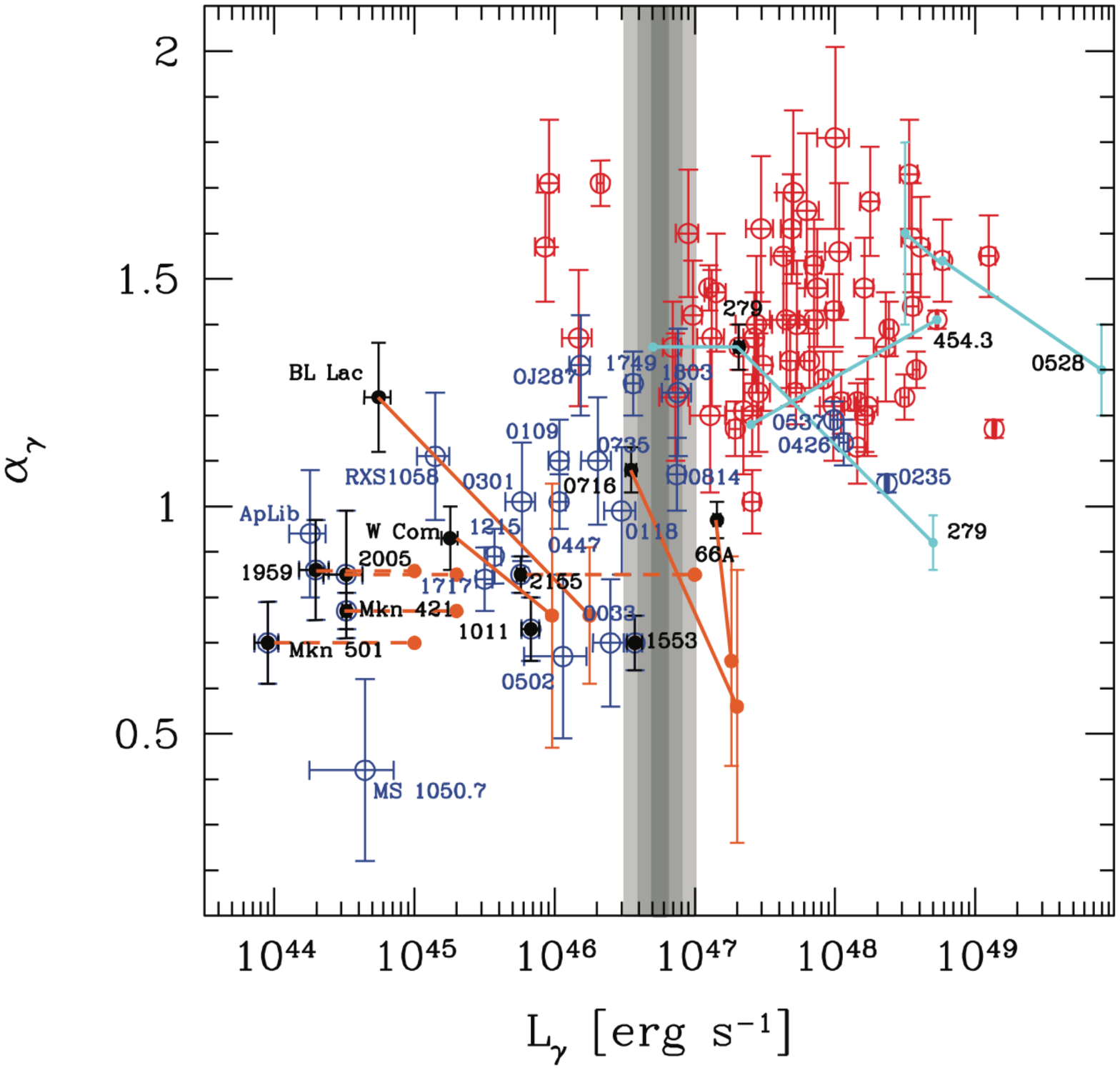}
\caption{The original plot of the {\it Fermi blazars' divide} corresponding to the spectral index ($\alpha_\gamma$) $vs$ $\gamma$-ray luminosity ($L_\gamma$) plane for all blazars listed in the LBAS sample. Blue and red points show the location of BL Lacs and FSRQs, respectively. Black points indicate the sources known as TeV emitters in 2009, when the {\it Fermi blazars' divide} was originally presented (they were all BL Lacs except 3C 279). For several  sources, the observed range of $\gamma$-ray luminosities and spectral indices obtained with using past EGRET or {\it AGILE} observations are also marked \citep[see][for more details]{ghisellini09}. The grey area at $\gamma$-ray luminosity of $\sim$10$^{47}$ erg s$^{-1}$ marks the distinction between BL Lacs and FSRQs. (Figure courtesy of Dr. G. Ghisellini, under the copyright permissions of the Monthly Notices of the Royal Astronomical Society).}
\label{fig:divide}
 \end{figure}

\begin{figure}[]
\includegraphics[height=8.cm,width=12.cm,angle=0]{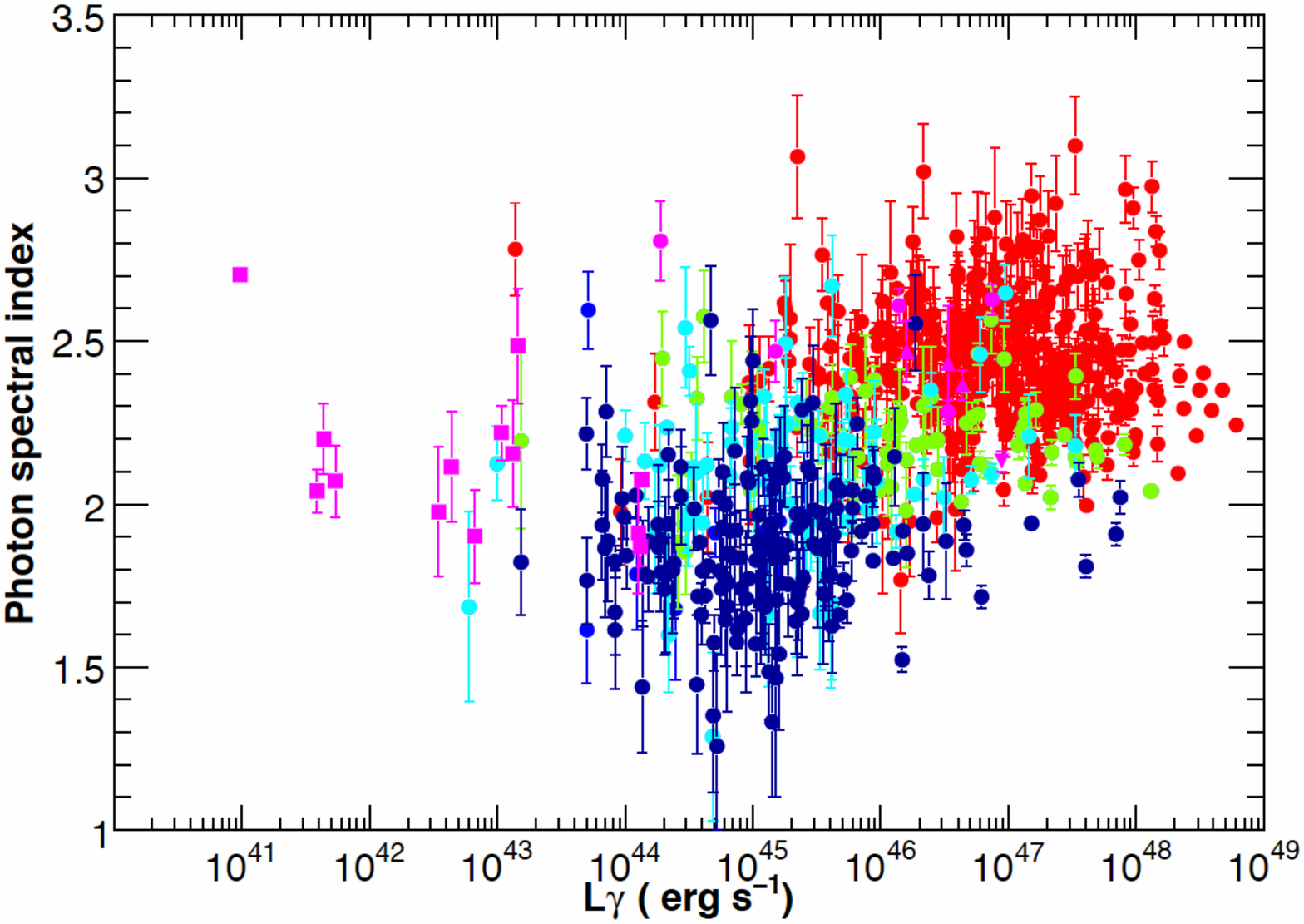}
\caption{The $\gamma$-ray photon index $vs$ $\gamma$-ray luminosity for 3LAC sources with known redshifts. FSRQs are shown in red while green is for LSP-BL Lacs, light blue for ISP-BL Lacs and dark blue for HSP-BL Lacs. Non-blazar AGNs are reported in magenta (circles: NLSy1s, squares: mis-aligned AGNs, up triangles: SSRQs, down triangles: other AGNs) \citep[see][for more details]{ackermann15a}. (Figure courtesy of the \fer\ Large Area Telescope Collaboration, under the copyright permissions of the Astrophysical Journal).}
\label{fig:divide3LAC}
 \end{figure}

With the advent of \fer\ observations a new trend has been identified for the blazars \citep{ghisellini09}. FSRQs and BL Lac objects appear to populate different regions in the $\gamma$-ray spectral index $vs.$ $\gamma$-ray luminosity plane. Such division, known as the {\it Fermi blazars' divide}, has been interpreted as due to different accretion regimes of the two blazar classes. Moreover the spectral separation into hard (BL Lacs) and soft (FSRQs) $\gamma$-ray sources could also be the result of a diverse radiative cooling. In Figure~\ref{fig:divide} we show the original plot of the {\it Fermi blazars' divide} derived with the analysis of the $\gamma$-ray sources in the LBAS sample after 3 months of \fer\ operations. The ``updated'' version for radio-loud $\gamma$-ray sources in the 3LAC is shown in Figure~\ref{fig:divide3LAC}.

\section{Gamma-ray emission from other radio-loud active galaxies}
\label{sec:radioloud}

\subsection{Radio galaxies}
\label{sec:rgs}

Although hints of $\gamma$-ray emission from radio galaxies were seen in the EGRET data, \fer--LAT clearly added these extragalactic sources to the list of $\gamma$-ray classes. These celestial objects actually constitute the second most numerous class of AGN emitting in the MeV-GeV energy range \citep[see also][]{kataoka11}.

Radio galaxies are the largest fraction of radio-loud AGN, characterized by radio luminosities exceeding those of normal galaxies by at least a factor of a few tens at GHz frequencies. In 1974 \citet{fanaroff74} introduced the first classification scheme for extragalactic radio sources with large scale structures (i.e., size greater than $\sim$15-20 kpc). They proposed to distinguish them into two main classes on the basis of the correlation between relative positions of regions of high and low surface brightness in their extended components. This scheme was based on the ratio $R_{FR}$ of the distance between the regions of highest surface brightness on opposite sides of the central galaxy to the total extent of the source up to the lowest brightness contour in the radio images. Sources with $R_{FR}$ $\leq$ 0.5 were placed in Class I (i.e., FR\,I) and sources with $R_{FR}$ $\geq$ 0.5 in Class II (i.e., FR\,II).  Then, at radio frequencies, FR\,Is show surface brightness higher toward their cores while FR\,IIs toward their edges \citep{fanaroff74}. This morphology-based classification scheme was also linked to the cosmological evolution of the radio sources, when Fanaroff and Riley found that all sources with luminosity $L_{\rm 178 MHz}$ $\leq$ 2 $\times$ 10$^{25}$ $h_{\rm 100}^{-2}$ W Hz$^{-1}$ str$^{-1}$ were classified as FR\,I while the brighter sources all were FR\,II. 

According to the standard unification scenario of radio-loud active galaxies \citep{urry95}, radio galaxies are interpreted as the parent population of blazars. BL Lac objects are believed to be the counterpart of the FR\,Is while the FSRQs those of the FR\,IIs, both viewed at small angle with respect to the line of sight. This model has led to radio galaxies within the sample of $\gamma$-ray emitters sometimes being called  ``misaligned'' AGN \citep{abdo10d}.
\begin{table*}
\label{tab:radiogalaxies}
\caption{Radio galaxies detected by \fer\ ordered by their cosmological distance.}
\begin{tabular}{|lllll|}
\hline
Common & $z$ & 1FGL & 2FGL & 3FGL \\ 
name & & name & name & name \\ 
\hline
\noalign{\smallskip}
FR\,Is & & & & \\
\hline
\noalign{\smallskip}
Centaurus A &  0.001825 & J1325.6-4300 & J1325.6-4300 & J1325.4-4301 \\  
M 87        &  0.004283 & J1230.8+1223 & J1230.8+1224 & J1230.9+1224 \\   
Centaurus B &  0.012916 &                  & J1346.6-6027 & J1346.6-6027 \\ 
NGC 1275    &  0.017559 & J0319.7+4130 & J0319.8+4130 & J0319.8+4130 \\  
IC 310      &  0.018940 &                  & J0316.6+4119 & J0316.6+4119 \\
4C+39.12    &  0.020591 &                  &                   & J0334.2+3915 \\
3C 264      &  0.021718 &                  &                   & J1145.1+1935 \\
NGC 6251    &  0.024710 & J1635.4+8228 & J1629.4+8236 & J1630.6+8232 \\   
3CR 78      &  0.028653 & J0308.3+0403 &                   & J0308.6+0408 \\                    
3C 189      &  0.042836 &                  &                   & J0758.7+3747 \\
3C 303      &  0.141186 &                  &                   & J1442.6+5156 \\
PKS 0625$-$35 &  0.054594 & J0627.3-3530 & J0627.1-3528 & J0627.0-3529 \\
\hline
\noalign{\smallskip}
FR\,IIs & & & \\
\hline
\noalign{\smallskip}
Fornax A    &  0.005871 &                  & J0322.4-3717 & J0322.5-3721 \\ 
3C 111      &  0.048500 & J0419.0+3811 &                   & J0418.5+3813c \\                      
Pictor A    &  0.035058 &                  &                   & J0519.2-4542 \\
\hline
\noalign{\smallskip}
CSSs & & & \\
\hline
\noalign{\smallskip}
3C 286      & 0.849934 & & & J1330.5+3023 \\
4C +39.23B & 1.210000 & & & J0824.9+3916 \\
\noalign{\smallskip}
\hline
\end{tabular}
\end{table*}

As summarized in Table~\ref{tab:radiogalaxies} there were 6 FR\,Is listed in the 1FGL catalog;  their number increased to 7 in the 2FGL and to 12 in the 3FGL. The radio galaxies in the first release of the \fer\ catalog were not the closest cases but, as noticed by \citet{dimauro14b}, the FR\,Is detected are those with the greater radio luminosity of the core. This study was based on the correlation found between the radio core luminosity at 5 GHz and that in the $\gamma$-rays, similar to the radio-$\gamma$-ray connection of the blazars that thus seems to be valid also for their parents.

The FR\,Is detected by \fer\ are characterized by power-law spectra and no signs of variability, with the exception of NGC 1275 \citep[e.g.][]{kataoka10}.  This variable radio galaxy was indicated as a tentative association in the COS-B survey \cite{strong83} but did not appear in any of the EGRET catalogs; then it was detected in the \fer--LAT bright source list (0FGL).   M 87 is known to be a variable X-ray \citep[see e.g.,][]{harris03,harris11} and TeV source \citep[see e.g.][]{acciari10,abramowski12}. \fer\ observations did not reveal any variation since its discovery \citep{abdo09i,hada14}. The lack of $\gamma$-ray variability in FR\,Is radio galaxies is consistent with the scenario in which the former class is the parent population of the latter simply viewed at larger angle along the line of sight. 

The emission of MeV-GeV radiation in radio galaxies is due to the non-thermal particles accelerated in their radio jets, the same interpretation as for blazars. These FR\,Is are generally included within the sample of misaligned AGN, due to the orientation of their kpc-scale jets, but there might be the chance that the inner regions of these jets are indeed still pointing along the line of sight and then bend in a different direction at larger distance from the core. This is also consistent with the scenario in which BL Lac objects are relatives of FR\,Is. The ratio of the number of known $\gamma$-ray BL Lacs and that of FR\,I radio galaxies has remained almost constant when comparing the 1FGL and the 3FGL catalogs, again in agreement with the idea that the former are relatives of the latter.  It is not possible to test this hypothesis for the $\gamma$-ray FR\,IIs, because their number is too low.  

The $\gamma$-ray emission from the radio galaxy IC 310 \citep[see e.g.,][]{neronov10a}, clearly showing a blazar-like behavior. The latter, originally classified as a head-tail galaxy \citep{sijbring98} shows a VLBI compact jet structure (Kadler et al. 2012) and rapid TeV variability (Aleksic et al. 2014) as short as minutes time scale, such as seen in BL Lac objects as PKS 2155$-$304 \citep{aharonian07} or Mrk 501 \citep{albert07}. 

Centaurus A was already listed as $\gamma$-ray source in the 0FGL, thus confirming the association with the EGRET object 3EG J1324$-$4314 \citep{hartman99,sreekumar99}.  This FR\,I-type radio galaxy has bright radio lobes located on the northern and southern sides with respect to the core, clearly visible and dominating the radio morphology at low frequencies (i.e.. below $\sim$1GHz). Thanks to the improved sensitivity and PSF of \fer\ with respect to previous $\gamma$-ray missions, and to the small distance of this radio galaxy, it was possible to separate the contribution of the Centaurus A extended structures from the nuclear radiation in the $\gamma$-ray band \citep{abdo10e,abdo10f}. This was definitely one of the major \fer\ discoveries in the extragalactic sky, showing that high-energy particles fill the lobes, either transported from the core region or accelerated in place. Even after the release of the 3FGL, Centaurus A is the only AGN with a significant detection of spatially extended $\gamma$-ray emission. Other nearby radio galaxies with large radio extensions, exceeding the size of the \fer\ PSF, were analyzed, including Centaurus B \citep{katsuta13} and NGC 6251 \citep{takeuchi12}, but no clear detection of extended $\gamma$-ray emission has yet been found. On the other hand,   extended $\gamma$-ray structures, similar to lobes of a radio galaxy, were also detected in our Galaxy, now known as ``Fermi Bubbles'' \citep{su10,ackermann14c}.
\begin{figure}[]
\includegraphics[height=12.cm,width=9.4cm,angle=-90]{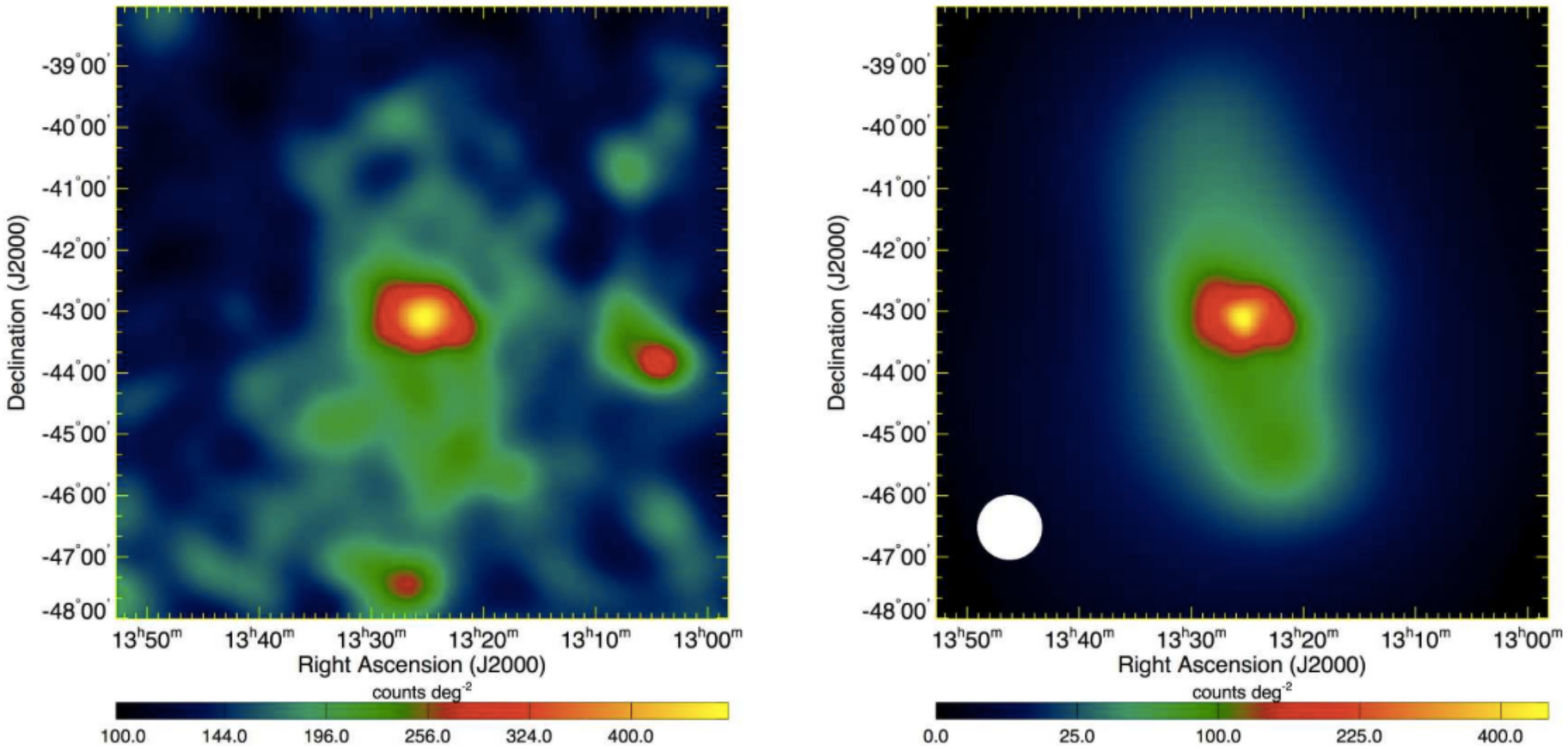}
\caption{\fer--LAT smoothed count maps centered on Centaurus A. In both panels,  the emission  from the Galactic and  isotropic $\gamma$-ray components were subtracted. The images are shown before (Left panel) and after (Right panel) subtraction of point sources in the field. The white spot represents the LAT PSF.  \citep{abdo10e}. (Figure courtesy of the \fer\ Large Area Telescope Collaboration, under the copyright permissions of Science).}
\label{fig:cenAlobes}
\end{figure}

The situation is quite different for the \fer\ FR\,II radio galaxies, first of all because their number is smaller than the  $\gamma$-ray FR\,Is. There was only one FR\,II detected in the 1FGL, namely 3C 111, not present in the 2FGL catalog, where the only FR\,II associated with a \fer\ source is Fornax A. Both are listed in the 3FGL, together with Pictor A.
These \fer\ FR\,II radio galaxies have very prominent blazar-like behavior, suggesting that their $\gamma$-ray emission is due to their inner jets.  They are, however, the radio sources with some of the brightest radio lobes in the whole sky, clearly detected in X-rays. Thus given the \fer\ detection of Centaurus A lobes it is possible that for the FR\,IIs the $\gamma$-ray emission is partially contaminated by that of their extended components. Of these, only Fornax A is close enough that extension might be detectable by \fer--LAT.  This scenario might be supported by a future detection of other FR\,IIs with bright radio-to-X-ray lobes, such as Cygnus A.

An additional type of radio galaxy has been also recently been added to the zoo of extragalactic $\gamma$-ray emitters: Compact Steep Spectrum (CSS) radio sources \citep{odea98}. These are compact, powerful radio sources with well-defined peaks in their radio spectra at $\sim$100 MHz,  entirely contained within their host galaxies, with size of the order of $\sim$15 kpc. CSSs are interpreted as young and evolving radio sources that could become large-scale radio galaxies  toward the end of their evolution. Their strong evolution suggests that at least some of them evolve to become lower luminosity FR\,Is. 

A preliminary investigation based on the largest sample of X-ray-detected young and compact extragalactic radio sources did not find any statistically significant detection of CSS sources in the \fer\ data, and the model-predicted $\gamma$ fluxes for the sample were consistent with the 2FGL flux threshold and compatible with the derived upper limits \citep{migliori14}. This situation changed with 3FGL/3LAC. Two potential CSSs have been associated in the 3LAC,  3C 286  and 4C +39.23B.  In the latter case the nearby blazar 4C +39.23 could be responsible for the $\gamma$-ray emission, making this association less reliable.  These sources have been also classified as Steep Spectrum Radio Quasars in the literature.

\subsection{Seyfert galaxies as $\gamma$-ray sources}
\label{sec:sey}

Although Seyfert galaxies as a class are not found to be $\gamma$-ray sources \citep{ackermann12e}, \fer\ has detected  $\gamma$-ray emission from several radio-loud narrow line Seyfert 1 galaxies \citep[RLNLSy1;][]{abdo09e}. Narrow-line Seyfert 1 galaxies \citep{komossa06} are a subclass of AGN exhibiting exceptional emission-line and continuum properties with respect to the ``classical'' Seyfert galaxies. They have all the properties of type 1 Seyfert galaxies but show characteristics of  Seyfert 2 galaxies, including the narrowest AGN Balmer lines, strongest AGN Fe II emission, and peculiar properties in  X-rays. A small fraction of sources classified as NLSy1 show strong radio emission and a flat radio spectrum; these represent the subclass called radio-loud narrow line Seyfert 1s (RLNLSy1). In addition to their optical and radio properties, these sources show extreme variability \citep{calderone11,foschini12}, and their IR colors are also similar to those of blazars \citep[ e.g.,][]{wibrals,foschini15}. 

The RLNLSy1s represent a third class of $\gamma$-ray AGN in addition to blazars and radio galaxies. At least some of the  $\gamma$-ray RLNLSy1 show rapid time variability at several frequencies, suggesting blazar-like relativistic jets \citep[e.g.,][]{paliya15,foschini15,dammando15a,dammando15b}.  A major difference to support the idea of a third class of active galaxies is the fact that NLSy1 are in general hosted in spiral galaxies, as occur for the Seyfert galaxies, rather than in elliptical galaxies as occurs for blazars and radio galaxies, although many of them are too distant to resolve their host galaxies easily.   

The presence of a relativistic jet in RLNLSy1 stands in contrast to the paradigm that the formation of relativistic jets is a feature of ellipticals. This discovery opens new, intriguing questions on the nature of these objects and on the mechanisms of high-energy emission from radio-loud AGN. Apparent superluminal motion has been also discovered in SBS 0846+513 and is the first direct confirmation of the presence of boosting effects of jets present in RLNLSy1 \citep{dammando13b}. 

On the other hand, RLNLSy1s lacking optical imaging information could be misclassified as narrow-line radio galaxies, since the only difference between these two classes is in their host galaxies. This implies that obtaining precise information about their hosts is crucial. This information has been somewhat hard to determine in one of the first detected cases: 1H 0323+342 \citep[see e.g.,][for a details on the host galaxy]{zhou07,leontavares14}. 

\begin{figure}[]
\includegraphics[height=12cm,width=9.cm,angle=-90]{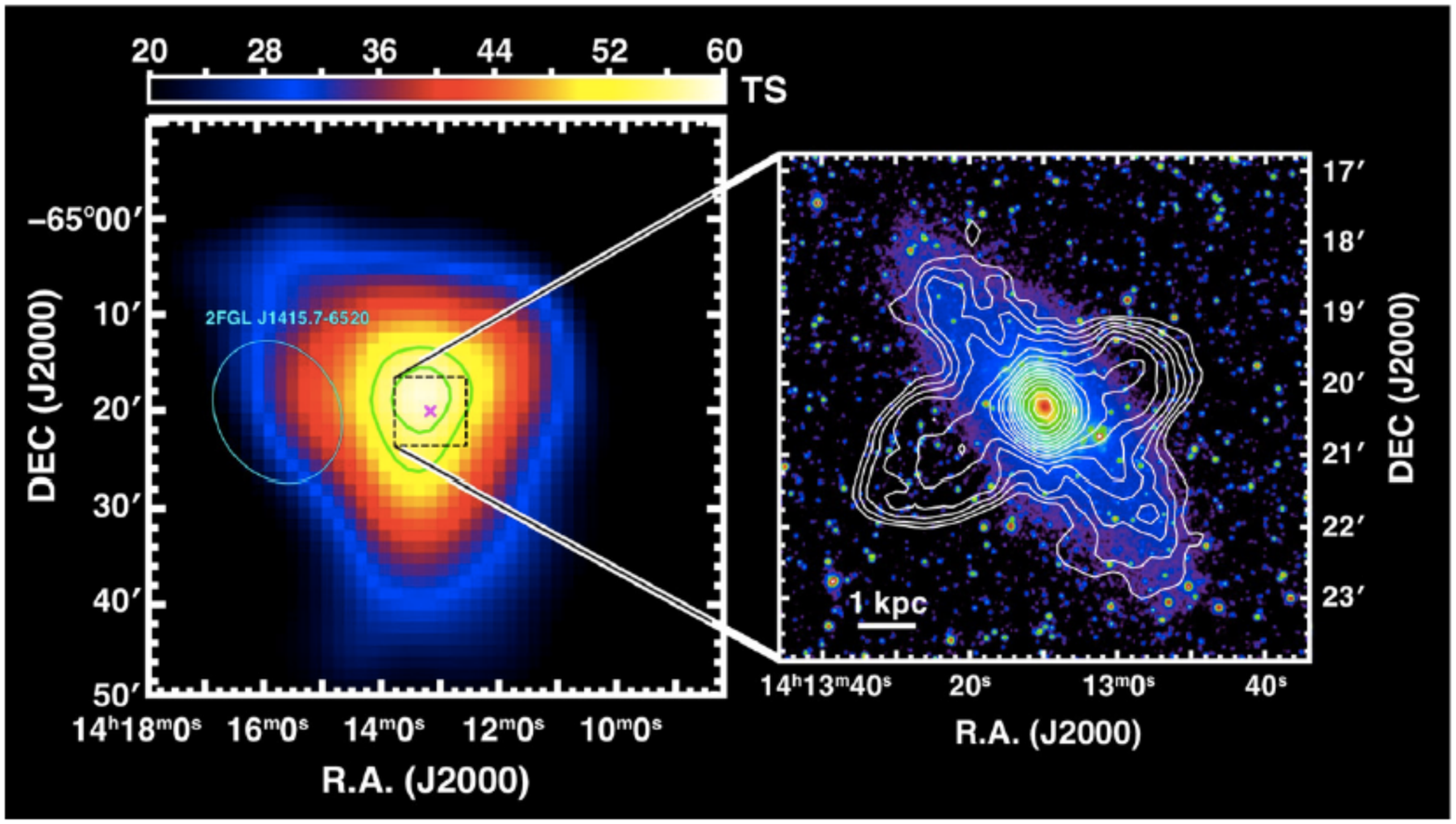}
\caption{Left)  Test Statistic (significance) map for the $\gamma$-ray excess centered on the Circinus Galaxy. Green lines indicate the \fer\ positional uncertainty at 68\% and 95\% level of confidence, respectively. The magenta cross marks the position of the Circinus Galaxy core, while the cyan ellipse corresponds to the positional uncertainty region of the nearby \fer\ source 2FGL J1415.7-6520. The black square corresponds to the area of the right panel, which shows the Australia Telescope Compact Array (ATCA) 1.4 GHz radio contours \citep{elmouttie98}  superposed with the 2MASS H-band color image \citep{jarrett03}. (Figure courtesy of D. M. Hayashida, under the copyright permissions of the Astrophysical Journal)}
\label{fig:circinus}
\end{figure}

Recently, the discovery of a different $\gamma$-ray Seyfert galaxy has been presented: the Circinus Galaxy \citep[Figure \ref{fig:circinus}; ][]{hayashida13}. This is one of the nearest Seyfert 2 galaxies and belongs to the Compton-thick class, since its nuclear X-ray emission is obscured up to about 10 keV. Although the observed $\gamma$-ray spectrum is similar to that of other \fer\--detected star-forming galaxies, its $\gamma$-ray luminosity exceeds by a factor of $\sim$5 the emission expected from the interaction of cosmic rays with the interstellar medium if Circinus were considered as a star-forming galaxy. Since this scenario is still unclear given the large uncertainty in the estimates of the star formation rate and in the expected cosmic-ray power released in the interstellar medium, another possible interpretation was proposed for its \fer\ detection: the $\gamma$ rays in the Circinus galaxy could be due to the particles accelerated in its extended radio lobes, similar to the case of Centaurus A.

The other closest Seyfert 2, Compton-thick, galaxy, NGC 1068, has been also detected by \fer\ \citep[ e.g.,][]{lenain10}, but in this case the origin of its $\gamma$-ray emission is most probably due to the significant amount of star formation present in its host galaxy rather than to the sort of non-thermal processes occurring in the other Seyferts.

\subsection{Physical Implications of gamma-ray emission from Seyfert and radio galaxies}
\label{sec:otherAGN}

The non-blazar AGN seen in the \fer\ data all share one feature: they are radio-loud.  This feature is not surprising, since the same energetic particles required to produce $\gamma$ rays will almost always interact with any local magnetic fields to produce radio emission. What is less obvious is that jets are found in nearly all these sources.  Extensions of the modeling used for the blazar sources are likely candidates to interpret the $\gamma$-ray emission from these AGN.  A remaining puzzle is the lobes seen for Cen A (and hinted at in some other sources), in particular how such energetic particles can be found so far from the central engine. The sample of these non-blazar $\gamma$-ray AGN is small enough that far-reaching conclusions do not seem justified.  As more such sources appear during the ongoing \fer\ mission, more detailed analysis may be possible.

\section{Normal galaxies and Starburst galaxies in the MeV-GeV energy range}
\label{sec:starbursts}

The first known cosmic source of high-energy $\gamma$ rays was our Galaxy.  The $\gamma$-ray-producing interactions of energetic cosmic-ray particles with interstellar matter and photon fields in the Milky Way set the stage for studies of other galaxies with populations of stars whose endpoints were likely to produce cosmic rays. The diffuse emission of our Galaxy at energies above $\sim$100 MeV is due to bremsstrahlung, pion production and decay, and inverse-Compton scattering, and its investigation provides insights into the origin and transport of cosmic rays.

Before \fer\ was launched the only nearby galaxy known as a $\gamma$-ray emitter was the Large Magellanic Cloud, seen by {\it CGRO}/EGRET as a spatially extended source with $\gamma$-ray flux compatible with a cosmic-ray population similar to that of the Milky Way.  \fer--LAT has added to our knowledge of star-forming galaxies in two ways. 

First,  the \fer\ discovery of  high-energy $\gamma$-ray emission from two nearby starburst galaxies was reported \citep{abdo10h}. Steady point-like emission above 200 MeV was  detected for sources spatially coincident with the starburst galaxies M82 and NGC 253. Their $\gamma$-ray fluxes were also found in agreement with the theoretical expectations of $\gamma$-ray emission  from the interaction of cosmic rays with local interstellar gas and radiation fields, similar to processes in our Galaxy. This was the first evidence for a link between massive star formation and $\gamma$-ray emission in star-forming galaxies.

Second, the Small Magellanic Cloud (SMC), the Large Magellanic Cloud (LMC) and the Andromeda galaxy were all detected  as spatially extended $\gamma$-ray sources
by the \fer--LAT \citep[][respectively]{abdo10i,abdo10j,abdo10k}. The observed $\gamma$-ray flux was  used to set an upper limit on the average  density  of cosmic-ray nuclei in the SMC of $\sim$15\% of the value measured locally in the Milky Way. High-energy pulsars in the SMC might account for a substantial fraction of the $\gamma$-ray flux, making the inferred density of cosmic rays even lower. The analysis of the LMC revealed that the massive star-forming region known as 30 Doradus is a bright $\gamma$-ray source in the LMC. The high-luminosity pulsars PSR J0540$-$6919 and PSR J0537$-$6910 lie in this region and may contribute to the $\gamma$-ray emission.  The Andromeda galaxy shows a $\gamma$-ray flux about half that of the Milky Way, implying a lower cosmic-ray density there. 

\begin{figure}[]
\begin{center}
\includegraphics[height=8.1cm,width=8.3cm,angle=0]{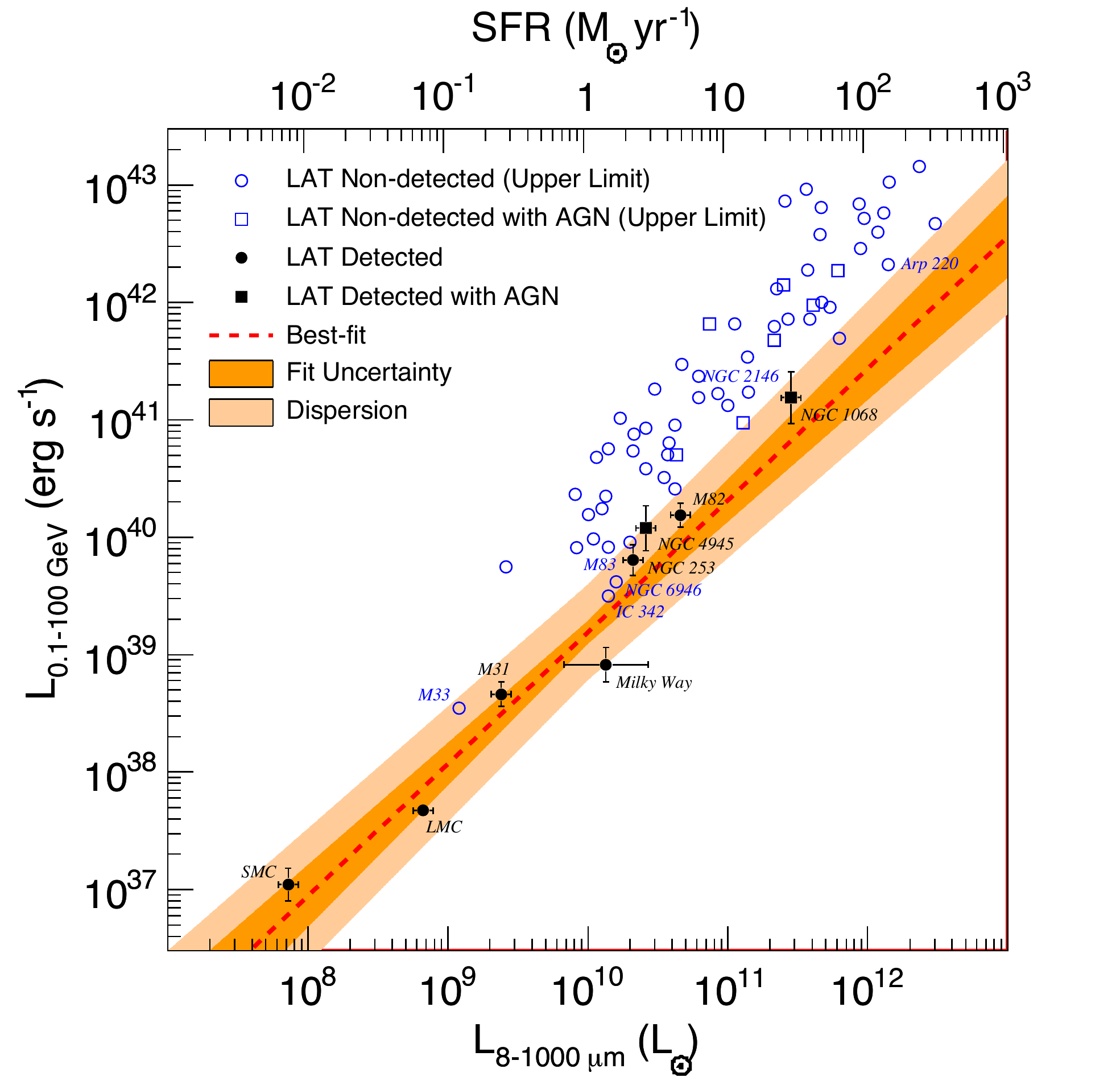}
\caption{Gamma-ray luminosity (0.1--100 GeV) versus total IR
luminosity (8--1000 $\mu$m) for normal and starburst galaxies \citep{ackermann12d}. Galaxies significantly detected by the LAT are indicated with filled
symbols whereas galaxies with $\gamma$-ray flux upper limits (95\% confidence level) are marked with open symbols. Galaxies
hosting \textit{Swift}-BAT AGN are shown with square markers.  IR luminosity uncertainties for the non-detected galaxies are omitted for clarity, but are
typically $\sim0.06$ dex. The upper
abscissa indicates SFR estimated from the IR luminosity. (Figure courtesy of the \fer\ Large Area Telescope Collaboration, under the copyright permissions of the Astrophysical Journal).}
\label{fig:SFR}
\end{center}
 \end{figure}

In all these galaxies, the cosmic-ray density inferred from the $\gamma$-ray flux is nearly proportional to the estimated Star Formation Rate (SFR), supporting the concept that cosmic rays are accelerated in massive star-forming regions as a result of the large amounts of kinetic energy deposited by stellar winds and supernova explosions of massive stars into the interstellar medium. A selected sample of nearby 69 dwarf, spiral, and luminous and ultraluminous IR galaxies has been investigated to search for $\gamma$-ray emission. A combined statistical analysis including detected sources and flux upper limits found further evidence for quasi-linear scaling  between $\gamma$-ray luminosity and both radio continuum luminosity and total iIR luminosity in both quiescent galaxies of the Local Group and low-redshift starburst galaxies \citep{ackermann12d}. Figure~\ref{fig:SFR} shows one representation of this correlation over several orders of magnitude, using IR flux as a proxy for SFR.

Upper limits for other galaxies in the Local Group are also compatible with expectations  from the interaction of  cosmic rays with the interstellar medium within these galaxies. The estimates  in the cases of M 33 and M 83 appear to indicate that if the \fer\ mission extends beyond 10 years we should detect both of them \citep[see e.g.,][]{abdo10k}. 

There is an additional important consequence arising from this \fer\ discovery.
Star-forming galaxies provide a considerable contribution to the extragalactic $\gamma$-ray background, since they are guaranteed reservoirs of cosmic rays that interact with the interstellar medium to produce GeV emission (see section \ref{sec:egb} for more details).

\section{Extragalactic $\gamma$ rays and cosmology}
\label{sec:cosmo}
The extragalactic $\gamma$-ray sky offers a unique opportunity to explore phenomena at cosmological distances. In particular, \fer\ allows us to obtain: (i) independent information about the evolution  of $\gamma$-ray sources, computing their contribution to the extragalactic $\gamma$-ray background (EGB) (ii) an indirect estimate of the optical/UV EBL and (iii) a measurement or constraint on the intergalactic magnetic field.

\subsection{Source populations and their contribution to the extragalactic gamma-ray background}
\label{sec:egb}

Despite the dramatic increase in both numbers and classes of extragalactic $\gamma$-ray sources, there remains a nearly isotropic EGB flux left when all sources and the Galactic diffuse radiation have been subtracted from the \fer\ all-sky map.  A recent analysis shows the EGB energy spectrum between 0.1 and 820 GeV to be well represented by a power law with an exponential cutoff above $\sim$ 300 GeV \citep{ackermann15b}.  The high-energy cutoff is consistent with the effect of the EBL attenuation described in \ref{sec:ebl}.  Although the spectral shape is consistent, the normalization of the EGB varies by $\pm$ 15\% depending on the details of the Galactic foreground model. At energies above 100 GeV, half or more of the EGB is resolved into known sources, primarily BL Lac objects. 

The question remains whether all the EGB  results from unresolved sources or whether some truly diffuse (and possibly exotic) process is needed.  Not surprisingly, the answer is model dependent. The extensive \fer\  results on blazars have now provided well-defined luminosity functions for both Bl Lacs and FSRQs  \citep{ajello15}.  In addition to the uncertainty in the EGB itself, however, the $\gamma$-ray luminosity functions of some other potential classes (such as star-forming galaxies) are not well determined due to the small number of examples, and the cosmological evolution of the various classes must be estimated.  Blazars clearly contribute, with earlier estimates ranging from a fairly small fraction \citep{abdo10g,dimauro14a,venters09} to most of the EGB \citep[see e.g.][]{stecker11}.  Examples of other classes include star-forming galaxies \citep{fields10,makiya11,ackermann12d} and misaligned AGN \citep{inoue11, dimauro14b} as well as a potential contribution arising from lobes of FR\,II radio galaxies \citep[see e.g.][]{massaro11c}. 

\begin{figure}[]
\includegraphics[height=12.cm,width=8.cm,angle=-90]{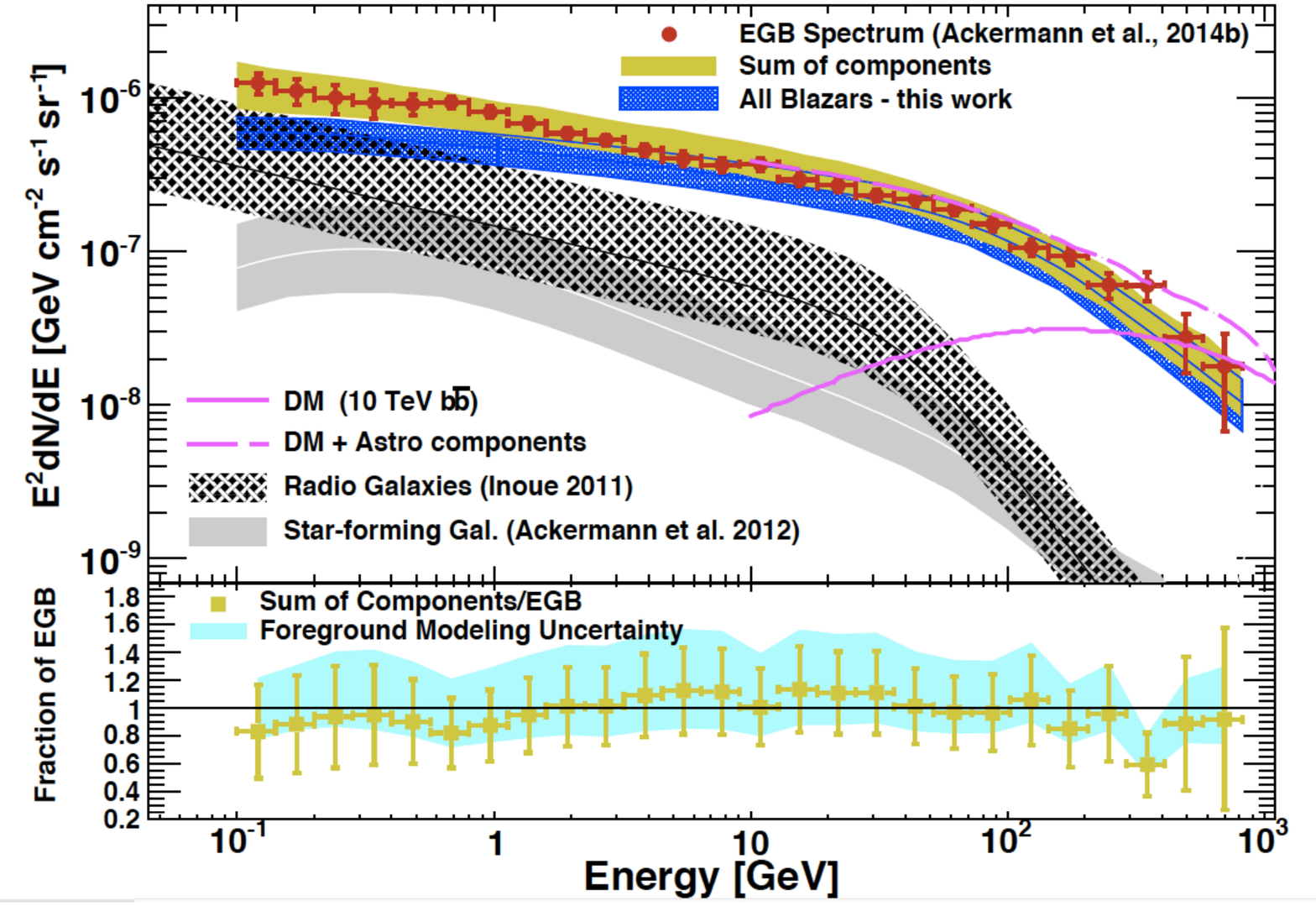}
\caption{Integrated $\gamma$-ray contributions to the EGB from blazars \citep{ajello15}, star-forming galaxies \citep{ackermann12d}, and radio galaxies \citep{inoue11},   compared to the measured EGB intensity \citep{ackermann14a}. An example of a hypothetical DM-induced 
$\gamma$-ray signal ruled out by the analysis is shown by the solid pink line, and summed with the other contributions
(long-dashed pink line). The signature of the EBL absorption is also visible as a high energy cutoff. (Figure courtesy of Prof. M. Ajello, under the copyright permissions of the Astrophysical Journal).}
\label{fig:EGB}
 \end{figure}

Figure \ref{fig:EGB} shows a recent analysis \citep{ajello15}, including the EGB spectrum itself, with the foreground modeling uncertainty,  plus several of the possible components discussed above.  Within the assumptions and uncertainties, the EGB can be fully explained as the sum of resolved plus unresolved sources.  The important implication of these results is that little phase space is left for diffuse processes such as DM interactions. \citet{ajello15} conclude that this modeling sets upper limits on DM self-annihilation cross sections that are similar to those from independent types of analysis \cite[e.g.][]{ackermann14a}. 

\subsection{Probing the extragalactic background light with $\gamma$-ray sources}
\label{sec:ebl}

The light emitted by stars and galaxies and their cosmological evolution through the history of the Universe is imprinted in the EBL. The UV EBL is of particular interest, because such photons were involved in the reionization of the early Universe.  Direct measurement of the optical/UV EBL by deep and large surveys  is challenging due to foreground sources such as the zodiacal light \citep[ e.g.,][]{stecker07,reimer07b}.  The local Universe is completely transparent to $\gamma$ rays, but high-energy photons can be attenuated by interactions with optical and UV photons via the pair production process:  $\gamma$ + $\gamma$ $\rightarrow$  $e^+$ +  $e^-$, \cite[e.g.][]{salamon98}. The $\gamma$-ray spectra of \fer\ sources at moderate redshifts (i.e., $z>0.1$) should show a depletion at the upper end of the LAT energy range due to this opacity effect, enabling an indirect measurement of the EBL.  This effect has been seen in low-redshift blazars at TeV energies by ground-based  telescopes (e.g.  H.E.S.S reference), but these higher-energy $\gamma$ rays interact primarily with the IR \citep{abramowski13}.
\begin{figure}[]
\begin{center}
 \includegraphics[height=7.2cm,width=6.cm,angle=0]{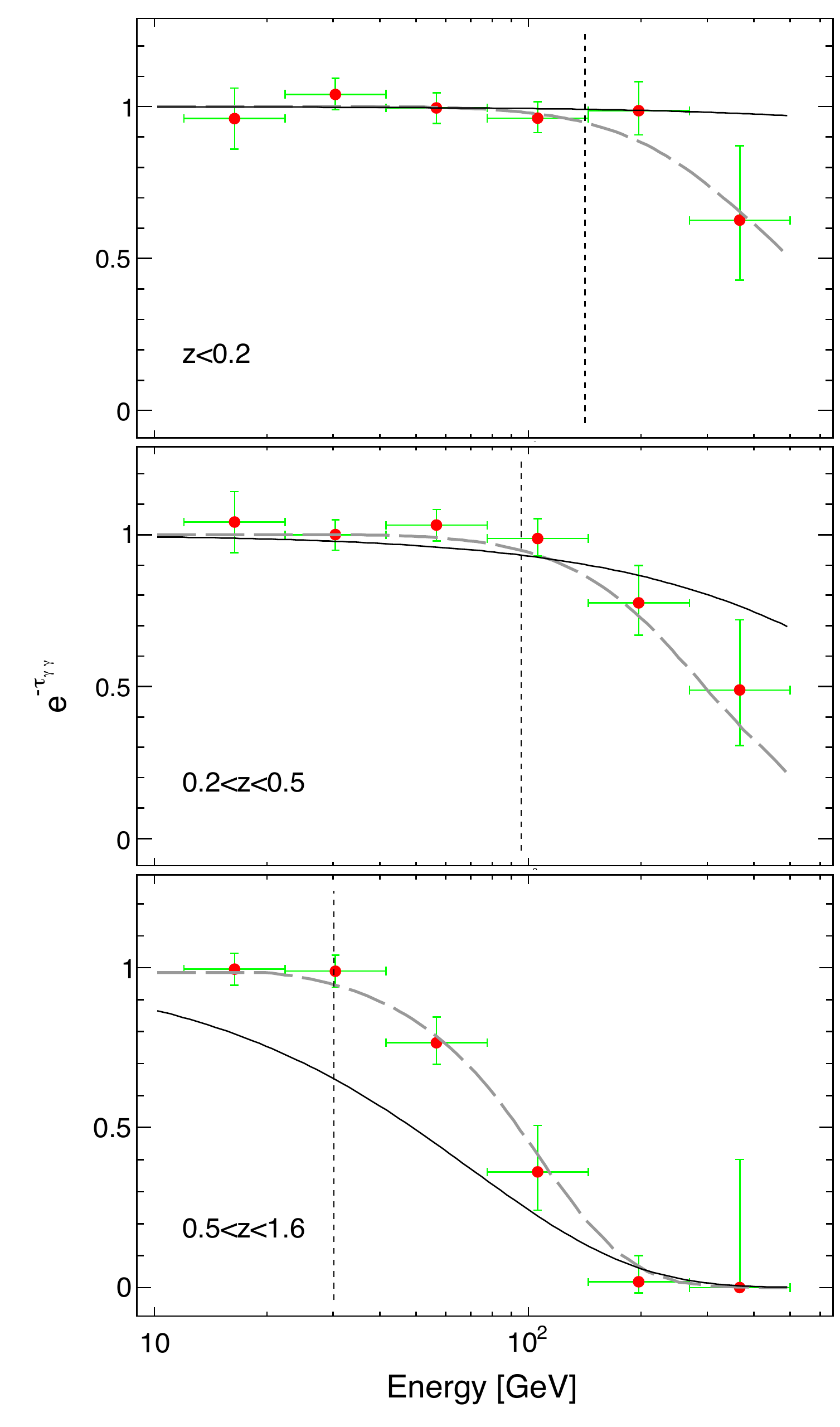}
 \caption{Absorption feature in the spectra of BL Lacertae objects as a function of increasing redshift (data points, from top to bottom, from \citealt{ackermann12c}). The dashed curves show the attenuation expected for the sample of sources by averaging, in each redshift and energy bin, the opacities of the sample and multiplying this average by the best-fit scaling parameter obtained independently in each redshift interval. The vertical line shows the critical energy  below which $\sim$5\% of the source photons are absorbed by the EBL. The thin solid curve represents the best-fit model assuming that all the sources have an intrinsic exponential cut-off. The absorption model is a significantly better representation of the data. (Figure courtesy of Prof. M. Ajello, under the copyright permissions of Science).}
\label{fig:opacity}
\end{center}
 \end{figure}

Because this absorption is expected at the upper end of the \fer--LAT energy range, observations of single  objects lack the statistics to reveal such EBL absorption features \citep[e.g. the case of TXS 0536+135][]{orienti14}. Instead, a different approach has been used. A composite analysis of 150 BL Lacs at different redshifts out to z = 1.6 showed a significant improvement when the $\gamma$-ray spectral fits included a high-energy absorption feature consistent with that expected by the EBL effect \citep{ackermann12c}.  The signature of the absorption feature in the BL Lac spectra as a function of increasing redshift is shown in Figure \ref{fig:opacity}.
This indirect measurement of the EBL flux density in the optical-UV energy range was close to the limits obtained from direct measurements, constraining the maximum redshift of low-metallicity stars in the early Universe. Recently, the EBL absorption feature has been also clearly observed in the integrated EGB spectrum, as discussed in the previous section \citep[see Figure \ref{fig:EGB} and e.g.,][]{venters11}.

The first statistically significant detection of the cosmic $\gamma$-ray horizon (CGRH) estimated independently of any EBL model was also recently reported \citep{dominguez13a}. This fundamental cosmological parameter provides an estimate of the opacity of the Universe. This CGRH estimate has been computed via modeling quasi-simultaneous SEDs of a blazar sample to determine the expected unattenuated $\gamma$-ray fluxes.  These were compared with the observations. The estimated CGRH is compatible with the current knowledge of the EBL.

A measurement of the expansion rate of the universe (i.e., Hubble constant, $H_0$) was also estimated using the $\gamma$-ray absorption observed in the spectra of \fer\ blazars \citep{dominguez13b}. The Hubble constant determined with this approach is compatible with present-day measurements using well-established methods such as local distance ladders and cosmological probes. Both the measurement of the expansion rate of the Universe using $\gamma$ rays and the detection of the CGRH were listed as potential aims of the \fer\ mission \citep{hartmann07} and have been successfully achieved.

\subsection{\fer\ constraints on the intergalactic magnetic field}
\label{sec:igmf}
Nearby \fer\ blazars with hard $\gamma$-ray spectra are often detected at TeV energies. These TeV photons, when emitted at moderately high redshifts (i.e., z$>$0.2-0.5) and interacting with  background photons  in the infrared to the optical band, could efficiently convert into electron-positron pairs with Lorentz factors up to 10$^{6-7}$. $Gamma$-ray emission arising from these particles interacting with the cosmic microwave background could be detected by \fer\ at GeV energies. The $\gamma$-ray spectrum and flux have the imprint of the intergalactic magnetic field (IGMF) \citep[e.g.,][and references therein]{neronov10a,tavecchio10,tavecchio11} but are also affected by the EBL model.
The possibility to observe the IGMF feature in the \fer\ blazar spectra \citep[see also][]{dermer11,dolag11,bonnolli15} is still controversial, because the cascade might not occur if the pairs lose energy to plasma beam instabilities \citep[ e.g.,][]{broderick12,chang12}. Arguments against \citep[ e.g.,][]{miniati13,sironi14} and in favor \citep[ e.g.,][]{schlickeiser12,chang14} of these plasma processes are still under debate. 

Another effect of such a cascade, depending on the IGMF, is the possibility of measuring a ``pair halo'' around a TeV blazar.  When the TeV photons convert into electron-positron pairs, the IGMF can cause their trajectories to diverge from their original paths.  As these particles Compton scatter  CMB photons to GeV energies, these $\gamma$ rays form a spatially extended halo around the original source.  If these haloes could be measured by \fer\--LAT, they would provide a direct measurement of the IGMF.  Careful comparisons of the LAT PSF with observed TeV blazars have shown no convincing evidence of such extended emission, thus placing constraints on the IGMF to be greater than $\sim 10^{-15}$ gauss \citep{ackermann13b,neronov10b}.

An IGMF imprint might also be seen in the angular anisotropy of the EGB \citep{venters13}. A high value of the IGMF could deflect electron-positron pairs produced in the cascades, making GeV photons more isotropic. Thus the EGB spectrum of the anisotropies could probe the IGMF strength.

\section{Unidentified Sources in the $\gamma$-ray Sky: A Persistent Puzzle}
\label{sec:ugs}

\subsection{The population of unidentified gamma-ray sources}
\label{sec:ugsdescr}

One of the main scientific objectives of the \fer\ mission, identified well before its launch, was ``resolve the $\gamma$-ray sky: unveiling the origin of the unidentified $\gamma$-ray sources''. As highlighted by several authors after the end of the {\it CGRO} mission, in the 3EG catalog well over half the $\gamma$-ray sources had no likely counterpart at low energies (e.g. Reimer 2005, Thompson 2008). The main interest in identifying/associating these sources with their low energy counterparts is the chance to discover something new and unexpected such as: (i) a new class of sources  or (ii) an unknown behavior of known $\gamma$-ray emitters.

Although \fer--LAT has confirmed nearly all the EGRET sources that had associations or identifications, it has not resolved the nature of most of the 170 EGRET UGS, in many cases not even confirming their existence.  The 120 3EG sources with LAT counterparts are indicated in the 3FGL catalog \citep{acero15}, with pulsars representing the largest addition to the associations with known objects. Three factors appear to account for the absence of more of the EGRET sources in the LAT catalogs: (i) The LAT analysis includes a major improvement in understanding the diffuse Galactic $\gamma$-ray emission.  The LAT model, for example, contains a component known as ``dark gas''  \citep{grenier05}, whose properties are inferred from IR measurements of dust rather than from direct observations.  Many of the EGRET UGS are now recognized as  cosmic-ray interactions with these gas clouds.  (ii) Many of the EGRET UGS were transient detections, seen in only one of the {\it CGRO} viewing periods. (iii) A number of the EGRET UGS were detections just above the EGRET threshold and could have been spurious. \citet{hartman99} estimated their number to be between 11 and 21. 

The \fer--LAT team built on the EGRET legacy by developing the systematic process to search for source associations described in Section \ref{sec:hunt}, using selected catalogs as a reference.  The significant improvements of \fer--LAT  in source localization have clearly simplified this challenging search. Thanks to more LAT sources, continual improvements in $\gamma$-ray analysis, more detailed background modeling, and expanding catalogs, this system has produced a larger number of associated sources with each catalog release.  In each catalog, however, there remains a substantial fraction of UGS:  33\% of the sources in 3FGL.  The absolute number of UGS remains large, potentially hiding new discoveries.  In the following section, we describe some of the methods that have been used and continue to be used to find new associations for \fer\ sources.

\subsection{Methods for resolving unidentified gamma-ray sources}
\label{sec:ugssearch}

When the standard LAT association pipeline fails to find a candidate counterpart, we have to look beyond the reference catalogs.  Three approaches to searching for new associations are:

\begin{enumerate}
\item Follow-up observations at one or more wavelengths may be able to find fainter or time-variable objects of familiar types that do not appear in the catalogs. 

Radio follow-up observations of the \fer\ UGS, mainly driven by the radio $\gamma$-ray connection of the \fer\ blazars, have been performed since the beginning of the mission \cite[e.g.,][]{kovalev09,hovatta12,petrov13,hovatta14,schinzel15}. 
A tight connection between the IR sky seen by the {\it Wide-Field Infrared Survey Explorer} \cite[\wse;][]{wright10} and the \fer\ one has been recently discovered for blazars \citep{paper1,paper2,ugs1}. This previously unknown connection significantly decreased the fraction of UGS with no assigned counterpart  \citep{paper3,paper4,ugs2,wibrals}. Additional studies have been also carried out with near-infrared observations \citep{raiteri14} as well as in the sub-millimeter range \cite[e.g.,][]{giommi12,lopez13}. 
A \swf\ X-ray survey for all the UGS listed in the \fer--LAT source catalogs is still ongoing\footnote{\underline{http://www.swift.psu.edu/unassociated/}} 
\cite[e.g.,][]{mirabal09,paggi13,takeuchi13,stroh13,acero13}. At the same time, additional X-ray pointed observations carried out with \chn\ and \suz\ have been very productive, in particular when performed on the confused region of the Galactic Plane \cite[e.g.,][]{maeda11,cheung12,kataoka12,takahashi12}, discovering several low-energy counterparts.

\item Unexpected relationships between $\gamma$-ray sources and those seen at other wavelengths, similar to the one found using the {\it WISE} colors, may emerge. 

Several recent low-frequency radio observations (i.e., below $\sim$1~GHz) revealed an unexpected spectral connection between some $\gamma$-ray blazars and low-frequency radio sources, simplifying the search for blazar-like sources \citep{ugs3,ugs6} for a limited fraction of the sky where radio observations with good resolution are accessible at MHz frequencies.  The extension of the radio-$\gamma$-ray connection to lower frequencies was found while searching for low-energy counterparts of UGSs using the radio survey by the Westerbork Synthesis Radio Telescope \citep{ugs3,ugs6}. 

Another new discovery suggests a population of radio-weak blazars. For two UGS listed in the 2FGL catalog, optical spectroscopic observations  of {\it WISE} sources  having blazar-like IR colors but with no radio counterpart listed in any of the major surveys \citep{paggi14,ricci15} revealed two radio-weak, but not radio quiet, sources with a featureless optical continuum as seen in BL Lacs. Since they do not show up in any of the comparison catalogs used to search for \fer\ counterparts, they were not associated in the 3FGL. This discovery, if confirmed, will change our view on the UGSs because it will allow the possible existence of $\gamma$-ray extragalactic sources with no bright radio counterpart.

\item The properties of the $\gamma$-ray sources as a statistical ensemble can be used to suggest the type of counterpart that might be found. 

In addition to the studies of individual $\gamma$-ray sources, statistical studies based on the spatial, spectral and temporal properties of the \fer\ UGS population help to suggest the nature of the potential counterparts for the UGS.

The sky distribution of the UGS is fairly concentrated at low Galactic latitudes. However if selecting only those sources with no  $\gamma$-ray analysis flags (which indicate uncertainties in the analysis) it becomes more isotropic. Thus the most promising class of sources that could be recognized among the UGS could be the AGN, in particular unknown BL Lacs, strongly supported by the recent optical spectroscopic campaigns that are finding these objects within the \fer\ positional uncertainty regions of the UGS sample. Other $\gamma$-ray source parameters that are useful for classification are the variability index (e.g., blazars often show strong variability) and the spectral shape (e.g., pulsars have power-law spectra with exponential cutoffs at a few GeV).  For example, \citet{mirabal12} applied a Random Forest data-mining technique to the 2FGL unassociated sources using a combination of the $\gamma$-ray observables.  Their analysis suggested that a large fraction of the unassociated sources are AGN-like, but some have characteristics of pulsars.  Similar analyses using a variety of techniques \cite[e.g.,][]{ackermann12b,hassan13,doert14} have concluded that most of the high-latitude UGS appear blazar-like, again suggesting that targeted follow-up observations should be seeking that type of source. 

\end{enumerate}

\begin{figure}[]
\includegraphics[height=7.8cm,width=5.8cm,angle=0]{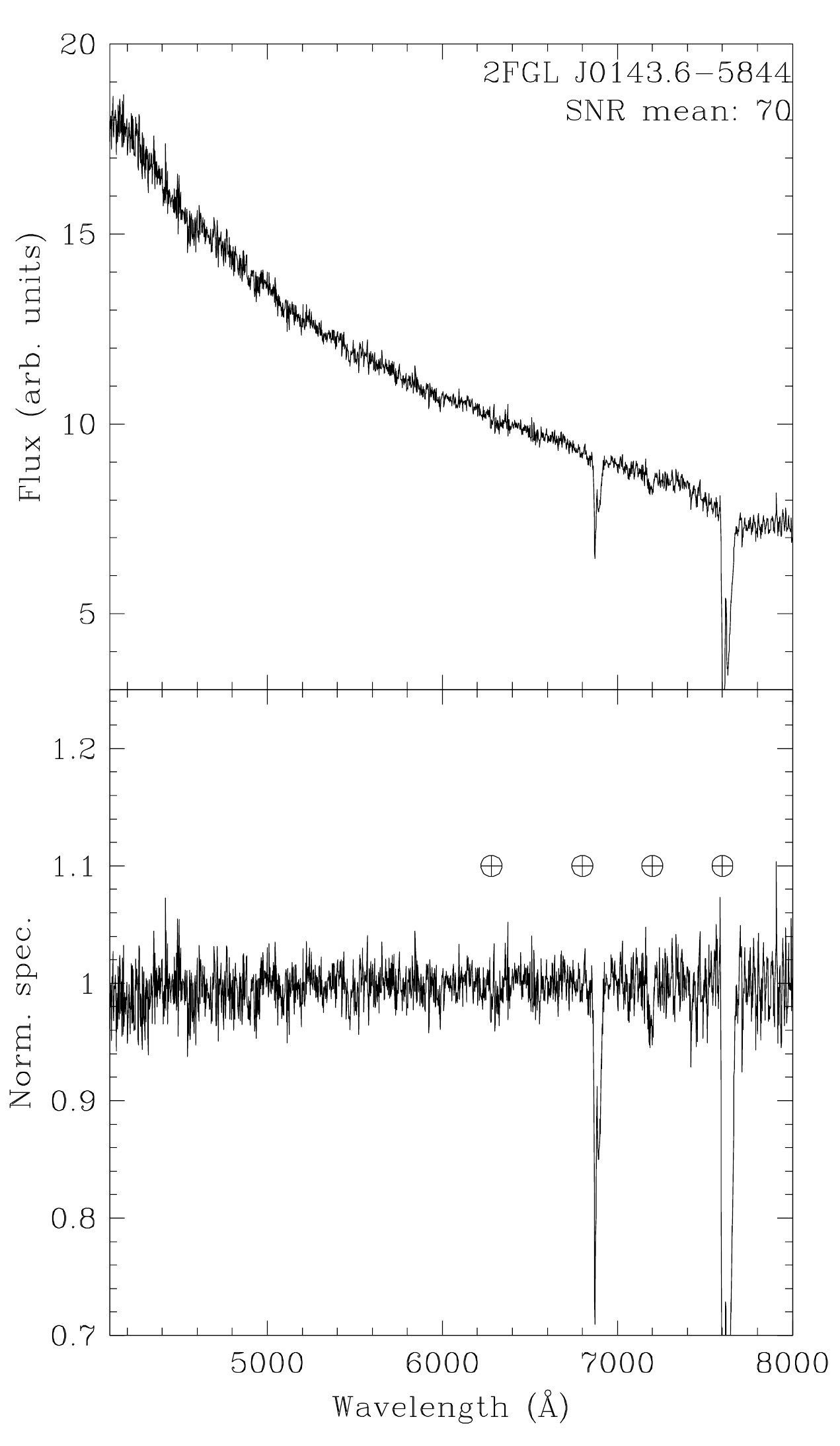}
\includegraphics[height=8.0cm,width=6.2cm,angle=0]{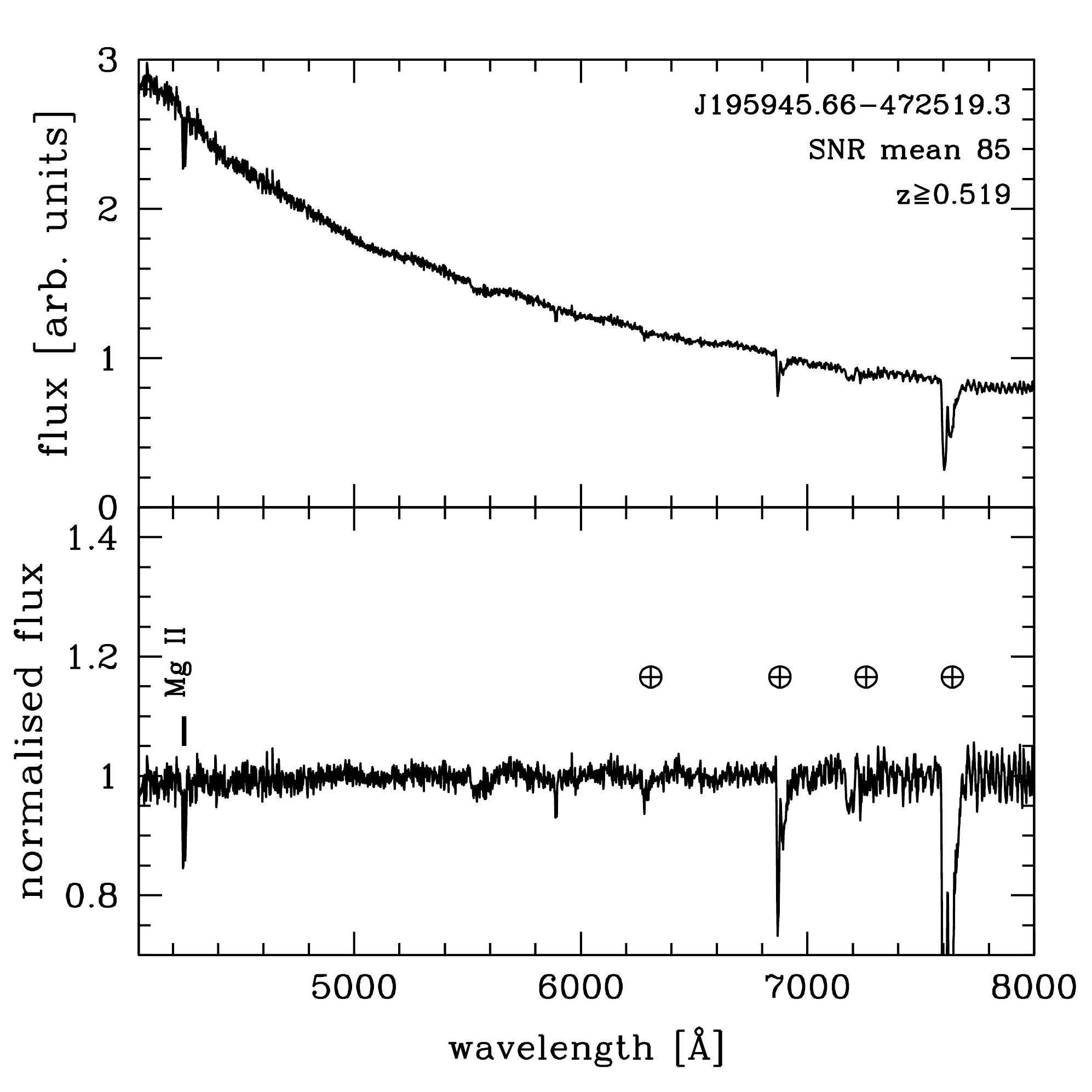}
\caption{Left) The optical spectrum of WISE J014347.41-584551.4 on the basis of the IR colors as potential counterpart for the unidentified $\gamma$-ray source: 2FGL J0143.6-5844. Right) The optical spectrum of WISE J195945.66-472519.3 associated in the 2LAC with 2FGL J1959.9-4727, a potential blazar-like source. Both spectra have been obtained at Southern Astrophysical Research Telescope (SOAR) with High Throughput Goodman Spectrograph (Figure courtesy of Dr. M. Landoni and Dr. F. Ricci, under the copyright permissions of the Astronomical Journal).}
\label{fig:optspec}
\end{figure}

In practice, the follow-up searches for UGS counterparts often use a multiwavelength approach.  The first step is to look for radio, IR, or X-ray sources with nonthermal characteristics, since many $\gamma$-ray sources have some level of emission at these wavelengths.  Finding one or more candidate sources in these bands provides a far more precise position than that of the original UGS. The next step is to perform optical spectroscopic observations on the candidate objects, if possible.  Spectroscopy can measure or set a limit on redshift, establishing a source as extragalactic.  If these optical observations  also confirm their blazar nature, they can be  included  in the Roma-BZCAT, so as to gain new associations in future releases of the \fer\ catalogs \citep[e.g.,][]{masetti13,paggi14,massaro14,massaro15a,landoni15,ricci15}. Figure \ref{fig:optspec} shows two \fer\ sources for which optical spectroscopic observations have been crucial to determine  the low energy counterparts: the first case is the UGS 2FGL J0143.6-5844, for which a potential counterpart was selected using  IR colors (left panel) while for 2FGL J1959.9-4727 (right panel) the nature of its associated counterpart was unknown leading to an AGU classification in the 2LAC and now they have been confirmed as BL Lac objects.

Almost 1/3 of the sources in each \fer--LAT catalog were unassociated at the time the catalog was released. Looking back at the 1FGL and  2FGL catalogs when the 3FGL was in preparation,  the fraction of UGSs had decreased to 10-20\% (depending on whether sources flagged because of potential issues in the $\gamma$-ray analysis were included).  This improvement indicates the value of follow-up observations for UGSs.  In addition, however, there are 3FGL sources corresponding to 2FGL objects that were not associated in the 2FGL release but now have a low-energy counterpart not because of changes in the comparison catalogs but due to variation in the $\gamma$-ray source density that increased the values of the association probability above the threshold. This result might suggest that the deliberately conservative methods adopted for the source associations could be improved (although at a potential cost of false associations)  using different approaches, such as those used in the radio or in hard X-ray surveys.

\subsection{The ``contamination'' of the extragalactic sky from a  population of Galactic sources}
\label{sec:galcontam}

Not all high-Galactic-latitude astrophysical sources are necessarily extragalactic.  A nearby population of Galactic sources can appear to have a nearly isotropic distribution on the sky.  Although such sources are scientifically interesting in their own right, for purposes of this review they may be considered a ``contamination'' of the extragalactic sky.  Without distance estimates from the $\gamma$-ray observations themselves, we rely on other $\gamma$-ray properties and finding longer-wavelength counterparts to recognize these interlopers. 

The discovery that millisecond pulsars can be bright $\gamma$-ray sources \citep{abdo09g} suggested just such a population.  A  \fer--LAT-affiliated group called the Pulsar Search Consortium carried out successful radio searches of \fer--LAT unassociated source error boxes for new millisecond pulsars \citep{ray12}.   As seen in Fig. 2 of the second \fer--LAT pulsar catalog \cite[2PC;][]{abdo13}, these nearby pulsars are scattered across the sky. These searches were based largely on two observed properties of $\gamma$-ray pulsars: (1) They are persistent sources, not showing the sort of flaring activity seen in the AGN; and (2) Their $\gamma$-ray energy spectra are well described by a power law with an exponential cutoff, a much more curved spectrum than seen for AGN. Using these observables, \citet{lee12} used a Gaussian mixture model to identify a sorted list of possible pulsars among the 2FGL sources.  Such analyses serve as a reminder that there are likely additional Galactic sources in the high-latitude sky.

\subsection{Dark matter constraints from unseen sources}
\label{sec:unseen}

Many models for the DM in the Universe predict $\gamma$-ray emission from interactions of hypothesized DM particles \citep[see e.g.][]{baltz08}.  Searches for $\gamma$ rays from sources thought to be rich in DM can test such predictions.  Two extragalactic source classes illustrate the application of this principle. 

Dwarf spheroidal satellite galaxies to the Milky Way are known to have high concentrations of DM.  They are also unlikely to contain high concentrations of cosmic rays or individual $\gamma$-ray-producing objects, making them ideal candidates for DM searches.  To date there have been no high-confidence indications of $\gamma$ rays from any of these dwarf spheroidals, and their absence in the \fer--LAT data has set strong constraints on possible DM particles with masses less than 100 GeV \citep{ackermann15c}. 

Clusters of galaxies are another reservoir of DM.  Unlike the dwarf spheroidals, galaxy clusters may contain other potential $\gamma$-ray sources, including cosmic-ray interactions and AGN \citep{ackermann14b}.  Like the dwarf spheroidals, however, clusters of galaxies remain undetected as a class in the \fer--LAT data, and their absence also constrains some properties of DM \citep{ackermann10}.

Finally, assuming that the UGS sample is not only constituted by unknown blazars, pulsars, or radio-weak sources, it is possible to obtain tight constraints on DM using the UGS distribution and their $\gamma$-ray spectral properties. The DM halo of the Milky Way is expected to contain a large number of smaller subhalos. These objects could emit $\gamma$ rays via DM annihilations and the nearby and massive ones might be detected by \fer\-LAT as point-like or spatially extended sources, without any low-energy counterpart, as occurs for the UGS.  Searches for such subhalos have been carried out by various authors, \cite[e.g][]{mirabal12}.  Using the cosmological simulations of the Aquarius numerical code Berlin \& Hooper (2013) and Bertoni, Hooper, \& Linden (2015) predicted the distribution of nearby DM subhalos and compared it to selected samples of \fer\ UGS. This comparison allowed them to derive limits on the DM annihilation cross section that in the ``mass'' range of DM particles lighter than $\sim$200 GeV is compatible with those derived from observations of dwarf galaxies,  the Galactic Center, or the EGB \citep[see also][]{rando15}.

\section{Summary}
\label{sec:summary}

Gamma-ray astrophysics is now living a golden age that started with the launch of the \fer\ mission in 2008. The discoveries made by \fer\ have been revolutionary for  high-energy astrophysics,  definitively opening a new window on the unexplored high-energy sky.  Our views of both the Galactic and the extragalactic sky significantly changed in recent years thanks to its all-sky survey. In this review we focused on the principal results  for extragalactic sources detected in the MeV-GeV energy range.

The major \fer\ results  presented in this review can be summarized as follows.

Confirming that blazars constitute the largest population of $\gamma$-ray sources, \fer\ showed that:
\begin{itemize}
\item Their $\gamma$-ray spectra appear to be curved;

\item The multifrequency behavior of their $\gamma$-ray flares is not ubiquitous. Only in some cases is the flaring activity also detected quasi-simultaneously at lower frequencies, allowing constraints on the size and the location of the emitting region. Simultaneous multifrequency observations performed at the peak of the $\gamma$-ray activity, when this is the trigger for the follow up campaigns, are not always available and could bias these results.

\item For B0218+357, the temporal delay measured between flares 
during a state of high $\gamma$-ray activity permitted us to detect the first conclusive signatures of a $\gamma$-ray gravitational lens.

\item Statistical studies on the blazar population have been performed with unprecedented significance, given the large number of them detected by \fer. New insights on the blazar sequence as well as significant improvements of the SED blazar modeling have already been obtained.

\item A more precise estimate of the blazar contribution to the EGB has been computed, and  absorption signatures of the EBL on the blazar spectra have been measured. Additional results came from EBL studies, such as the detection of the CGRH, constraints on the intergalactic magnetic field, and the estimate of the Hubble constant. 
\end{itemize}

The large number of follow-up multifrequency observations carried out on the \fer\ blazars also provides new clues on their broad-band behavior. While there is still a lack of a simple optical and/or  X-ray connection to the $\gamma$-ray sky, the relationship between the radio and  $\gamma$-ray emission of blazars has been clearly strengthened by  statistical analyses of a large sample of \fer\ sources,  extending to low frequencies. A new link between the IR and the $\gamma$-ray emission for the whole blazar population has also been  discovered,  related not only to their power but also to their spectral shape. Blazars detected by \fer\ at GeV energies have led to the discovery of new TeV sources, highlighting a valuable synergy with  Cherenkov Telescopes.

Radio galaxies have been confirmed to be the second largest class of $\gamma$-ray extragalactic objects. As a parent population of blazars they share a very similar $\gamma$-ray behavior. Also notable is the discovery of the first spatially extended AGN $\gamma$-ray emission,  from the lobes of Centaurus A.

Current \fer\ observations coupled with other multifrequency data seem to support the idea that the presence of a relativistic jet in RLNLSy1 is in contrast to the paradigm that jet formation occurs in elliptical galaxies. Additional multifrequency data are needed to confirm these results. In addition to RLNLSy1, also two nearby Seyfert 2 galaxies have been detected by \fer: Circinus galaxy and NGC 1068. The origin of the $\gamma$-ray emission from the former may be non-thermal emission arising from radio lobes, while for the latter it could be mostly related to its SFR.

A few nearby, star-forming galaxies detected by \fer, emit $\gamma$ rays due to the interaction of energetic cosmic-ray particles with interstellar matter and photon fields in their SFR.

The discovery that millisecond pulsars can be bright $\gamma$-ray sources at high Galactic latitudes did not simplify the source association tasks. New statistical analyses, in addition to those developed during the EGRET era, have been also carried out to address this problem. There is still a significant fraction of UGS in the $\gamma$-ray sky, but thanks to new methods developed and  synergies with other observatories as well as follow-up observations we are filling the gap.

\section{Future perspectives: what's next?}
\label{sec:future}

\fer\ is still successfully operating,  and since it is continuously scanning the $\gamma$-ray sky, more and more high-energy photons will be collected. Hoping that \fer\ will  have a ``long and prosperous'' future, we highlight some potential investigations that could be addressed in coming years.

New  \fer--LAT products (i.e., Pass 8) have recently been made available to the astronomical community. Thus a new \fer\ catalog is also expected. Improving the association methods to obtain more precise estimates of the fraction of  UGS and the extragalactic sources with a clear association could be one of the first goals for a new catalog. Given the high variability of many \fer\ sources,  \fer\ will be an active participant in the era of time domain astronomy. Extragalactic transient catalogs are expected to be released in the near future as already done with the first   \fer\ All-sky Variability Analysis catalog listing flaring $\gamma$-ray sources while searching for transients in our Galaxy \citep{ackermann13c}.

On the blazar phenomenon, a deep investigation of the X-ray and optical view of the $\gamma$-ray sky is probably the next step. This will help also to highlight new multifequency correlations/trends that could be used to search for counterparts of the remaining UGS. It will be also interesting to verify the presence of other radio-weak BL Lacs. A detailed study of the host galaxies and the environments of the \fer\ BL Lacs and FSRQs is also a promising avenue for study.  Major advances and improvements have been achieved in the \fer\ era both on theoretical and observational aspects of blazars. Some important questions still remain unanswered, such as: where does the $\gamma$-ray emission originate in jets? What is the physical origin of short time scale variability seen in the $\gamma$-rays during flaring states? What causes ``orphan flares''? Are hadronic models really needed to describe  hard $\gamma$-ray spectra of some blazars? Are jet particles accelerated by shocks or magnetic reconnection, and what is the role of turbulence and magnetohydrodynamic instabilities? Are the correlations between jet powers, luminosities and peak frequencies in blazars real or do selection effects play a major role? Do radio quiet BL Lacs really exist and emit $\gamma$-rays.

It will be important to search for additional radio galaxies that can be associated with \fer--LAT UGS or if  a new analysis of the \fer\ sky will reveal new objects that belong to this class. This will help to understand how closely their  $\gamma$-ray emission process is related to blazar-like properties or if there is still something unknown underlying their \fer\ detections. 
In addition to radio galaxies a deeper study of RLNLSy1  will also be crucial in order to clarify the origin of these sources and how they fit in the scenario of jet formation.

Solving the continuing mystery of the UGS remains a high priority.  New methods for associating longer-wavelength sources with both known and future $\gamma$-ray sources will be needed.  The potential for new discoveries remains high. The potential to improve DM constraints using these unknown $\gamma$-ray sources remains plausible.

Finally, the era of the Cherenkov Telescope Array (CTA), the next generation ground-based very-high-energy $\gamma$-ray facility, is approaching, and the CTA mini-array (the precursor of the CTA) should be operative  in 2017$-$2018. Assuming the \fer\ mission will be supported until the advent of CTA, there will be a perfect synergy between these two observatories, providing the low-energy extensions of the source spectra observed at TeV energies. In addition \fer\ is also an all-sky TeV monitor that will lead/motivate follow-up CTA observations of TeV transients and flaring objects, as has already been done with existing TeV telescopes. Combining these two facilities will definitively provide  deeper insight into the non-thermal and ``extreme'' high-energy universe.

\begin{acknowledgements}
This review would not have been possible without the dedicated efforts of scientists, engineers, and technicians who have made the {\it Fermi Gamma-ray Space Telescope} mission so successful. We extend thanks to all those who contributed. Special thanks to Justin Finke, Filippo D'Ammando and Seth Digel for valuable comments on the manuscript. 
F. Massaro wishes to thank M. Ajello, R. D'Abrusco, D. Gasparrini, M. Giroletti, L. Latronico, N. Masetti, A. Paggi, H. Smith and G. Tosti for their support during the last four years spent working on \fer\ blazars.
The work by is supported by the Programma Giovani Ricercatori - Rita Levi Montalcini - Rientro dei Cervelli (2012). This review is also supported by the NASA grants NNX12AO97G and NNX13AP20G. 
Part of this work is based on archival data, software or on-line services provided by the ASI Science Data Center.
This research has made use of data obtained from the high-energy Astrophysics Science Archive
Research Center (HEASARC) provided by NASA's Goddard Space Flight Center; 
the SIMBAD database operated at CDS,
Strasbourg, France; the NASA/IPAC Extragalactic Database
(NED) operated by the Jet Propulsion Laboratory, California
Institute of Technology, under contract with the National Aeronautics and Space Administration.
This publication makes use of data products from the Wide-field Infrared Survey Explorer, 
which is a joint project of the University of California, Los Angeles, and 
the Jet Propulsion Laboratory/California Institute of Technology, 
funded by the National Aeronautics and Space Administration.
TOPCAT\footnote{\underline{http://www.star.bris.ac.uk/$\sim$mbt/topcat/}} 
(Taylor 2005) for the preparation and manipulation of the tabular data and the images.
The Aladin Java applet\footnote{\underline{http://aladin.u-strasbg.fr/aladin.gml}}
was used to create the finding charts reported in this paper (Bonnarel 2000). 
It can be started from the CDS (Strasbourg - France), from the CFA (Harvard - USA), from the ADAC (Tokyo - Japan), 
from the IUCAA (Pune - India), from the UKADC (Cambridge - UK), or from the CADC (Victoria - Canada).
\end{acknowledgements}

\appendix
\section{Abbreviations}
\label{app:abbreviations}

\begin{tabbing}
\hspace{3.5cm} \= \hspace{10cm} \kill
AGN(s)                   \> Active Galactic Nucleus(i) \\
AGU                      \> Active Galaxy of uncertain type \\
BCU                      \> Blazar Candidate of uncertain type \\
CGRO                     \> Compton Gamma Ray Observatory \\
CSS                      \> Compact Steep Spectrum radio source \\
DM                       \> Dark Matter \\
EBL                      \> Extragalactic Gamma-ray Background light \\
EGB                      \> Extragalactic Gamma-ray Background \\
EC                       \> External Compton \\
FR                       \> Fanaroff - Riley radio source class \\
FSRQ                     \> Flat Spectrum Radio Quasar \\
GRB                      \> Gamma-Ray Burst \\
HBL                      \> High frequency peaked BL Lac object \\
HSP                      \> High synchrotron peaked BL Lac object \\
QSO                      \> Quasi Stellar Object (a.k.a quasar) \\
RG                       \> radio galaxy \\
RLNLSy1                  \> Radio Loud Narrow Line Seyfert of type 1 \\
SED                      \> Spectral Energy Distribution \\
SFR                      \> Star Formation Rate \\
SSC                      \> Synchrotron Self-Compton  \\
SSRQ                     \> Steep Spectrum Radio Quasar \\
UGS                      \> Unidentified/Unassociated Gamma-ray Source \\
\> \\
\hspace{3.5cm} \= \hspace{10cm} \kill
1FGL                     \> First Fermi Gamma-ray LAT point source catalog \\
2FGL                     \> Second Fermi Gamma-ray LAT point source catalog \\
3FGL                     \> Third Fermi Gamma-ray LAT point source catalog \\
1LAC                     \> First Fermi LAT AGN catalog \\
2LAC                     \> Second Fermi LAT AGN catalog \\
3LAC                     \> Third Fermi LAT AGN catalog \\
2PC                      \> Second Pulsar LAT catalog \\
3EG                      \> Third EGRET catalog \\
CRATES                   \>  the Combined Radio All-Sky Targeted Eight GHz Survey \\
LBAS                     \> LAT Bright AGN Sample \\
Roma-BZCAT               \> Roma - multifrequency Blazar Catalog \\
WGS                      \> WISE Gamma-ray Strip \\
WIBRaLS                  \> WISE Blazar-like Radio-Loud Source catalog \\
\> \\
\hspace{3.5cm} \= \hspace{10cm} \kill
AGILE                    \> Astrorivelatore Gamma ad Immagini LEggero \\
ATCA                     \> Australia Telescope Compact Array \\
BAT                      \> Burs Alert Telescope \\
CGRH                     \> Cosmic Gamma-ray Horizon \\
CGRO                     \> Compton Gamma-Ray Observatory \\
COS-B                    \> Celestial Observation Satellite - B \\
CTA                      \> Cherenkov Telescope Array \\
EGRET                    \> Energetic Gamma-ray Experiment Telescope \\
FAVA                     \> Fermi All-sky Variability Analysis \\
FoV                      \> Filed of View \\
GASP                     \> GLAST-AGILE Support Program \\
GBM                      \> Gamma-ray Burst Monitor \\
GLAST                    \> Gamma-Ray Large Area Space Telescope \\
HESS                     \> High Energy Stereoscopic System \\
IBIS                     \> Imager on-Board the INTEGRAL Satellite \\
INTEGRAL                 \> INTErnational Gamma-Ray Astrophysics Laboratory \\
IR                       \> Infrared energy range\\
LAT                      \> Large Area Telescope \\
LR                       \> Likelihood Ratio \\
MAGIC                    \> Major Atmospheric Gamma-ray Imaging Cherenkov Telescope \\
MOJAVE                   \> Monitoring Of Jets in Active galactic nuclei with VLBA Experiments \\
OSO-3                    \> Orbiting Solar Observatory - 3 \\
PSF                      \> Point Spread Function \\
SAS-2                    \> Small Astronomy Satellite - 2 \\
SMARTS                   \> Small and Medium Aperture Research Telescope System \\
SOAR                     \> Southern Astrophysical Research Telescope \\
UV                       \> Ultraviolet energy range \\
VERITAS                  \> Very Energetic Radiation Imaging Telescope Array System \\
VLA                      \> Very Large Array \\
VLBI                     \> Very Long Baseline Array \\
VLBI                     \> Very-long-baseline interferometry \\
WEBT                     \> Whole Earth Blazar Telescope \\
\end{tabbing}

\addcontentsline{toc}{section}{References}

\end{document}